\documentclass[5p]{elsarticle}
\usepackage[colorlinks=true]{hyperref}
\usepackage{amsmath}
\usepackage{amsfonts,amssymb}
\usepackage{natbib}
\usepackage{graphicx}
\usepackage{subcaption}
\usepackage{marginnote}
\usepackage{lineno}
\biboptions{authoryear}
\usepackage{multirow}
\usepackage{etoolbox}
\usepackage{amsthm}
\usepackage{longtable}
\usepackage{pdflscape}
\usepackage{etoolbox}
\usepackage{orcidlink}
\usepackage{bbding}
\usepackage[normalem]{ulem}
\usepackage{xcolor}
\usepackage[normalem]{ulem}


\journal{https://arxiv.org/}

\DeclareCaptionFormat{cont}{#1 (continued)#2#3\par}

\begin{document}
\sloppy
\begin{frontmatter}
\title{Federated Learning: A new frontier in the exploration of multi-institutional medical imaging data}

\author[SANO]{Dominika Ciupek\orcidlink{0000-0002-1411-3060}}

\author[SANO,AGH]{Maciej Malawski\orcidlink{0000-0001-6005-0243}}

\author[UVA,SANO]{Tomasz Pieciak\corref{cor1}\orcidlink{0000-0002-7543-3658}\href{https://twitter.com/PieciakTomasz}{\includegraphics[width=0.07in,height=0.07in]{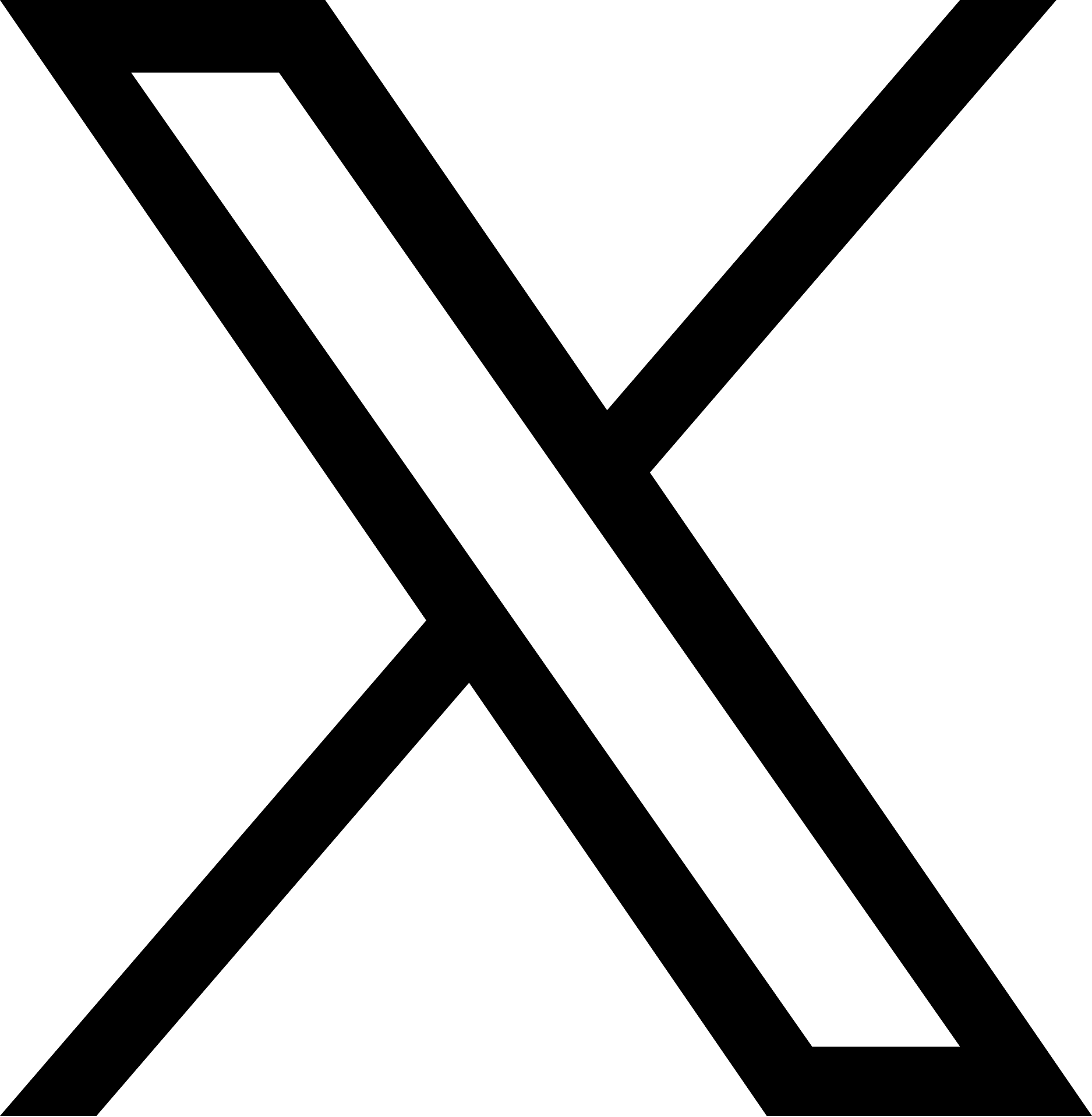}}}
\cortext[cor1]{Corresponding author}
\address[SANO]{Sano Centre for Computational Medicine, Krak\'{o}w, Poland}
\address[AGH]{AGH University of Krakow, Krak\'{o}w, Poland}
\address[UVA]{LPI, ETSI Telecomunicaci\'{o}n, Universidad de Valladolid, Valladolid, Spain}

\begin{abstract}
Artificial intelligence has transformed the perspective of medical imaging, leading to a genuine technological revolution in modern computer-assisted healthcare systems. However, ubiquitously featured deep learning (DL) systems require access to a considerable amount of data, facilitating proper knowledge extraction and generalization. Access to such extensive resources may be hindered due to the time and effort required to convey ethical agreements, set up and carry the acquisition procedures through, and manage the datasets adequately with a particular emphasis on proper anonymization. One of the pivotal challenges in the DL field is data integration from various sources acquired using different hardware vendors, diverse acquisition protocols, experimental setups, and even inter-operator variabilities. In this paper, we review the federated learning (FL) concept that fosters the integration of large-scale heterogeneous datasets from multiple institutions in training DL models. In contrast to a centralized approach, the decentralized FL procedure promotes training DL models while preserving data privacy at each institution involved. We formulate the FL principle and comprehensively review general and specialized medical imaging aggregation and learning algorithms, enabling the generation of a globally generalized model. We meticulously go through the challenges in constructing FL-based systems, such as data and model heterogeneities across the institutions, resilience to potential attacks on data privacy, and the variability in computational and communication resources among the entangled sites that might induce efficiency issues of the entire system. Finally, we explore the up-to-date open frameworks for rapid FL-based algorithm prototyping, comprehensively present real-world implementations of FL systems and shed light on future directions in this intensively growing field.
\end{abstract}

\begin{keyword} federated learning, deep learning, heterogeneous data, model heterogeneity, multi-institutions,  aggregation methods, malicious clients, privacy protection.
\end{keyword}

\end{frontmatter}
\tableofcontents

%% --------------------------------------------------------------------------------------------------
\begin{figure*}[t!]
\centering
\includegraphics[width=1.0\textwidth]{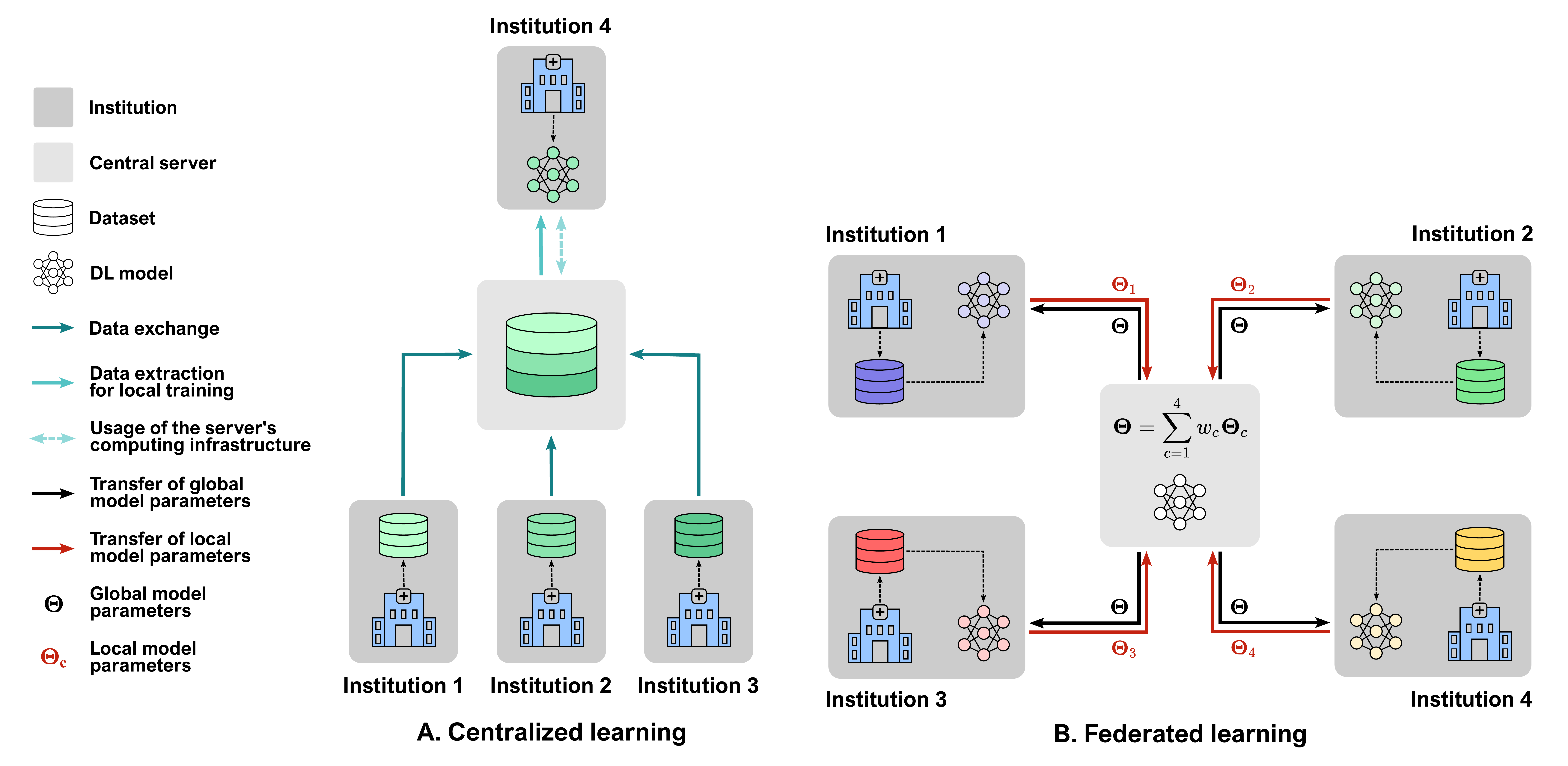}
\caption{Comparison between centralized and federated learning approaches: \textbf{A.} In a centralized architecture, the institutions (here, 1, 2, 3) transfer their local datasets to the central server. Other centers (Institution 4) extract datasets from the global server or use its computing infrastructure to train the DL models. \textbf{B.} Each institution's data remains locally preserved in a federated architecture while the parameters of locally trained models $\mathbf{\Theta}_c$ are transferred to the central server. The central server aggregates received parameters and sends back the parameters of a global model $\mathbf{\Theta}$ to each center.}
\label{fig:MLvsFL}
\end{figure*}

%% --------------------------------------------------------------------------------------------------
%% --------------------------------------------------------------------------------------------------
%% --------------------------------------------------------------------------------------------------
\section{Introduction}
\label{sec:introduction}
Medical imaging (MI) is an intrinsic part of modern healthcare systems and plays an indispensable role in clinical and medical applications nowadays. The versatility of medical image analysis systems and their advancement over the last decades have enabled to inspect, measure and visualize the structure and function of different organs, allowing for early detection, precise diagnosis and prognosis, and meticulous therapy planning \citep{duncan2000medical,bankman2008,frangi2023medical}. A variety of MI techniques and their derivatives founded on different phenomena, such as X-ray imaging, computed tomography (CT), magnetic resonance imaging (MRI) and positron emission tomography (PET), provide comprehensive and complementary information on macro- and microstructure \citep{lerch2017studying,alexander2019imaging,novikov2019quantifying}, as well as functional \citep{frangi2001three,deffieux2021functional}, physiological \citep{cohen2017computational} and molecular characteristics \citep{gambhir2002molecular}, resulting in a better understanding of natural processes that occur in the body \citep{cox2016ageing,Chen2019Functional,bethlehem2022brain}.

Over the last decade, we have been observing an authentic technological shift in medical image analysis and image-guided systems towards machine learning (ML) and predominantly deep learning (DL) techniques, leading to faster and more accurate data reconstruction, processing and inference  \citep{shen2017deep,litjens2017survey,Barragan2021Artificial}. DL-based procedures have already demonstrated extraordinary results in a broad spectrum of imaging protocols and tasks \citep{sahiner2019deep,Wang2020Deep,Ozturk2020Automated,wen2020convolutional,arabi2021promise,isensee2021nnu,sermesant2021applications}, and are considered as the future of robotic-assisted surgeries \citep{li2022machine,varghese2024artificial}. As the DL system can integrate data and improve inference by learning from large-scale observations, training and validating such models requires an extensive amount of diverse and labeled datasets \citep{fan2016mgh,bakas2018identifying,Simpson2019Large,littlejohns2020uk}. These datasets can be collected from multifarious sources, including diverse machine vendors, acquisition protocols, experimental setups, and the preprocessing pipelines employed. Gathering such broad-sided comprehensive datasets is challenging, mainly due to sensitive medical information that requires proper protection \citep{Nelson2015Practical, Rieke2020Future, Olatunji2022Review}. Although meticulous agreements are critical to ensure secure data sharing and adequate use by the community, the datasets must be compiled suitably, including anonymization, which goes beyond merely removing patient-identifying information and may include some interference with the raw data, such as face masking \citep{Nelson2015Practical, Olatunji2022Review,willemink2020preparing}. Therefore, it is unlikely that large and diverse datasets will be available shortly free of charge due to the considerable effort and time required for their collection, labeling and management \citep{Rieke2020Future}.

Federated learning (FL) is a concept that addresses the need to integrate multi-institutional heterogeneous data in training DL models. FL is a decentralized approach, which means that DL models are trained and validated locally without sharing private data between the involved sites \citep{McMahan2017FedAvg, Rieke2020Future, Sheller2020Federated, Li2020FedProx}. In contrast to a traditional centralized procedure that operates on the data collected from different institutions on a single  server, in the FL-based approach, only the characteristics of locally trained models are exported from local collaborators to generate a global model (see Fig.~\ref{fig:MLvsFL}). Developing an FL-based architecture requires carefully selected aggregation and learning strategies for a particular imaging modality and/or DL problem. The aggregation algorithm combines the parameters from locally trained models to yield a single global model \citep{Qi2023Model}. The quality of the new-created model vastly depends on the aggregation strategy chosen \citep{Reddi2020FedAdam, Li2020FedProx, Wicaksana2023FedMix, Hosseini2023PropFFL, Kalapaaking2023Block, Wu2023ModFed}.

Learning techniques determine the overall FL-based process, such as the training strategy \citep{Zhang2022SplitAVG, Feng2023FedMRI, Sanchez2023Memory, Jiang2023IOP-FL, Dalmaz2022OneModel}, information sharing procedure \citep{Guo2021FL-MR, Liu2021FedDG, Zhao2023ADI}, and data augmentation techniques for training phase \citep{Peng2023FedNI}. The appropriate selection of these methods plays a key role in the MI field, significantly affecting the accuracy and versatility of the final global model and its resilience to potential attacks on data privacy \citep{Li2019TumSeg, Feng2023FedMRI, Zhao2023ADI}. The latest results from the use of FL-based models in MI indicate that models trained in a federated approach can compete with those trained in a centralized way \citep{Li2020fMRI, Feki2021X-ray, Lu2022Giga, Dalmaz2022OneModel} and even outperform models trained on separate institutional data \citep{Roth2020BrDen, Bercea2022FedDis, Feng2023FedMRI}.

\begin{figure}[t!]
\centering
\includegraphics[scale=0.12]{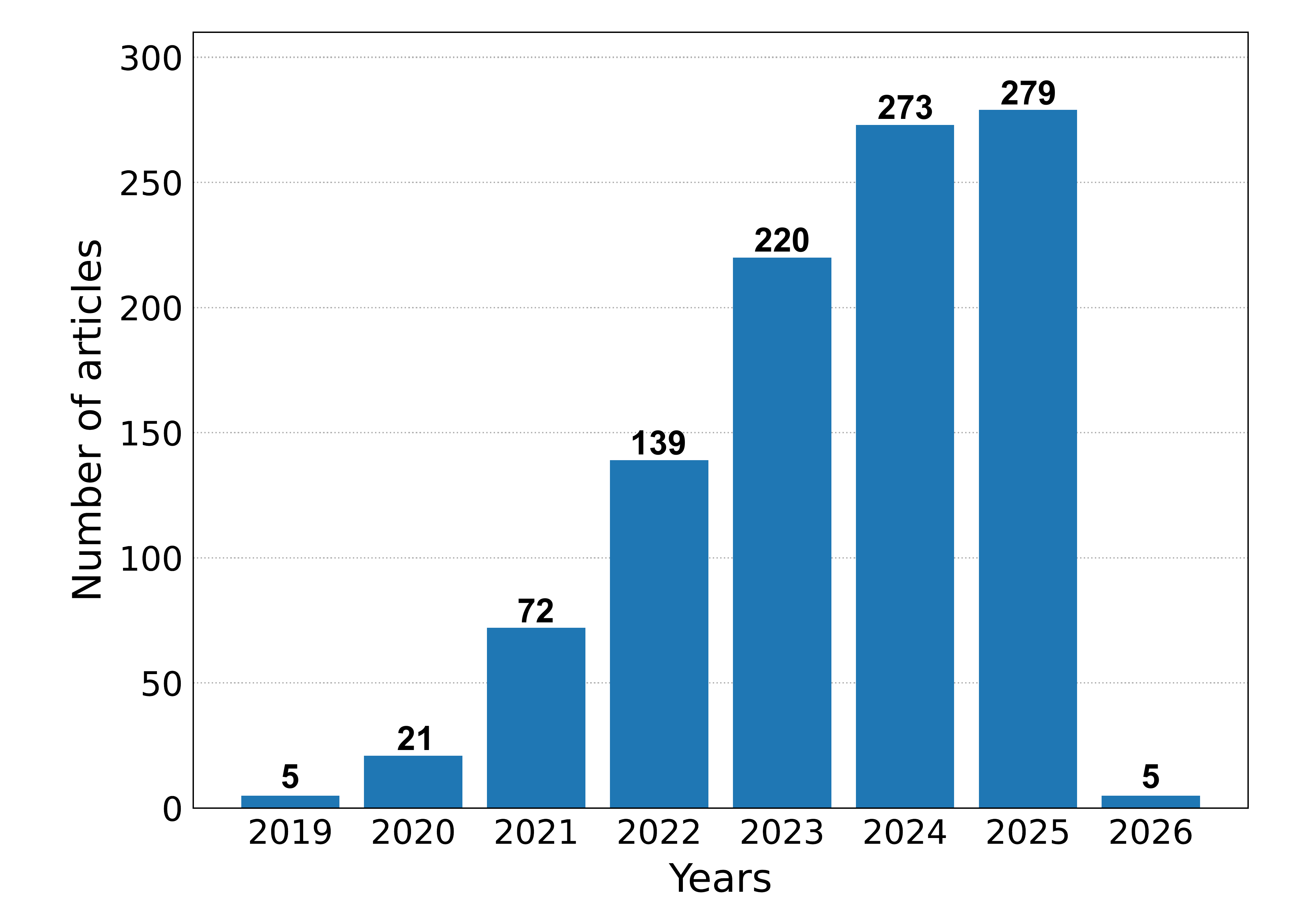}
\caption{The number of screened research articles on applications of federated learning in medical imaging per year. The results were obtained from the MEDLINE and Google Scholar databases and followed the procedure presented in section~\ref{sec:search_strategy}. Data is up to date as of November 2025.}
\label{fig:Data_years}
\end{figure}

However, implementing the FL-based architecture demands further technical advancements to ensure optimal accuracy and versatility without compromising data security. In recent years, there has been immense growth in interest in federated techniques among the MI community (see Fig.~\ref{fig:Data_years}), leading to the development of several unique procedures intended to address specific MI limitations and constraints, such as data heterogeneity \citep{Li2020fMRI, Feng2023FedMRI, Zhao2023ADI}, model heterogeneity \citep{weng2025fedskd, nezhad2025generative} or label deficiency \citep{Wu2021FCL, Peng2023FedNI}.

\subsection{Related works}
\label{sec:related_works}
Several review articles covering the FL concept in the context of MI have been published so far, as summarized in Table \ref{table:reviews}.
Some of them demonstrate the application perspective in a targeted scope, including clinical oncology and cancer research \citep{ Ankolekar2025Advancing, Bechar2025Federated},
radiomics \citep{Guzzo2023Data}, ocular imaging \citep{Nguyen2022FederatedOcular, Lim2024Artificial}, or radiology in general \citep{Rehman2023Federated}. In other reviews, the FL has been exposed as a part of the study with particular emphasis on a specific problem considered in MI such as brain tumor segmentation \citep{Ahamed2023Review}, diagnosis of common retinal diseases \citep{Lim2024Artificial}, or COVID-19 detection \citep{Mondal2023Deep}.
Alternatively, these reviews address specific FL-related challenges, e.g., the security and privacy preservation of medical data \citep{Kaissis2020Secure,Koutsoubis2024PrivacyPreserving},
uncertainty estimation \citep{Koutsoubis2024PrivacyPreserving} or lack of large sets of labeled data \citep{Ng2021Federated}. Other reports present the computational platforms and tools allowing for collaborative data analysis or rapid prototyping of the FL-based systems \citep{Rootes2022Federated,Riviera2023Felebrities}. 

Only a few reports summarize the applications and challenges of FL in MI area without presenting strict mathematical formulations of aggregation and learning methods \citep{Da2023Federated,Sandhu2023Medical,Nazir2023Federated,Guan2024Federated,Bechar2025Federated,Raza2025Federated}. Instead, these reviews present a general summary of FL methods without categorizing them by their dedicated problem-solving capabilities or concentrate solely on the applications of standard FL-based aggregation and learning algorithms to various imaging modalities, body organs, or DL tasks.

%% --------------------------------------------------------------------------------------------------
\subsection{Contributions}
\label{sec:contributions}
This work goes beyond the existing reviews and offers a comprehensive overview of FL principles and the applications of dedicated procedures to various imaging modalities and DL tasks. The paper bridges the gap between previously published articles that superficially present the FL on top of DL-related topics and addresses the challenges of implementing the federated architecture for MI, such as data heterogeneity across the involved sites, the occurrence of inadvisable malicious clients, security and privacy preservation and limited data labeling. Besides, we revise various system architectures and communication efficiency problems in the federated-based systems.
Compared to the previous reviews focusing on FL in MI, we provide a strict theoretical and unified mathematics to illustrate FL aggregation and learning algorithms and follow the formulations across the paper by presenting MI dedicated methodologies. Finally, the article reviews computational frameworks for rapid prototyping of algorithms in FL frameworks and presents future directions in the field.

The researchers can use this review to select appropriate aggregation and learning algorithms to address specific problems while implementing the FL architecture for a MI task. The paper might also serve as a starting point for exploring the prospective development requirements of FL algorithms.

Our contributions can be summarized as follows:
\begin{enumerate}
\item we provide a comprehensive overview of the theoretical aspects of FL, including recent advances in architectures, aggregation and learning approaches, and fundamental techniques used in privacy protection for local data in MI,

\item we give an up-to-date overview of FL applications in the field of MI, with a particular emphasis on different imaging modalities and tasks,

\item we analyze distributions of aggregation strategies and learning methods used, categorized by imaging modality and deep learning tasks,

\item we discuss open issues in FL and recent practices used to address them, particularly in the context of real-world applications,

\item provide an issue–method–effect analysis that illustrates how different aggregation strategies and learning methods address key challenges in federated learning,

\item we provide an overview of open-source tools and frameworks created to facilitate the rapid prototyping of FL-based algorithms,

\item finally, we define still unsolved challenges in FL and possible directions for further development.
\end{enumerate}

%% --------------------------------------------------------------------------------------------------
\subsection{Paper organization}
\label{sec:paper_organization}
In the next section, we define our methodology for collecting the literature used in the study and illustrate the search results from existing databases, focusing on miscellaneous imaging modalities and DL applications. Section~\ref{sec:fl} covers the fundamental concepts of FL, starting with its definition and then smoothly introducing the basic optimization and aggregation techniques, the learning techniques, and the most common data privacy protection methods. Then, section~\ref{sec:broader_overview} delves deeper into FL, presenting various topics and problems that must be adequately clarified before developing a practical architecture. In sections \ref{sec:fl_aggregation} and~\ref{sec:fl_learning}, we present an in-depth analysis of aggregation and learning methods used in MI. Section~\ref{sec:frameworks} reviews actively developed open-source FL frameworks, while section~\ref{sec:real-world} presents examples of deploying federated learning in clinical settings. Finally, section~\ref{sec:discussion} discusses the challenges of FL and proposes strategies to address them, guiding future developments in the field.

%% --------------------------------------------------------------------------------------------------
\section{Research methodology}
\label{sec:research_method}

\subsection{Search strategy}
\label{sec:search_strategy}

The sample comes from two publicly available search engines, namely MEDLINE (\url{https://www.nlm.nih.gov/medline/}) and Google Scholar (\url{https://scholar.google.com/}). Each search query combined two terms: (1) ``federated learning'' and (2) an imaging modality (see Table~\ref{table:search_query} for details). In the case of Google Scholar, we limited each search result to the first ten pages (i.e., 100 results).

\begin{figure}[t!]
\centering
\includegraphics[scale=0.12]{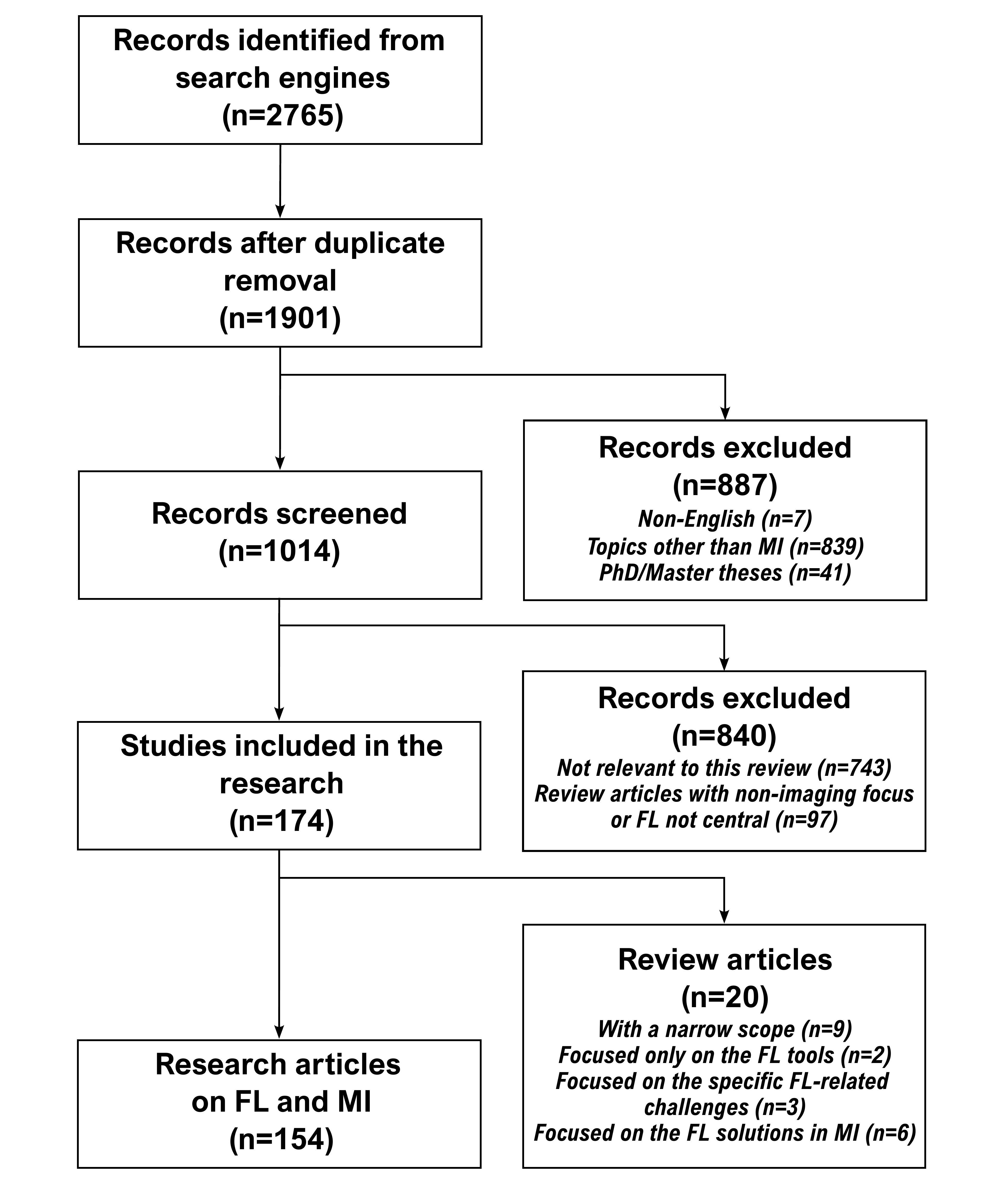}
\caption{The flow chart presents the screening procedure used for collecting the research papers.}
\label{fig:Search_strategy}
\end{figure}

The initial search returned 2765 papers, of which 1901 were screened after removing duplicates. Among these, 887 articles did not meet the inclusion criteria based on their titles and abstracts. Seven papers were not written in English, and the remaining articles were unrelated to FL applications in MI (839) or were Ph.D./fig/Master theses (41). In the next round of screening, we excluded another 840 articles. These papers were either short conference abstracts, unavailable, did not present advances in aggregation or learning techniques, or reviews that were not focused on MI or FL. A total of 174 articles were considered for this study, of which 20 were review papers included in Section \ref{sec:related_works}. The remaining records are the research papers discussed in the main body of this article. Fig.~\ref{fig:Search_strategy} illustrates the flow chart of the screening procedure used in our examination.

%% --------------------------------------------------------------------------------------------------

\begin{landscape}
\begin{table}[!t]
\centering
\scriptsize
\begin{tabular}{p{3.0cm}| p{2.2cm}| p{2.4cm}| p{2.4cm}| p{1.6cm}| p{2.0cm}| p{1.5cm}| p{1.6cm}| p{1.5cm}}
\hline\hline
\multicolumn{1}{c|}{\multirow{2}{*}{\bf{Reference}}} & \multicolumn{1}{c|}{\bf{MI modality}} & \multicolumn{1}{c|}{\bf{Mathematical}} & \multicolumn{1}{c|}{\bf{Dedicated to}} & \multicolumn{1}{c|}{\bf{Technical}} & \multicolumn{1}{c|}{\bf{FL challenges}} & \multicolumn{1}{c|}{\bf{FL challenges}} & \multicolumn{1}{c|}{\bf{FL frameworks}} & \multicolumn{1}{c}{\bf{Real-world}} \\
 & \multicolumn{1}{c|}{\bf{coverage}} & \multicolumn{1}{c|}{\bf{principles}} & \multicolumn{1}{c|}{\bf{MI algorithms}} & \multicolumn{1}{c|}{\bf{presentation}} & \multicolumn{1}{c|}{\bf{presentation}} & \multicolumn{1}{c|}{\bf{solutions}} & \multicolumn{1}{c|}{\bf{presentation}} & \multicolumn{1}{c}{\bf{examples}} \\ \hline

\cite{Kaissis2020Secure} & Multiple & \XSolidBrush & \XSolidBrush & \XSolidBrush & Narrow & \XSolidBrush & \XSolidBrush & \XSolidBrush \\\hline

\cite{Ng2021Federated} & MRI & \CheckmarkBold & \XSolidBrush & Descriptive & Comprehensive & \XSolidBrush & \XSolidBrush & \XSolidBrush \\\hline

\cite{Darzidehkalani2022FederatedI} & Multiple & \CheckmarkBold & \XSolidBrush & \XSolidBrush & \XSolidBrush & \XSolidBrush & \XSolidBrush & \XSolidBrush \\\hline

\cite{Nguyen2022FederatedOcular} & OCT,\newline retinal images & \CheckmarkBold & \CheckmarkBold & Descriptive & Comprehensive & \XSolidBrush & \XSolidBrush & \XSolidBrush \\\hline

\cite{Rootes2022Federated} & MRI & \XSolidBrush & \XSolidBrush & \XSolidBrush & \XSolidBrush & \XSolidBrush & \CheckmarkBold & \XSolidBrush \\\hline

\cite{Ahamed2023Review} & MRI & \CheckmarkBold & \XSolidBrush & \XSolidBrush & \XSolidBrush & \XSolidBrush & \XSolidBrush & \XSolidBrush \\\hline

\cite{Guzzo2023Data} & Multiple & \CheckmarkBold & \CheckmarkBold & Descriptive & Comprehensive & \CheckmarkBold & \XSolidBrush & \XSolidBrush \\\hline

\cite{Mondal2023Deep} & Multiple & \XSolidBrush & \CheckmarkBold & Descriptive & Comprehensive & \CheckmarkBold & \XSolidBrush & \XSolidBrush \\\hline

\cite{Nazir2023Federated} & Multiple & \CheckmarkBold & \CheckmarkBold & Descriptive & Comprehensive & \CheckmarkBold & \XSolidBrush & \XSolidBrush \\\hline

\cite{Rehman2023Federated} & Multiple & \XSolidBrush & \CheckmarkBold & Detailed & Comprehensive & \CheckmarkBold & \XSolidBrush & \XSolidBrush \\\hline

\cite{Riviera2023Felebrities} & \XSolidBrush & \CheckmarkBold & \XSolidBrush & \XSolidBrush & Comprehensive & \XSolidBrush & \CheckmarkBold & \XSolidBrush \\\hline

\cite{Sandhu2023Medical} & Multiple & \CheckmarkBold & \CheckmarkBold & Descriptive & Comprehensive & \CheckmarkBold & \XSolidBrush & \XSolidBrush \\\hline

\cite{Da2023Federated} & Multiple & \CheckmarkBold & \CheckmarkBold & Descriptive & Comprehensive & \CheckmarkBold & \XSolidBrush & \XSolidBrush \\\hline

\cite{Sohan2023Systematic} & Multiple & \CheckmarkBold & \XSolidBrush & \XSolidBrush & Comprehensive & \XSolidBrush & \XSolidBrush & \XSolidBrush \\\hline

\cite{Guan2024Federated} & Multiple & \CheckmarkBold & \CheckmarkBold & Detailed & Comprehensive & \CheckmarkBold & \CheckmarkBold & \XSolidBrush \\\hline

\cite{Koutsoubis2024PrivacyPreserving} & Multiple & \CheckmarkBold & \CheckmarkBold & Detailed & Narrow & \CheckmarkBold & \CheckmarkBold & Mentioned \\\hline

\cite{Lim2024Artificial} & OCT,\newline retinal images & \XSolidBrush & \CheckmarkBold & \XSolidBrush & Comprehensive & \XSolidBrush & \XSolidBrush & \XSolidBrush \\\hline

\cite{Ankolekar2025Advancing} & Multiple & \XSolidBrush & \CheckmarkBold & Descriptive & Comprehensive & \XSolidBrush & \XSolidBrush & \XSolidBrush \\\hline

\cite{Bechar2025Federated} & Multiple & \CheckmarkBold & \CheckmarkBold & Descriptive & Comprehensive & \CheckmarkBold & \CheckmarkBold & \XSolidBrush \\\hline

\cite{Raza2025Federated} & Multiple & \CheckmarkBold & \CheckmarkBold & Descriptive & Comprehensive & \CheckmarkBold & \XSolidBrush & \XSolidBrush \\\hline\hline

This review & Multiple & \CheckmarkBold & \CheckmarkBold & Detailed & Comprehensive & \CheckmarkBold & \CheckmarkBold & \CheckmarkBold \\

\hline\hline
\end{tabular}
\caption{Summary of review articles on the application of federated learning in medical imaging. Columns explanations: \textbf{(2) MI modality coverage} -- number of medical imaging modalities discussed (\emph{multiple} -- more than two); \textbf{(3) Mathematical principles} -- does the article coverage mathematical and algorithmic foundations of federated learning?; \textbf{(4) Dedicated to MI algorithms} -- does the article focus on federated learning methods tailored to medical imaging?; \textbf{(5) Technical presentation} – level of technical presentation (\emph{descriptive} -- high-level discussion without mathematical detail, \emph{detailed} -- covers mathematical formulations of the methods); \textbf{(6) FL challenges presentation} -- scope and depth of discussion of federated learning challenges such as data heterogeneity or malicious clients (\emph{narrow} -- superficial discussion on challenges only, \emph{comprehensive} -- coverage of multiple challenges in detail); \textbf{(7) FL challenges solutions} -- does the article present discussion on the methods addressing specific federated learning challenges?; \textbf{(8) FL frameworks presentation} -- does the article present the frameworks supporting the implementation of federated learning?; \textbf{(9) Real-world examples} -- does the article discuss real-world deployment of federated learning in clinics? (\emph{mentioned} -- examples are only mentioned without details).}
\label{table:reviews}
\color{black}
\end{table}
\end{landscape}

\begin{table*}[t!]
\centering
\footnotesize
\begin{tabular}{p{9cm}| p{2.5cm}| p{2.5cm}| p{2cm}}
\hline\hline
\multicolumn{1}{c|}{\multirow{2}{*}{\bf{Search query}}} & \multicolumn{1}{c|}{\multirow{2}{*}{\bf{MEDLINE}}} & \multicolumn{1}{c|}{\multirow{2}{*}{\bf{Google Scholar}}} & \multicolumn{1}{c}{\bf{Total w/o}} \\
& & & \multicolumn{1}{c}{\bf{duplicates}}  \\
\hline
``federated learning"  \texttt{and}  ``medical imaging" & 153 & 100 & 226 \\\hline
``federated learning"  \texttt{and}  ``neuroimaging" & 38 & 100 & 126 \\\hline
``federated learning"  \texttt{and}  ``MRI" & 84 & 98 & 165 \\\hline
``federated learning"  \texttt{and}  ``magnetic resonance imaging" & 88 & 95 & 165 \\\hline
``federated learning"  \texttt{and}  ``CT" & 79 & 100 & 165 \\\hline
``federated learning"  \texttt{and}  ``computed tomography" & 46 & 100 & 135 \\\hline
``federated learning"  \texttt{and}  ``tomography" & 66 & 100 & 151 \\\hline
``federated learning"  \texttt{and}  ``PET" & 19 & 100 & 116 \\\hline
``federated learning"  \texttt{and}  ``positron emission tomography" & 12 & 100 & 108 \\\hline
``federated learning"  \texttt{and}  ``x-ray" & 61 & 100 & 144 \\\hline
``federated learning"  \texttt{and}  (``histopathology"   \texttt{or}   ``histopathological") & 22 & 100 & 113 \\\hline
``federated learning"  \texttt{and}  (``dermatoscopy"   \texttt{or}   ``dermatoscopic") & 0 & 100 & 100 \\\hline
``federated learning"  \texttt{and}  ``retinal" & 25 & 100 & 115 \\\hline
``federated learning"  \texttt{and}  ``OCT" & 235 & 100 & 235 \\\hline
``federated learning"  \texttt{and}  ``optical coherence tomography" & 13 & 100 & 106 \\\hline
``federated learning"  \texttt{and}  ``ultrasound" & 23 & 100 & 122 \\\hline
``federated learning"  \texttt{and}  ``mammography" & 6 & 100 & 102 \\\hline
``federated learning"  \texttt{and}  ``microscopic" & 2 & 100 & 100 \\
\hline\hline
\end{tabular}
\caption{The list of queries used in MEDLINE and Google Scholar engines to retrieve the preliminary list of research articles related to federated learning in medical imaging. The results from Google Scholar were limited to the first 100 entries from each query. Data is up to date as of November 2025.}
\label{table:search_query}
\end{table*}

\begin{figure*}[t!]
\centering
\includegraphics[scale=0.115]{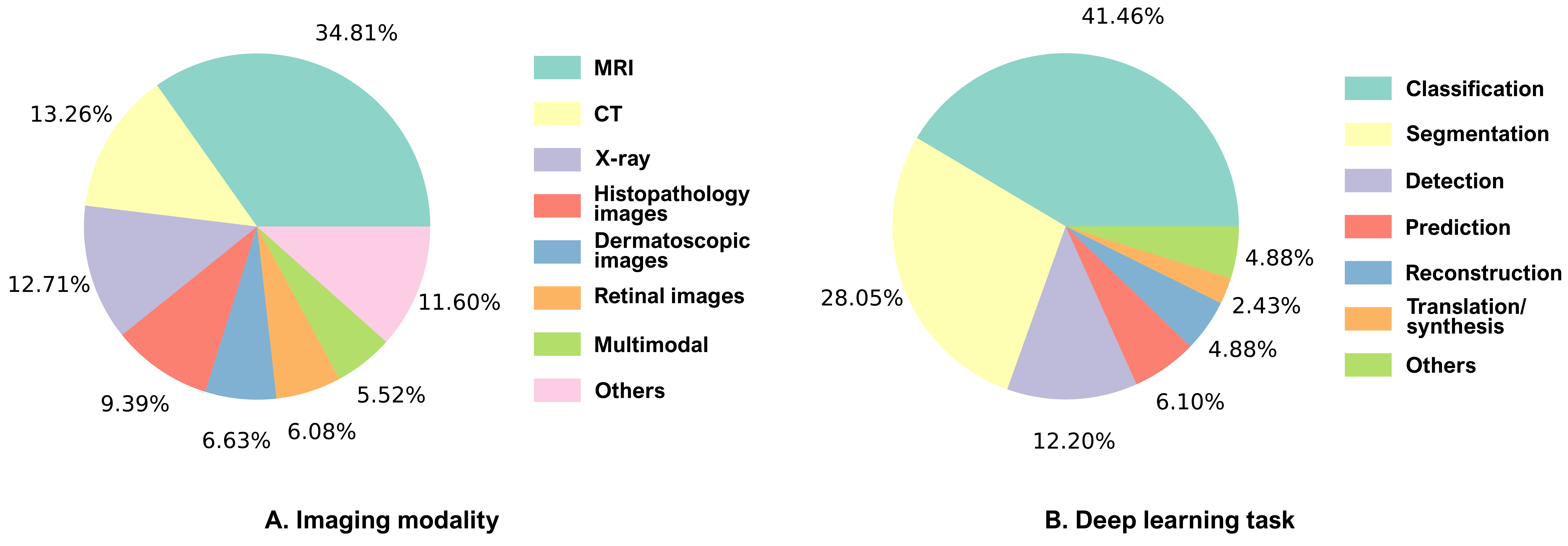}
\caption{\textbf{A.} Distribution of imaging modalities and \textbf{B.} deep learning tasks in the examined federated learning research papers in this review ($n=154$).}

\label{fig:Data}
\end{figure*}

\begin{figure*}[t!]
\centering
\includegraphics[scale=0.115]{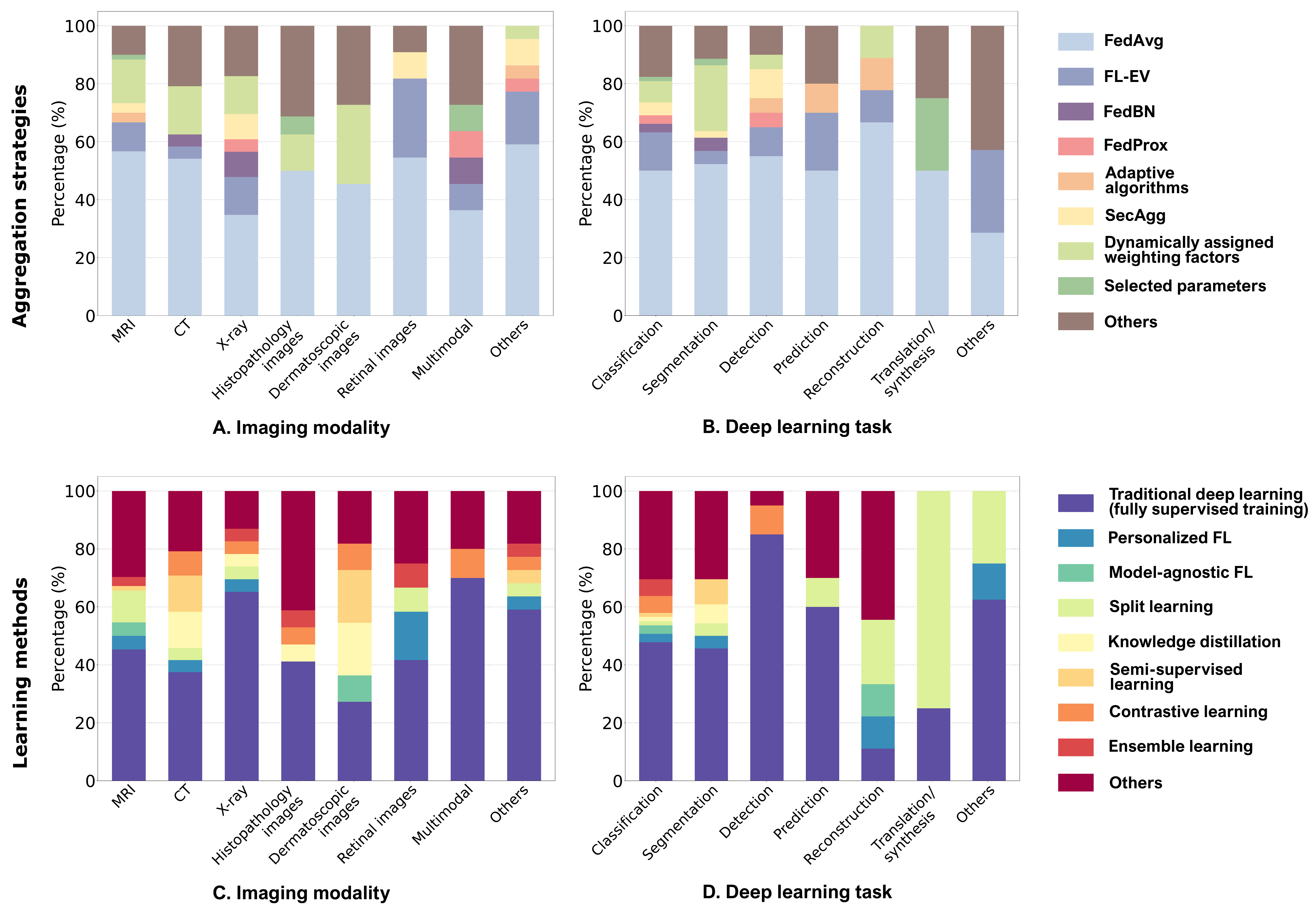}
\caption{\textbf{A,B} Distribution of aggregation strategies and \textbf{C,D} learning methods, categorized by imaging modality (left diagrams) and deep learning task (right diagrams) across $n=154$ FL-based research articles considered in this study.}
\label{fig:Data_task_aggregation}
\end{figure*}

%% ----------------------------------------------
\subsection{Descriptive statistical analysis}
\label{sec:stat_analysis}

The review integrates applications of FL with a wide range of MI modalities, particularly 63 papers related to MRI, 24 to CT, 23 to X-ray imaging, and 17 articles linked to histopathological imaging.
In particular, 10 articles examined multimodal imaging techniques that involve CT and MRI, CT and PET, CT and X-ray, MRI and PET, ultrasonography and mammography, or CT, MRI, ultrasonography and histopathological images. The remaining articles covers PET, OCT, mammography, dermatoscopic, retinal, endoscopic and microscopy imaging. Fig.~\ref{fig:Data}A. provides an exact distribution of articles involving specific topics.
Fig.~\ref{fig:Data}B shows a distribution of the papers examined according to the task they address. 
Among them, 68 articles focus on classification, 46 on segmentation, and 20 on detection. The rest of the papers broadly address prediction, prognosis, reconstruction, translation, restoration, normalization, denoising, screening, and super-resolution, with a few considering multiple tasks. Figure~\ref{fig:Data_task_aggregation} summarizes the most commonly used aggregation and learning methods, categorized by imaging modality and deep learning task in the reviewed articles. Notably, the simplest and most established aggregation technique, Federated Averaging (FedAvg), is commonly employed across all modalities and deep learning tasks considered in this review. Considering the learning procedure, the traditional deep learning approaches remain predominant across modalities and medical imaging tasks. A detailed description of aggregation and learning techniques can be found in sections \ref{sec:optimization_and_aggregation}, \ref{sec:fl_aggregation} and \ref{sec:fl_learning}. Additionally, Table~\ref{table:Tasks_FL} presents a comparative and detailed information gathered from the evaluated articles, including tasks, employed datasets with references, the presence of heterogeneous data, and the number of clients involved in the analysis.

%---------------------------- BIG TABLE:   PART I ----------------------------
\begin{table*}[t!]
\centering
\footnotesize
\begin{tabular}{p{3.3cm}| p{2.0cm}| p{7.1cm}| p{2.3cm}| p{1.5cm}}
\hline\hline
\multicolumn{1}{c|}{\multirow{2}{*}{\bf{Reference}}} & \multicolumn{1}{c|}{\multirow{2}{*}{\bf{Task}}} & \multicolumn{1}{c|}{\multirow{2}{*}{\bf{Used datasets}}} & \multicolumn{1}{c|}{\bf{Heterogeneous}} & \multicolumn{1}{c}{\bf{Number}} \\
 & & & \multicolumn{1}{c|}{\bf{data}} & \multicolumn{1}{c}{\bf{of clients}} \\
\hline

\multicolumn{5}{c}{\bf{Magnetic resonance imaging}} \\ \hline

\cite{Roy2019BrainTorrent} & Segmentation & MALC \citep{Landman2012Miccai} & N/A & 5, 7, 10, 20 \\
\cite{Li2019TumSeg} & Segmentation & BraTS2018 \citep{Menze2014Brats} & Yes & 13 \\
\cite{Parekh2021Cross-domain} & Segmentation & BraTS2017 \citep{Menze2014Brats}, own dataset & Yes & 2 \\
\cite{Roth2021Prost} & Segmentation & Medical Segmentation Decathlon \citep{Simpson2019Large}, PROMISE12 \citep{Litjens2014Evaluation}, NCI-ISBI2013, ProstateX \citep{Litjens2014Computer} & Yes & 4 \\
\cite{Tian2021Privacy} & Segmentation & Spinal Cord Grey Matter Segmentation Challenge dataset \citep{Prados2017SpinalCord} & Yes & 3 \\
\cite{Liu2021FedDG} & Segmentation & PROMISE12 \citep{Litjens2014Evaluation}, I2CVB \citep{Lemaitre2015Computer}, NCI-ISBI2013, dataset from article \citep{Liu2020MS-Net} & Yes & 6 \\
\cite{Wu2021FCL} & Segmentation & ACDC \citep{Bernard2018DeepLT} & Yes & 10 \\
\cite{Sarma2021Federated} & Segmentation & SPIE-AAPM-NCI PROSTATEx \citep{Armato2018PROSTATEx}, own dataset & Yes & 3 \\
\cite{Khan2022SimAgg} & Segmentation & FeTS2021 \citep{Pati2021Federated} & Yes & 17, 22 \\
\cite{Machler2022FedCostWAvg} & Segmentation & FeTS2021 \citep{Pati2021Federated} & Yes & 17 \\
\cite{He2022CosSGD} & Segmentation & BraTS2018 \citep{Menze2014Brats} & Yes & 10 \\
\cite{Zhang2022SplitAVG} & Segmentation & BraTS2017 \citep{Menze2014Brats} & Yes & 10 \\
\cite{Yang2022RosFL} & Segmentation & ACDC \citep{Bernard2018DeepLT} & Yes & 4 \\
\cite{Bercea2022FedDis} & Segmentation & OASIS-3 \citep{LaMontagne2018OASIS3LN}, ADNI-3 \citep{Weiner2017Alzheimer}, MSLUB \citep{Lesjak2017ANP}, MSISBI \citep{Carass2017LongitudinalMS}, own dataset & Yes & N/A \\
\cite{Huang2022Continual} & Segmentation & Dataset from articles \citep{Putz2020FSRTVS, Oft2021Volumetric} & Yes & 2, 7 \\
\cite{Khan2023RegSimAgg} & Segmentation & FeTS2022 \citep{Pati2021Federated} & Yes & 23, 33 \\
\cite{Machler2023FedPIDAvg} & Segmentation & FeTS2022 \citep{Pati2021Federated} & Yes & N/A \\
\cite{Mushtaq2023FAT} & Segmentation & FeTS2021 \citep{Pati2021Federated} & Yes & 13 \\
\cite{Liu2023MultSkle} & Segmentation & MSSEG2016 \citep{Commowick2018Objective}, own dataset & Yes & 3, 4 \\
\cite{Jiang2023IOP-FL} & Segmentation & NCI-ISBI2013, PROMISE12 \citep{Litjens2014Evaluation}, I2CVB \citep{Lemaitre2015Computer}, dataset from article \citep{Liu2020MS-Net} & Yes & 6 \\
\cite{Yang2023DynAggr} & Segmentation & PROMISE12 \citep{Litjens2014Evaluation}, I2CVB \citep{Lemaitre2015Computer}, NCI-ISBI2013 & Yes & 6 \\
\cite{Kumar2024Privacy} & Segmentation & BraTS2020 \citep{Menze2014Brats} & Yes & 10 \\
\cite{Liang2024ACFL} & Segmentation & Dataset from articles \citep{Liu2020MS-Net, Liu2020Shape} & Yes & 6 \\

\cite{cheng2025federated} & Segmentation & Dataset from article \citep{kuijf2019standardized}, own dataset & Yes & 7 \\
\cite{narmadha2025enhanced} & Segmentation & Brain MRI segmentation, Brain Tumor Segmentation datasets & N/A & 5, 7, 10 \\

\cite{Sheller2020Federated} & Classification & BraTS2017 \citep{Menze2014Brats}, own dataset & Yes & 10 \\
\cite{Li2020fMRI} & Classification & ABIDE \citep{DiMartino2014Autism} & N/A & 4 \\
\cite{Huang2021CML} & Classification & ADNI \citep{Jack2008Alzheimer}, OASIS-1 \citep{Marcus2007Open}, AIBL \citep{Ellis2009Australian} & Yes & 3 \\
\cite{Kumar2022DataVal} & Classification & BraTS2019 \citep{Menze2014Brats}, IXI & Yes & 4 \\
\cite{Linardos2022FL-EV} & Classification & M\&M \citep{Campello2021Multi}, ACDC \citep{Bernard2018DeepLT} & Yes & 4 \\
\cite{Islam2022Effectiveness} & Classification & UK Data Service brain tumor dataset \citep{Pernet2016Structural} & No & 50 \\
\cite{Zhao2023ADI} & Classification & Alzheimer MRI dataset & N/A & 3 \\
\cite{Rajagopal2023Federated} & Classification & Own dataset & Yes & 2 \\
\cite{Zhou2024DistributedFL} & Classification & Brain Tumor Classification (MRI) dataset & N/A & 5 \\
\cite{Gupta2024Enhancing} & Classification & ABIDE \citep{DiMartino2014Autism} & N/A & 10 \\
\cite{Albalawi2024Integrated} & Classification & Brain Tumor MRI dataset & Yes & 10 \\

\cite{li2025fedcmc} & Classification & Own dataset & Yes & 3 \\

\hline\hline

\multicolumn{5}{l}{NCI-ISBI2013 dataset: \url{https://www.cancerimagingarchive.net/analysis-result/isbi-mr-prostate-2013/}} \\
\multicolumn{5}{l}{IXI dataset: \url{https://brain-development.org/ixi-dataset/}} \\
\multicolumn{5}{l}{Brain MRI segmentation dataset: \url{https://www.kaggle.com/datasets/mateuszbuda/lgg-mri-segmentation}} \\
\multicolumn{5}{l}{Brain Tumor Segmentation dataset: \url{https://www.kaggle.com/datasets/nikhilroxtomar/brain-tumor-segmentation}} \\
\multicolumn{5}{l}{Alzheimer MRI dataset: \url{https://www.kaggle.com/datasets/legendahmed/alzheimermridataset}} \\
\multicolumn{5}{l}{Brain Tumor Classification (MRI) dataset: \url{https://www.kaggle.com/datasets/sartajbhuvaji/brain-tumor-classification-mri}} \\
\multicolumn{5}{l}{Brain Tumor MRI dataset: \url{https://www.kaggle.com/datasets/masoudnickparvar/brain-tumor-mri-dataset}} \\

\end{tabular}
\caption{Summary of federated learning frameworks employed in various medical imaging tasks and imaging modalities. The column ``Heterogeneous data'' indicates whether heterogeneous data were used in the aggregation procedure. The last column refers to the number of clients involved in the deep learning task. All links were verified as of December 19, 2025.}
\label{table:Tasks_FL}
\end{table*}

%---------------------------- BIG TABLE:   PART II ----------------------------
\begin{table*}[t!]
\ContinuedFloat
\captionsetup{list=off,format=cont}
\centering
\footnotesize
\begin{tabular}{p{3.3cm}| p{2.0cm}| p{7.1cm}| p{2.3cm}| p{1.5cm}}
\hline\hline
\multicolumn{1}{c|}{\multirow{2}{*}{\bf{Reference}}} & \multicolumn{1}{c|}{\multirow{2}{*}{\bf{Task}}} & \multicolumn{1}{c|}{\multirow{2}{*}{\bf{Used datasets}}} & \multicolumn{1}{c|}{\bf{Heterogeneous}} & \multicolumn{1}{c}{\bf{Number}} \\
 & & & \multicolumn{1}{c|}{\bf{data}} & \multicolumn{1}{c}{\bf{of clients}} \\
\hline

\cite{weng2025fedskd} & Classification & ABIDE \citep{DiMartino2014Autism} & N/A & 4 \\
\cite{song2025developing} & Classification & ADNI \citep{Jack2008Alzheimer}, AIBL \citep{Ellis2009Australian}, own dataset & Yes & 3 \\
\cite{zhao2025federated} & Classification & REST-meta-MDD \citep{yan2019reduced} & Yes & 4 \\
\cite{onaizah2025fl} & Classification & Dataset from \cite{cheng2015enhanced} & N/A & 10, 20, 30, 50 \\
\cite{appasami2025federated} & Classification & Own dataset & N/A & 4 \\

\cite{Casella2023Merge} & Detection & ADNI \citep{Jack2008Alzheimer} & Yes & 3 \\

\cite{Pati2022BigData} & Detection & Own dataset & Yes & 71 \\
\cite{Mitrovska2024Secure} & Detection & ADNI \citep{Jack2008Alzheimer} & Yes & 2, 3, 4 \\
\cite{Yamada2024Investigation} & Detection & OpenBTAI \citep{Ocana2023Comprehensive}, own datasets & Yes & 3 \\

\cite{Stripelis2021Scali} & Prediction & UK Biobank \citep{Miller2016Multimodal} & No & 8 \\
\cite{Stripelis2021SecureNeuro} & Prediction & UK Biobank \citep{Miller2016Multimodal} & No & 8 \\

\cite{Stripelis2022Secure} & Prediction & ADNI \citep{Jack2008Alzheimer}, OASIS-3 \citep{LaMontagne2018OASIS3LN}, AIBL \citep{Ellis2009Australian} & Yes & 3, 4, 5 \\

\cite{Peng2023FedNI} & Prediction & ABIDE \citep{DiMartino2014Autism}, ADNI \citep{Jack2008Alzheimer} & Yes & 5 \\
\cite{Denissen2023Transf} & Prediction & Own dataset & N/A & 3 \\
\cite{Guo2021FL-MR} & Reconstruction & fastMRI \citep{Knoll2020fastMRI}, BraTS \citep{Menze2014Brats}, IXI, HPKS \citep{Jiang2019Identifying} & Yes & 3, 4 \\
\cite{Feng2023FedMRI} & Reconstruction & fastMRI \citep{Knoll2020fastMRI}, BraTS \citep{Menze2014Brats}, SMS \citep{Feng2021Multi}, uMR \citep{Feng2021Multi} & Yes & 4 \\
\cite{Elmas2023FedGIMP} & Reconstruction & fastMRI \citep{Knoll2020fastMRI}, BraTS \citep{Menze2014Brats}, IXI, own dataset & Yes & 3 \\
\cite{Wu2023ModFed} & Reconstruction & fastMRI \citep{Knoll2020fastMRI}, CC359 \citep{Beauferris2021MultiCoil}, own dataset & Yes & 3 \\
\cite{Levac2023EndToEnd} & Reconstruction & fastMRI \citep{Knoll2020fastMRI}, Stanford dataset, NYU knee dataset & Yes & 10 \\
\cite{Yan2024Cross} & Reconstruction & fastMRI \citep{Knoll2020fastMRI}, BraTS \citep{Menze2014Brats}, own dataset & Yes & 2, 3, 6 \\
\cite{nezhad2025generative} & Reconstruction & fastMRI \citep{Knoll2020fastMRI}, dataset from  \cite{Elmas2023FedGIMP, souza2018open} & Yes & 3 \\
\cite{ahmed2025fedgraphmri} & Reconstruction & fastMRI \citep{Knoll2020fastMRI}, IXI & Yes & 3, 5, 10 \\

\cite{Dalmaz2022MRITrans} & Translation & BraTS \citep{Menze2014Brats}, IXI, MIDAS \citep{Bullitt2005Vessel}, OASIS-3 \citep{LaMontagne2018OASIS3LN} & Yes & 4 \\
\cite{fiszer2025validation} & Translation & BraTS \citep{Menze2014Brats}, OASIS-3 \citep{LaMontagne2018OASIS3LN}, HCP WU-Minn \citep{Van2013Wu}, HCP MGH \citep{fan2016mgh}, UCSF-PDGM-v3 \citep{calabrese2022university} & Yes & 6 \\

\cite{Dalmaz2022OneModel} & Synthesis & BraTS \citep{Menze2014Brats}, IXI, MIDAS \citep{Bullitt2005Vessel}, OASIS-3 \citep{LaMontagne2018OASIS3LN} & Yes & 4 \\ 

\cite{Liu2024Multi} & Super-resolution & MPMRI \citep{Chen2020Trusted}, ADNI \citep{Jack2008Alzheimer}, Own dataset & Yes & 4 \\ 

\hline

\multicolumn{5}{c}{\bf{Computed tomography}} \\ \hline
\cite{Wang2020Autom} & Segmentation & Own dataset & Yes & 2 \\
\cite{Shen2021DWA} & Segmentation & TCIA Pancreas-CT \citep{Roth2015Deeporgan}, Medical Segmentation Decathlon \citep{Simpson2019Large}, BTCV Abdomen CT \citep{Landman2015MICCAI} & Yes & 3 \\
\cite{Ziller2021Differentially} & Segmentation & Medical Segmentation Decathlon \citep{Simpson2019Large} & Yes & 3 \\
\cite{Yang2022FedZaCt} & Segmentation & Own dataset & N/A & 4 \\
\cite{Guo2022Auto-FedRL} & Segmentation & Medical Segmentation Decathlon \citep{Simpson2019Large}, TCIA Pancreas-CT \citep{Roth2015Deeporgan}, BTCV Abdomen CT \citep{Landman2015MICCAI}, TCIA Images in COVID-19 \citep{Harmon2020Artificial} & Yes & 3 \\
\cite{Misonne2022HeartSeg} & Segmentation & NSCLC-Radiomics \citep{Aerts2014Decoding}, Pediatric-CT-SEG \citep{Jordan2022Pediatric}, LCTSC2017 \citep{Yang2018Autosegmentation} & Yes & 3 \\

\cite{Kanhere2023SegViz} & Segmentation & Medical Segmentation Decathlon \citep{Simpson2019Large}, BTCV Abdomen CT \citep{Landman2015MICCAI}, KiTS19 \citep{Heller2020Kits19} & Yes & 4 \\

\hline\hline

\multicolumn{5}{l}{Stanford dataset: \url{http://mridata.org/}} \\
\multicolumn{5}{l}{NYU knee dataset: \url{https://fastmri.med.nyu.edu/}} \\

\end{tabular}
\caption{Part II.}
\end{table*}

%---------------------------- BIG TABLE:   PART III ----------------------------
\begin{table*}[t!]
\ContinuedFloat
\captionsetup{list=off,format=cont}
\centering
\footnotesize
\begin{tabular}{p{3.3cm}| p{2.0cm}| p{7.1cm}| p{2.3cm}| p{1.5cm}}
\hline\hline
\multicolumn{1}{c|}{\multirow{2}{*}{\bf{Reference}}} & \multicolumn{1}{c|}{\multirow{2}{*}{\bf{Task}}} & \multicolumn{1}{c|}{\multirow{2}{*}{\bf{Used datasets}}} & \multicolumn{1}{c|}{\bf{Heterogeneous}} & \multicolumn{1}{c}{\bf{Number}} \\
 & & & \multicolumn{1}{c|}{\bf{data}} & \multicolumn{1}{c}{\bf{of clients}} \\
\hline

\cite{Mushtaq2023FAT} & Segmentation & KiTS19 \citep{Heller2020Kits19} & N/A & 6 \\
\cite{Yang2023DynAggr} & Segmentation & COVID-19-CT, COVID-19-1110 \citep{Ma2020Mosmeddata}, COVID-19-9, MS COVID-19-CT & Yes & 4 \\

\cite{Kim2024Federated} & Segmentation & BTCV Abdomen CT \citep{Landman2015MICCAI}, Medical Segmentation Decathlon \citep{Simpson2019Large}, KiTS19 \citep{Heller2020Kits19}, LiTS17 \citep{Bilic2023Liver} & Yes & 7, 21 \\
\cite{Zheng2024Federated} & Segmentation & Own datasets & Yes & 3 \\
\cite{Wang2024FedDUS} & Segmentation & RIDER \citep{Zhao2009Evaluating}, INTEROBS, NSCLC-Radiomics \citep{Aerts2014Decoding}, own datasets & Yes & 4 \\

\cite{Luo2022FedSLD} & Classification & OrganMNIST \citep{Yang2021Medmnist} & N/A & 12 \\
\cite{Lu2022DistributionFree} & Classification & OrganMNIST \citep{Yang2021Medmnist}  & N/A & N/A \\

\cite{Wu2023FedIIC} & Classification & RSNA ICH Detection \citep{Flanders2020Construction} & Yes & 20 \\

\cite{Lai2024Bilateral} & Classification & RSNA ICH Detection \citep{Flanders2020Construction} & N/A & 4 \\

\cite{li2025retrospective} & Classification & OrganMNIST \citep{Yang2021Medmnist}, own dataset & Yes & 12 \\

\cite{Parekh2021Cross-domain} & Detection & Own dataset & N/A & 2 \\
\cite{Dou2021FederatedDL} & Detection & Own dataset, CORONACASES & Yes & 3 \\

\cite{Florescu2022Pre-train} & Detection & Radiopaedia, Lung-PET-CT-Dx, Radiology Assistant, Harvard Dataverse, dataset from article \citep{Yan2020Automatic}, own dataset & N/A & 3 \\

\cite{meda2025dkcn} & Detection & LIDC-IDRI & N/A & N/A \\

\cite{Shiri2024Differential} & Prediction & Datasets from articles \citep{Ning2020Open, Shiri2022Covid} & Yes & N/A \\

\cite{Yang2022RosFL} & Restoration &  NIH-AAPM-Mayo Low-Dose CT \citep{McCollough2016TU-FG-207A-04} & N/A & 4 \\

\cite{yang2025patient} & Denoising & NIH-AAPM-Mayo Low-Dose CT \citep{McCollough2016TU-FG-207A-04} & N/A & 8 \\

\hline

\multicolumn{5}{c}{\bf{Positron Emission Tomography}} \\ \hline

\cite{Vo2024Comparison} & Prediction & NSCLC \citep{Bakr2018Radiogenomic}, own dataset & Yes & 4 \\

\cite{Zhou2023FedFTN} & Denoising & Dataset from article \citep{Xue2021Cross}, own dataset & Yes & 3 \\ \hline

\multicolumn{5}{c}{\bf{X-ray}} \\ \hline

\cite{Slazyk2022CXR-FL} & Segmentation, Classification & Chest X-Ray Images (Pneumonia) dataset, COVID-19 Image Data Collection (IEEE) \citep{Cohen2020ImageData}, RSNA2018 \citep{Shih2019Augmenting} & N/A & 3 \\
\cite{Sun2024FKD} & Segmentation & Datasets from article \citep{Jaeger2014Two} & N/A & 3 \\

\cite{Cetinkaya2021Commu} & Classification & Chest X-Ray Images (Pneumonia) dataset, COVID-19 Radiography Database \citep{Chowdhury2020Can, Rahman2021Exploring}, Figure 1 COVID-19 Chest X-ray, COVID-19 Chest X-Ray Image Repository, COVID-19 Image Data Collection (IEEE) \citep{Cohen2020ImageData} & N/A & 20 \\

\cite{Alkhunaizi2022DOS} & Classification & CheXpert \citep{Irvin2019Chexpert} & N/A & 10 \\
\cite{Elshabrawy2022Ensemble} & Classification & Chest X-Ray Images (Pneumonia) dataset, Covid-19 Image dataset, COVID-19 Image Data Collection (IEEE) \citep{Cohen2020ImageData}, dataset from articles \citep{Tartaglione2020Unveiling, Maguolo2021Critic}, own dataset & N/A & 4 \\

\cite{Kulkarni2023FedFBN} & Classification & ChestX-ray8 \citep{Wang2017Chestx}, CheXpert \citep{Irvin2019Chexpert}, MIMIC-CXR-JPG \citep{Johnson2019Mimic, Goldberger2000Physiobank} & N/A & 2 \\

\hline\hline

\multicolumn{5}{l}{COVID-19-CT dataset: \url{https://www.kaggle.com/datasets/andrewmvd/covid19-ct-scans}} \\
\multicolumn{5}{l}{COVID-19-9: \url{https://figshare.com/authors/MedSeg/9940190}} \\
\multicolumn{5}{l}{MS COVID-19-CT: \url{https://sirm.org/COVID-19/}} \\
\multicolumn{5}{l}{CORONACASES: \url{https://coronacases.org/}} \\
\multicolumn{5}{l}{Radiopaedia: \url{https://radiopaedia.org/}} \\
\multicolumn{5}{l}{Lung-PET-CT-Dx: \url{https://www.cancerimagingarchive.net/collection/lung-pet-ct-dx/}} \\
\multicolumn{5}{l}{RadiologyAssistant: \url{https://radiologyassistant.nl/chest/covid-19/covid19-imaging-findings}} \\
\multicolumn{5}{l}{Harvard Dataverse: \url{https://dataverse.harvard.edu/dataset.xhtml?persistentId=doi:10.7910/DVN/6ACUZJ}} \\
\multicolumn{5}{l}{LIDC-IDRI dataset: \url{https://www.kaggle.com/datasets/zhangweiled/lidcidri}} \\
\multicolumn{5}{l}{Chest X-Ray Images (Pneumonia) dataset: \url{https://data.mendeley.com/datasets/rscbjbr9sj/2}} \\
\multicolumn{5}{l}{Figure 1 COVID-19 Chest X-ray dataset: \url{https://github.com/agchung/Figure1-COVID-chestxray-dataset}} \\
\multicolumn{5}{l}{\parbox{16cm}{COVID-19 Chest X-Ray Image Repository: \url{https://figshare.com/articles/dataset/COVID-19_Chest_X-Ray_Image_Repository/12580328}}} \\
\multicolumn{5}{l}{Covid-19 Image dataset: \url{https://www.kaggle.com/datasets/pranavraikokte/covid19-image-dataset}} \\

\end{tabular}
\caption{Part III.}
\end{table*}

%---------------------------- BIG TABLE:   PART IV ----------------------------
\begin{table*}[t!]
\ContinuedFloat
\captionsetup{list=off,format=cont}
\centering
\footnotesize
\begin{tabular}{p{3.3cm}| p{2.0cm}| p{7.1cm}| p{2.3cm}| p{1.5cm}}
\hline\hline
\multicolumn{1}{c|}{\multirow{2}{*}{\bf{Reference}}} & \multicolumn{1}{c|}{\multirow{2}{*}{\bf{Task}}} & \multicolumn{1}{c|}{\multirow{2}{*}{\bf{Used datasets}}} & \multicolumn{1}{c|}{\bf{Heterogeneous}} & \multicolumn{1}{c}{\bf{Number}} \\
 & & & \multicolumn{1}{c|}{\bf{data}} & \multicolumn{1}{c}{\bf{of clients}} \\
\hline

\cite{Noman2023Block} & Classification & COVID-19 Radiography Database \citep{Chowdhury2020Can, Rahman2021Exploring}, datasets from articles \citep{Jaeger2014Two, Rahman2020Reliable} & N/A & 5 \\

\cite{Naumova2024MyThisYourThat} & Classification & CheXpert \citep{Irvin2019Chexpert} & N/A & 4 \\
\cite{Babar2024Investigating} & Classification & Chest X-ray (Covid-19 \& Pneumonia) dataset, COVID 19 XRay and CT Scan Image dataset, COVIDx \citep{Wang2020Covid} & N/A & 10, 20 \\
\cite{Ahmed2024Efficient} & Classification & Covid19-Pneumonia-Normal Chest X-Ray Images \citep{Shastri2022CheXImageNet} & N/A & 50, 100, 120 \\
\cite{Gupta2024Blockchain} & Classification &  Lungs Disease Dataset (4 types) dataset & N/A & 2 \\

\cite{Myrzashova2024Safeguarding} & Classification & ChestX-ray8 \citep{Wang2017Chestx} & N/A & 6 \\
\cite{karmakar2025convolutional} & Classification & Lungs Disease Dataset (4 types) dataset, COVID-19 Radiography Database & N/A & 3, 5, 7 \\
\cite{lotfinia2025boosting} & Classification & CheXpert \citep{Irvin2019Chexpert}, PadChest \citep{bustos2020padchest}, ChestX-ray14 \citep{Wang2017Chestx}, VinDr-CXR \citep{Nguyen2022Vindr}, PediCXR \citep{pham2023pedicxr} & Yes & 5 \\

\cite{Yan2021Experiments} & Detection & COVIDx \citep{Wang2020Covid} & N/A & 5 \\
\cite{Dong2021Federated} & Detection & CheXpert \citep{Irvin2019Chexpert}, ChestX-ray8 \citep{Wang2017Chestx}, VinDr-CXR \citep{Nguyen2022Vindr} & N/A & 3, 6 \\
\cite{Giuseppi2022FedLCon} & Detection & COVID-19, Pneumonia and Normal Chest X-ray PA dataset & N/A & 7 \\
\cite{Nguyen2022GenAdv} & Detection & DarkCOVID \citep{Ozturk2020Automated}, ChestCOVID \citep{Afshar2020Covid} & N/A & 5 \\

\cite{Kandati2023Federated} & Detection & Covid-19 Image dataset, Chest X-Ray Images (Pneumonia) dataset & N/A & 10 \\

\cite{Makkar2023SecureFedFL} & Detection & Figure 1 COVID-19 Chest X-ray, COVID-19 Radiography Database \citep{Chowdhury2020Can, Rahman2021Exploring}, Actualmed COVID-19 Chest X-ray & N/A & 5, 10, 20, 30, 50, 100 \\

\cite{Zhang2022SplitAVG} & Prediction & RSNA Bone Age dataset & N/A & 4 \\

\cite{Feki2021X-ray} & Screening & COVID-19 Image Data Collection (IEEE) \citep{Cohen2020ImageData}, dataset from article \citep{Jaeger2014Two} & N/A & 4 \\

\cite{Casella2023Merge} & Prognosis & Dataset from article \citep{Soda2021aiforcovid} & Yes & 6 \\ \hline

\multicolumn{5}{c}{\bf{Histopathological images}} \\ \hline

\cite{Zhang2024Fedsoda} & Segmentation & CoNSeP \citep{Graham2019Hover}, CPM-17 \citep{Vu2019Methods}, CRAG \citep{Graham2019MILD}, CryoNuSeg \citep{Mahbod2021CryoNuSeg}, Glas \citep{Sirinukunwattana2017Gland}, TNBC \citep{Naylor2018Segmentation}, dataset from article \citep{Kumar2017Dataset} & Yes & 7 \\

\cite{Gunesli2021FedDropoutAvg} & Classification & TCGA CRC-DX \citep{Weinstein2013TCGA} & N/A & 11 \\
\cite{Lu2022DistributionFree} & Classification & PathMNIST \citep{Yang2021Medmnist}  & N/A & N/A \\

\cite{Li2022Breast} & Classification & BreakHis \citep{Spanhol2015Dataset} & Yes & 11 \\
\cite{Adnan2022Federated} & Classification & TCGA Lung \citep{Weinstein2013TCGA} & Yes & 4, 8, 16, 32 \\

\cite{Lu2022Giga} & Classification, Prediction & TCGA Breast, Kidney \citep{Weinstein2013TCGA}, own dataset & Yes & 4 \\

\cite{Baid2022Federated} & Classification & TCGA Breast, Cervix, Colon, Lung, Pancreas, Prostate, Rectum, Skin, Stomach, Uterus, Uvea of the eye \citep{Weinstein2013TCGA} & Yes & 8 \\

\cite{Hosseini2023PropFFL} & Classification & TCGA Kidney, Lung \citep{Weinstein2013TCGA} & Yes & 4, 6 \\
\cite{Kalra2023Decen} & Classification &  Camelyon-17 \citep{Bandi2018Detection} & Yes & 4 \\
\cite{Lin2023Hyper} & Classification & PANDA \citep{Bulten2022Artificial}, TCGA Lung \citep{Weinstein2013TCGA} & Yes & 3, 5 \\

\cite{Lai2024Bilateral} & Classification & Camelyon-17 \citep{Bandi2018Detection} & Yes & 3 \\
\cite{Hossain2024Collaborative} & Classification & LC25000 \citep{Borkowski2019LC25000} & N/A & 5 \\

\hline

\hline\hline

\multicolumn{5}{l}{Chest X-ray (Covid-19 \& Pneumonia) dataset: \url{https://www.kaggle.com/datasets/prashant268/chest-xray-covid19-pneumonia}} \\
\multicolumn{5}{l}{\parbox{13cm}{COVID 19 XRay and CT Scan Image dataset: \url{https://www.kaggle.com/datasets/ssarkar445/covid-19-xray-and-ct-scan-image-dataset}}} \\
\multicolumn{5}{l}{Lungs Disease Dataset (4 types) dataset: \url{https://www.kaggle.com/datasets/omkarmanohardalvi/lungs-disease-dataset-4-types}} \\
\multicolumn{5}{l}{COVID-19 Radiography Database: \url{https://www.kaggle.com/datasets/kumcs2004/covid-19-radiography-database}} \\
\multicolumn{5}{l}{COVID-19, Pneumonia and Normal Chest X-ray PA dataset: \url{https://data.mendeley.com/datasets/jctsfj2sfn/1}} \\
\multicolumn{5}{l}{Actualmed COVID-19 Chest X-ray dataset: \url{https://github.com/agchung/Actualmed-COVID-chestxray-dataset}} \\
\multicolumn{5}{l}{RSNA Bone Age dataset: \url{https://www.kaggle.com/datasets/kmader/rsna-bone-age}} \\

\end{tabular}
\caption{Part IV.}
\end{table*}

%---------------------------- BIG TABLE:   PART V ----------------------------
\begin{table*}[t!]
\ContinuedFloat
\captionsetup{list=off,format=cont}
\centering
\footnotesize
\begin{tabular}{p{3.3cm}| p{2.0cm}| p{7.1cm}| p{2.3cm}| p{1.5cm}}
\hline\hline
\multicolumn{1}{c|}{\multirow{2}{*}{\bf{Reference}}} & \multicolumn{1}{c|}{\multirow{2}{*}{\bf{Task}}} & \multicolumn{1}{c|}{\multirow{2}{*}{\bf{Used datasets}}} & \multicolumn{1}{c|}{\bf{Heterogeneous}} & \multicolumn{1}{c}{\bf{Number}} \\
 & & & \multicolumn{1}{c|}{\bf{data}} & \multicolumn{1}{c}{\bf{of clients}} \\
\hline

\cite{Deng2024Feddbl} & Classification & Multi-center CRC \citep{Kather2019Predicting, Zhao2020Artificial}, BCSS \citep{Amgad2019Structured} & Yes & 3, 4 \\
\cite{Kong2024Federated} & Classification & PANDA \citep{Bulten2022Artificial}, DiagSet \citep{Koziarski2024DiagSet}, own datasets & Yes & 6, 7 \\
\cite{kumari2025cnn} & Classification & BreakHis \citep{Spanhol2015Dataset} & Yes & 5 \\
\cite{gupta2025applying} & Classification & BreakHis \citep{Spanhol2015Dataset} & Yes & 5 \\

\cite{Ke2021Style} & Normalization & TCGA Colon, Rectum \citep{Weinstein2013TCGA}, CRC-VAL-HE-7K, NCT-CRC-HE-100K \citep{Kather2019Predicting} & Yes & 8 \\
\hline

\multicolumn{5}{c}{\bf{Dermatoscopic images}} \\ \hline

\cite{Yang2022FedZaCt} & Segmentation & HAM10000 \citep{Tschandl2018HAM10000} & N/A & 4 \\
\cite{Wicaksana2023FedMix} & Segmentation & HAM10000 \citep{Tschandl2018HAM10000} & N/A & 4 \\

\cite{Tian2021Privacy} & Classification &  HAM10000 \citep{Tschandl2018HAM10000}, Dermofit \citep{Ballerini2013Color}, Derm7pt \citep{Kawahara2018Seven}, PH2 \citep{Mendoncca2013PH}, ISIC2017 \citep{Codella2018Skin} & Yes & 6 \\

\cite{Alkhunaizi2022DOS} & Classification & HAM10000 \citep{Tschandl2018HAM10000} & N/A & 10 \\
\cite{Lu2022DistributionFree} & Classification & DermaMNIST \citep{Yang2021Medmnist}  & N/A & N/A \\
\cite{Liu2023Class} & Classification & ISIC2018 \citep{Codella2018Skin} & N/A & 7 \\
\cite{Wu2023FedIIC} & Classification & HAM10000 \citep{Tschandl2018HAM10000}, PH2 \citep{Mendoncca2013PH}, Interactive atlas of dermoscopy, ISIC2019 \citep{Codella2018Skin, Tschandl2018HAM10000, Combalia2019BCN20000} & Yes & 5, 10 \\

\cite{Lai2024Bilateral} & Classification & HAM10000 \citep{Tschandl2018HAM10000}, ISIC \citep{Codella2018Skin} & Yes & 4 \\
\cite{Deng2024Federated} & Classification & HAM10000 \citep{Tschandl2018HAM10000}, ISIC2017 \citep{Codella2018Skin} & Yes & 4 \\

\cite{weng2025fedskd} & Classification & Derm7pt \citep{Kawahara2018Seven} & N/A & 3 \\

\hline

\multicolumn{5}{c}{\bf{Retinal imaging}} \\ \hline

\cite{Liu2021FedDG} & Segmentation & REFUGE \citep{Orlando2020Refuge}, DRISHTI-GS1 \citep{Sivaswamy2015Comprehensive}, RIM-ONE \citep{Fumero2011RIM-ONE} & Yes & 4 \\
\cite{Jiang2023IOP-FL} & Segmentation & REFUGE \citep{Orlando2020Refuge}, DRISHTI-GS1 \citep{Sivaswamy2015Comprehensive}, RIM-ONE \citep{Fumero2011RIM-ONE} & Yes & 4 \\
\cite{Wang2023FedDP} & Segmentation & REFUGE \citep{Orlando2020Refuge}, DRISHTI-GS1 \citep{Sivaswamy2015Comprehensive}, RIM-ONE \citep{Fumero2011RIM-ONE} & Yes & 4 \\

\cite{Liang2024ACFL} & Segmentation & Dataset from articles \citep{Wang2020Dofe, Wang2022Personalizing} & Yes & 4 \\

\cite{Malekzadeh2021Dopamine} & Classification & DR \citep{Choi2017Multi} & No & 10 \\
\cite{Lu2022DistributionFree} & Classification & RetinaMNIST \citep{Yang2021Medmnist} & N/A & N/A \\
\cite{Kumar2022DataVal} & Classification & i-ROP \citep{Lu2022Federated} & Yes & 5 \\
\cite{Zhang2022SplitAVG} & Classification & Diabetic Retinopathy Detection dataset & Yes & 4 \\
\cite{Zhou2022Comm-effi} & Classification & DR \citep{Choi2017Multi} & No & 20 \\

\cite{Kaushal2023Eye} & Classification & Ocular Disease Recognition dataset & Yes & 2 \\

\cite{dib2025decentralized} & Classification & Diabetic Retinopathy Detection dataset & Yes & 2-10 \\
\cite{mahadi2025kbfl} & Classification & Eye Diseases Classification Dataset (EDC) \citep{wahab2023artificial} & Yes & 5 \\

\hline

\multicolumn{5}{c}{\bf{Optical Coherence Tomography}} \\ \hline

\cite{Lo2021Federated} & Segmentation, Classification & SFU \citep{Heisler2020Ensemble}, OHSU \citep{Zang2020DcardNet}, dataset from article \citep{Heisler2019DLVessel} & Yes & 2, 4 \\

\cite{Kalapaaking2023Block} & Classification & OCTMNIST \citep{Yang2021Medmnist} & N/A & 10 \\

\hline

\multicolumn{5}{c}{\bf{Ultrasonography}} \\ \hline

\cite{Wicaksana2023FedMix} & Segmentation & BUS \citep{Al2020Dataset}, BUSIS \citep{Zhang2022BUSIS}, UDIAT \citep{Yap2017Automated} & Yes & 3 \\
\cite{Xiang2024Federated} & Segmentation & Own datasets & Yes & 3 \\

\hline\hline

\multicolumn{5}{l}{Interactive atlas of dermoscopy: \url{https://espace.library.uq.edu.au/view/UQ:229410}} \\
\multicolumn{5}{l}{SIIM-ISIC Melanoma Classification dataset: \url{https://kaggle.com/competitions/siim-isic-melanoma-classification}} \\
\multicolumn{5}{l}{Diabetic Retinopathy Detection dataset: \url{https://kaggle.com/competitions/diabetic-retinopathy-detection}} \\
\multicolumn{5}{l}{Ocular Disease Recognition dataset:  \url{https://www.kaggle.com/datasets/andrewmvd/ocular-disease-recognition-odir5k}} \\

\end{tabular}
\caption{Part V.}
\end{table*}

%---------------------------- BIG TABLE:   PART VI ----------------------------
\begin{table*}[t!]
\ContinuedFloat
\captionsetup{list=off,format=cont}
\centering
\footnotesize
\begin{tabular}{p{3.3cm}| p{2.0cm}| p{7.1cm}| p{2.3cm}| p{1.5cm}}
\hline\hline
\multicolumn{1}{c|}{\multirow{2}{*}{\bf{Reference}}} & \multicolumn{1}{c|}{\multirow{2}{*}{\bf{Task}}} & \multicolumn{1}{c|}{\multirow{2}{*}{\bf{Used datasets}}} & \multicolumn{1}{c|}{\bf{Heterogeneous}} & \multicolumn{1}{c}{\bf{Number}} \\
 & & & \multicolumn{1}{c|}{\bf{data}} & \multicolumn{1}{c}{\bf{of clients}} \\
\hline

\cite{gupta2025applying} & Classification & BUS \citep{Al2020Dataset} & No & 5 \\
\cite{hu2025addressing} & Classification & Own dataset & N/A & 5 \\

\cite{Qi2024Simulating} & Detection & Dataset of B-mode fatty liver ultrasound images \citep{Byra2018Transfer}, dataset from article \citep{Vianna2023Comparison} & Yes & 2 \\

\cite{fiorentino2025contrastive} & Detection & Dataset from \cite{burgos2020evaluation, sendra2023generalisability} & 2 & 5 \\

\cite{Lee2021Federated} & Prediction & Own dataset & Yes & 6 \\

\hline

\multicolumn{5}{c}{\bf{Mammography}} \\ \hline

\cite{Muthukrishnan2022MammoDL} & Segmentation, Classification & Dataset from article \citep{Maghsoudi2021Deep} & N/A & 2 \\

\cite{Roth2020BrDen} & Classification & Dataset from article \citep{Pisano2005Diagnostic}, own dataset & Yes & 7 \\

\cite{Tan2023Transf} & Classification & DDSM \citep{Heath1998Current}, CBIS-DDSM \citep{Lee2017Curated} & Yes & 3 \\
\cite{Sanchez2023Memory} & Classification & INBreast \citep{Moreira2012Inbreast}, own datasets & Yes & 3 \\

\cite{Alsalman2024Federated} & Detection & VinDr-Mammo \citep{Nguyen2023VinDr}, CMMD \citep{Cai2023Online}, INBreast \citep{Moreira2012Inbreast} & Yes & 2, 5 \\
\cite{Khan2024Bilevel} & Detection & DDSM \citep{Heath1998Current} & Yes & 3 \\
\hline

\multicolumn{5}{c}{\bf{Microscopy images}} \\ \hline

\cite{Lu2022DistributionFree} & Classification & BloodMNIST, TissueMNIST \citep{Yang2021Medmnist} & N/A & N/A \\
\cite{Kalapaaking2023Block} & Classification & TissueMNIST \citep{Yang2021Medmnist} & N/A & 10 \\
\cite{pervez2025towards} & Classification & Own dataset & N/A & 4 \\

\hline

\multicolumn{5}{c}{\bf{Endoscopic images}} \\ \hline

\cite{Wang2023FedDP} & Segmentation & CVC-ColonDB \citep{Bernal2012Towards}, CVC-ClinicDB \citep{Bernal2015WM-DOVA}, Kvasir-SEG \citep{Jha2020Kvasir-SEG}, dataset from article \citep{Silva2014Toward} & Yes & 4 \\

\hline

\multicolumn{5}{c}{\bf{Multimodal imaging}} \\ \hline

\cite{Bernecker2022FedNorm} & Segmentation &  CT \& MRI: LiTS17 \citep{Bilic2023Liver}, 3D-IRCADb, SLIVER07 \citep{Heimann2009Comparison}, Multi-Atlas \citep{Landman2015MICCAI}, CHAOS19 CT, MRI \citep{Kavur2021Chaos}, KORA \citep{Holle2005Kora} & Yes & 2, 3 \\
\cite{ma2025feature} & Segmentation &  CT \& MRI: CHAOS19 CT, MRI \citep{Kavur2021Chaos} & Yes & 2 \\

\cite{raggio2025fedsynthct} & Synthesis &  CT \& MRI: Own dataset & Yes & 4 \\

\cite{Subramanian2022Effect} & Classification & CT \& MRI: multiple Kaggle datasets & Yes & 10 \\

\cite{Parekh2021Cross-domain} & Detection & CT \& PET: Own dataset & Yes & 2 \\

\cite{Shiri2022DecentralizedCM} & Attenuation correction & CT \& PET: Own dataset & Yes & 6 \\

\cite{Zhang2021Dynamic} & Detection & CT \& X-ray: Dataset from article \citep{Yang2020CovidCTDataset}, Figure 1 COVID-19 Chest X-ray, COVID-19 Radiography Database \citep{Chowdhury2020Can, Rahman2021Exploring} & Yes & 3 \\

\cite{Abbas2024Multidisciplinary} & Classification & CT \& MRI \& Ultrasonography \& Histopathological images: Multi Cancer dataset & Yes & 3 \\

\cite{wang2025clusmfl} & Classification & MRI \& PET: ADNI \citep{Jack2008Alzheimer} & Yes & 10 \\

\cite{Tan2024Joint} & Segmentation & Ultrasonography \& Mammography: DDSM \citep{Heath1998Current}, CBIS-DDSM \citep{Lee2017Curated}, MIAS Digital Mammogram Database \citep{Suckling1994Mammographic}, INBreast \citep{Moreira2012Inbreast}, BUS \citep{Al2020Dataset} & Yes & 5 \\

\hline\hline

\multicolumn{5}{l}{Multi Cancer dataset:  \url{https://www.kaggle.com/datasets/obulisainaren/multi-cancer}} \\

\end{tabular}
\caption{Part VI.}
\end{table*}

%% --------------------------------------------------------------------------------------------------
\section{General overview of federated learning}
\label{sec:fl}

In this section, we provide a general theoretical overview of FL, including basic aggregation methods, learning techniques, and data privacy concerns.

%% --------------------------------------------------------------------------------------------------
\subsection{Theory}
\label{Theory}
The main FL principle can be schematically illustrated as a problem of optimizing the global model parameters $\mathbf{\Theta}\in\mathbb{R}^p$ that minimize the global objective function, defined as \citep{McMahan2017FedAvg, gafni2022federated}:
\begin{equation}
\min_{\mathbf{\Theta}} \mathcal{L}(\mathbf{\Theta};\mathcal{D}) \quad \textrm{with} \quad \mathcal{L}(\mathbf{\Theta};\mathcal{D}) = \sum_{c=1}^C w_c \mathcal{L}_c(\mathbf{\Theta};\mathcal{D}_c),
\label{eq:FL_objective}
\end{equation}

\noindent
where $\mathcal{L}$ denotes the global loss function that is computed by combining $C$ local losses $\{ \mathcal{L}_c\colon c \in \{1,2,...,C \} \}$, each contributing under normalized weight $w_c \ge 0$, and $\mathcal{D}_c=\left\{ \mathbf{x}_n, \mathbf{y}_n \right\}_{n=1}^{n_c}$ is the private dataset available only for $c$--th client with training data $\mathbf{x}_n$ and corresponding labels $\mathbf{y}_n$.

The global model parameters are iteratively updated based on the contributions of individual collaborators, each minimizing local objective function. In practice, this process is typically carried out in the following stages \citep{McMahan2017FedAvg, Rieke2020Future}:

\begin{enumerate}
\item The central server initializes the global model parameters $\mathbf{\Theta}$;
\item Global model parameters  $\mathbf{\Theta}$ are sent to all clients;
\item All $C$ clients train the model locally using their private datasets $\mathcal{D}_c$ for a selected number of local epochs;
\item The parameters of locally trained models $\mathbf{\Theta}_c$ are sent back to the central server;
\item The server aggregates the parameters $\mathbf{\Theta}_c$ in order to obtain an updated global model.
\end{enumerate}

\noindent
Steps 2--5 are repeated for a predetermined number of global rounds or until a global model obtains a sufficiently high accuracy for all the clients involved.

%% --------------------------------------------------------------------------------------------------
\subsection{Federated model accuracy, versatility and efficiency}

Before we go deeper into the FL, some essential terms must be adequately defined. We start with the accuracy of the FL-based model, as it can be interpreted twofold. In the former scenario, the accuracy of the global model in MI is interpreted in terms of a chosen metric, such as the mean squared error, Dice coefficient or Structural Similarity Index Measure computed by all clients with their testing datasets and then averaged across the involved sites \citep{Gunesli2021FedDropoutAvg, Baid2022Federated, Yang2023DynAggr, Wicaksana2023FedMix, Liang2024ACFL}. In the latter case, particularly present in simulation experiments, the accuracy is considered across the so-called global set formed by combining test datasets from all clients involved in the training process \citep{Slazyk2022CXR-FL, Lin2023Hyper}. The metric used to assess the global model should be carefully chosen given the specific MI problem. For instance, in a clinically oriented problem, the mean squared error might be suboptimal to assess the accuracy of the DL model \citep{aja2023validation}. Overall, a model with higher accuracy is considered more acceptable in clinical scenarios than one with lower accuracy \citep{Sohan2023Systematic}.

Another characteristic is versatility, which refers to the accuracy of the global model applied to datasets from institutions that do not participate in the training process of the global model \citep{Guo2021FL-MR, Adnan2022Federated, Wang2024FedDUS, Deng2024Feddbl}.

Lastly, the FL-based procedure requires extensive communication between involved sites. This process can be time-consuming, computational resources and high energy consumption demanding \citep{Shahid2021Communication}. The efficiency relates to the capabilities of the involved sites to train the global model at a specified accuracy level in a given time regime.

\begin{figure*}[t!]
\centering
\includegraphics[scale=0.12]{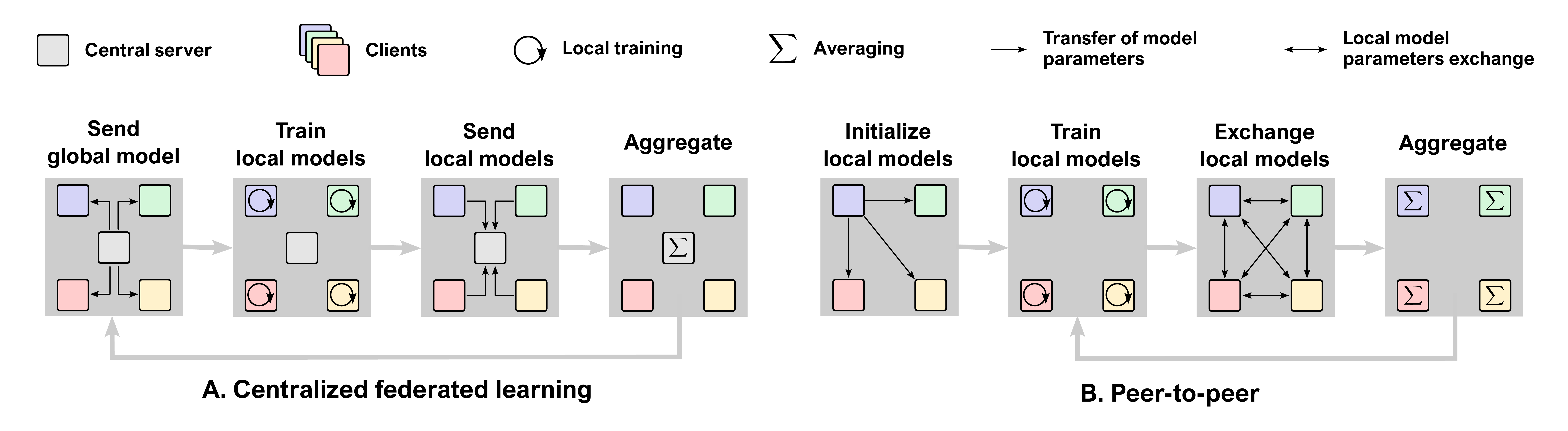}
\caption{Overview of the two most widely used approaches in federated learning: \textbf{A.} centralized architecture with a central (global) server and \textbf{B.} decentralized peer-to-peer methodology.}
\label{fig:CFLvsDFL}
\end{figure*}

%% ----------------------------------------------
\subsection{Optimization and aggregation}
\label{sec:optimization_and_aggregation}

The basic algorithm used to solve the problem defined in Eq. \eqref{eq:FL_objective} is the \textbf{FederatedSGD (FedSGD)} \citep{McMahan2017FedAvg}, which is equivalent to minibatch gradient descent over all data. The FedSGD applies the large-batch synchronous stochastic gradient descent (SGD) to the FL architecture. Each client $c$ calculates the average gradient of the cost function using its local dataset, $\nabla \mathcal{L}_c(\mathbf{\Theta}^r;\mathcal{D}_c)$, where $\mathbf{\Theta}^r$ denotes the parameters of the global model in the federated (global) round $r$. The calculated gradients are then aggregated by the central server given the equation:

\begin{equation}
\mathbf{\Theta}^{r+1} = \mathbf{\Theta}^r - \eta \sum_{c=1}^{C} w_c \nabla \mathcal{L}_c(\mathbf{\Theta}^r;\mathcal{D}_c),
\label{eq:FedSGD_agg}
\end{equation}

\noindent
where $w_c$ is the weight of the aggregation, usually defined as $w_c = n_c/N$ with $n_c$ being the number of examples in each client $c$ and $N$ a total number of examples, $N = \sum_{c=1}^C n_c$.

Assuming a local (single) update for client $c$--th performed during one local epoch is given by $\mathbf{\Theta}_c^r = \mathbf{\Theta}^r - \eta \nabla \mathcal{L}_c(\mathbf{\Theta}^r;\mathcal{D}_c)$, the Eq.~\eqref{eq:FedSGD_agg} can be rewritten as:
\begin{equation}
\mathbf{\Theta}^{r+1} = \sum_{c=1}^{C} w_c \mathbf{\Theta}_c^r.
\label{eq:FedAvg_agg}
\end{equation}

An extension of FedSGD with more than a single local epoch is the \textbf{FederatedAveraging (FedAvg)} \citep{McMahan2017FedAvg}, currently the most commonly used aggregation technique due to its simplicity of implementation. To aggregate the results, FedAvg uses a weighted average given by Eq.~\eqref{eq:FedAvg_agg}, or simply the method assumes $w_c=1/C$, which is the so-called FL \textbf{Equal Voting (FL-EV)} \citep{Linardos2022FL-EV}.

The capabilities of each institution in the FL workflow may differ
due to variations in system-level characteristics, such as
hardware (e.g., central or graphics processing units, memory) and network capacity. In heterogeneous systems, FedAvg ignores the clients unable to complete a certain number of local epochs within a specific time period. \textbf{FedProx} \citep{Li2020FedProx} generalizes FedAvg by allowing participants to perform varying amounts of work (i.e., different number of local epochs) depending on the system resources available locally. Consequently, the clients may obtain the parameters of their local models at different levels of accuracy.  To address the problem, one may include a regularization factor in the cost function.
\begin{equation}
\min_{\mathbf{\Theta}_c^r} \, \mathcal{L}_c(\mathbf{\Theta}_c^r;\mathcal{D}_c) + \frac{\mu}{2} || \mathbf{\Theta}_c^r - \mathbf{\Theta}^r ||_2^2
\label{eq:FedProx_loss}
\end{equation}
with the regularization parameter $\mu > 0$. Adding a proximal term to the local loss function serves two purposes. Firstly, it helps address statistical heterogeneity by ensuring that the local model updates are closer to the global model, eliminating the need to manually set the number of local epochs. Secondly, it enables the effective incorporation of the different amounts of local work that result from the heterogeneity of the system. Once the global round $r$ is finished, the model aggregation process is carried out in the same manner as it is for the FedAvg algorithm (see Eq.~\eqref{eq:FedAvg_agg}). 
Another modification of FedAvg for heterogeneous data is \textbf{FedBN} \citep{Li2021FedBN}, which assumes that local models incorporate batch normalization layers. The FedBN algorithm excludes these parameters from the aggregation process: 

\begin{equation}
\mathbf{\Theta}^{r+1}_l = \sum_{c=1}^{C} w_c \mathbf{\Theta}_{c,l}^r,
\label{eq:FedBN_agg}
\end{equation}

\noindent
where $\mathbf{\Theta}_{c,l}^r$ refers to the parameters at a global round $r$ from $l$--th non-normalization layer in client $c$.

Standard federated learning methods are often difficult to tune and may demonstrate unfavorable convergence behavior. Adaptive algorithms for gradient-based optimization, such as \textbf{FedAdaGrad}, \textbf{FedYogi}, or \textbf{FedAdam} \citep{Reddi2020FedAdam}, are used to address this issue. In such a case, each client updates the parameters of its model locally with the SGD algorithm. Once the number of local epochs has been reached, it calculates the absolute difference between the local current and global model parameters.

\begin{figure*}[t!]
\centering
\includegraphics[scale=0.12]{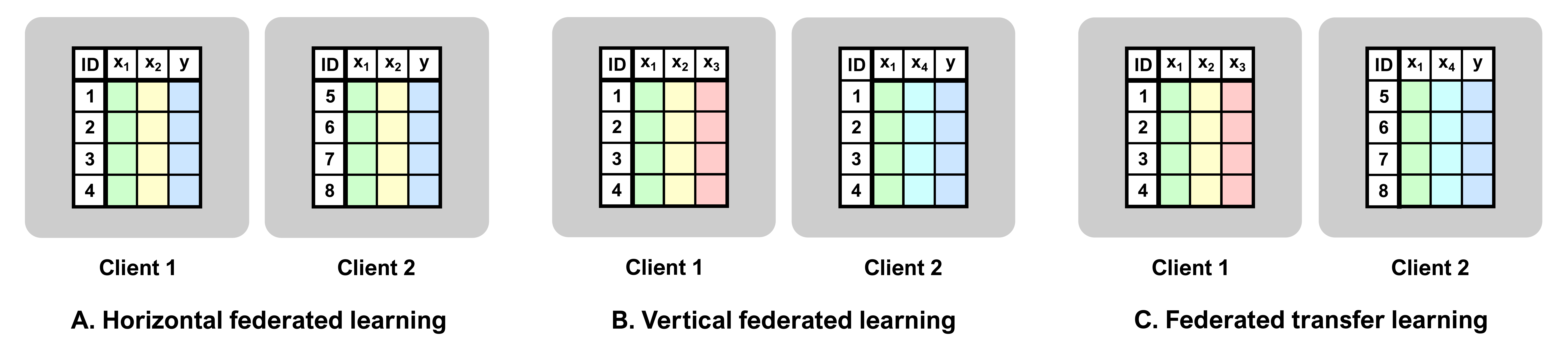}
\caption{Classification of federated learning based on the distribution patterns of the sample and data feature spaces. \textbf{A.} Horizontal federated learning uses data that share the same function or feature space while having non-overlapping identifiers. \textbf{B.} Vertical federated learning involves datasets with different feature spaces but overlapping identifiers. \textbf{C.} Federated transfer learning utilizes data with different identifiers and feature space.}
\label{fig:HFL,VFL,FTL}
\end{figure*}

\noindent
The global model parameters are updated then using the formula
\begin{equation}
\mathbf{\Theta}^{r+1} = \mathbf{\Theta}^r + \eta \frac{\mathbf{m}^r}{\sqrt{\mathbf{v}^r}+\tau}
\label{eq:FedAdam_agg}
\end{equation}

\noindent
and the parameter $\mathbf{m}^r$ can be defined in general as
\begin{equation}
\mathbf{m}^r = \beta_1 \mathbf{m}^{r-1} + (1-\beta_1) \mathbf{\Delta}^r \quad \textrm{with} \quad \mathbf{\Delta}^r = \frac{1}{C} \sum_{c=1}^C (\mathbf{\Theta}^r_c - \mathbf{\Theta}^r)
\label{eq:FedAdam_mr}
\end{equation}

\noindent
with $\mathbf{v}^r$ defined for the FedYogi algorithm as
\begin{equation}
\mathbf{v}^r = \mathbf{v}^{r-1} - (1-\beta_2) (\mathbf{\Delta}^r)^2 \text{sign}(\mathbf{v}^{r-1} - (\mathbf{\Delta}^r)^2),
\label{eq:FedYogi_vr}
\end{equation}
or for the FedAdam variant as follows
\begin{equation}
\mathbf{v}^r = \beta_2\mathbf{v}^{r-1} + (1-\beta_2) (\mathbf{\Delta}^r)^2
\label{eq:FedAdam_vr}
\end{equation}
with $\beta_1, \beta_2 \in [0,1)$ being the decay parameters. Alternatively, if one assumes $\beta_2=1$ in Eq.~\eqref{eq:FedAdam_vr}, it reduces to the FedAdaGrad case \citep{Reddi2020FedAdam}.

The aggregation methods presented in this section are the most fundamental and commonly used techniques in FL. They are straightforward to implement and are naturally considered as references for newly developed techniques. Besides, some of them allow addressing some challenges of FL, such as system heterogeneity (e.g., FedProx) or data heterogeneity (e.g., FedBN), and generally, improving the convergence (e.g., adaptive algorithms).

%% ----------------------------------------------

\subsection{Learning}
\label{sec:FL_intro_learning}

There are two main approaches used to implement the FL architecture, namely the \textbf{centralized} and \textbf{fully decentralized} \citep{Rieke2020Future}, as presented in Fig.~\ref{fig:CFLvsDFL}. Although the fundamental idea behind the FL is based on the latter procedure, the former requires a central server (see Fig.~\ref{fig:CFLvsDFL}A), also known as a manager. This server is typically a powerful computing unit located at the network's edge or in the cloud. The central server generates a global model based on updates of locally trained models and manages communication between the central server and individual clients \citep{Rieke2020Future, Mothukuri2021Survey, Brecko2022Federated}. This server must remain technically stable to prevent the generation of incorrect global models and to avoid network failures that could cause a bottleneck problem, stopping the entire process \citep{Brecko2022Federated}.

The decentralized FL eliminates the requirement for a central server to coordinate the learning process. Instead, clients communicate with each other using two primary communication modes: \textbf{peer-to-peer} (see Fig.~\ref{fig:CFLvsDFL}B) and \textbf{blockchain} \citep{Brecko2022Federated}. In peer-to-peer technology, all clients are treated equally and have equal privileges. Each client exchanges locally trained models with other clients and conducts the aggregation of its model \citep{Rieke2020Future, Mothukuri2021Survey, Brecko2022Federated}. 

On the other hand, the blockchain is a decentralized distributed database that stores all transaction records (in the case of FL, model parameters) generated by collaborating participants. These records are chained in blocks and are recorded over time. In FL, blocks store model updates and other basic information such as block version, previous block hash, and timestamp \citep{Li2022Blockchain, Gammar2023Securing}. The FL via blockchain is a more complex procedure than centralized and peer-to-peer approaches. It involves many additional steps to send local parameters and perform the aggregation. The essential components for blockchain functionality include smart contracts, consensus protocols, and miners. The smart contract is a self-executing program that automates the actions required in a blockchain transaction, allowing clients to codify agreements without the need for a trusted third party \citep{Gammar2023Securing}. In the context of FL, smart contracts are used for various purposes, including registering participants, coordinating model training, evaluating the contribution of each client, and awarding rewards \citep{Liu2020Secure, Behera2022FLusingsmart}. The consensus protocols are rules that determine which information can be allowed onto a blockchain and which must be discarded, thus compensating for the lack of a central server. These protocols are key to verifying the accuracy of a transaction and ensuring network security \citep{Gammar2023Securing}. In FL, they can be used, for example, to protect against malicious clients and enhance system scalability \citep{Li2020Blockchain}. Miners play a crucial role in the blockchain network by processing and adding new transaction records. They can be personal computers, standby servers, or cloud-based nodes that use specialized computer hardware to validate and verify transactions \citep{Gammar2023Securing}.

The primary stages of an FL blockchain procedure are as follows \citep{Gammar2023Securing}:
\begin{enumerate}
\item Clients train their models locally using their private datasets.
\item The clients register through the smart contract and transfer their local model updates to the miners on the blockchain.
\item The miners verify and authenticate the updates of local models based on the consensus protocol.
\item The miners aggregate the updates from all verified models.
\item Each miner runs the consensus algorithm until it receives a newly generated block from other miners. Afterward, the new block is broadcast to all other miners.
\item A new block is added to the blockchain network.
\item Clients can download the global model and continue the training procedure.
\end{enumerate}

The FL can be categorized with respect to the distribution patterns of the sample space and the data feature space into three categories: \textbf{horizontal FL}, \textbf{vertical FL}, and \textbf{federated transfer learning} \citep{Brecko2022Federated, Li2022Blockchain}, as illustrated in Fig.~\ref{fig:HFL,VFL,FTL}.
The horizontal FL relies on data that share the same function or feature space and has non-overlapping identifiers (e.g., medical record number, which allows for distinguishing records about one patient from records about another patient). Although institutions may gather data from distinct sources, the samples may look alike and thus represent a similar or identical feature space. In such cases, the FL is most commonly implemented in a traditional configuration with clients and a central server (discussed in section \ref{Theory}) to increase the number of training samples and improve the model precision \citep{Mothukuri2021Survey, Brecko2022Federated}. For example, the sites involved may work on a specific data segmentation problem, and hence, the medical data used by the institutions are characterized by a similar feature space, although they come from different patients.

Vertical FL, also known as heterogeneous FL, handles different feature spaces from the data (e.g., various medical modalities) but with overlapping identifiers, meaning that the same patient appears in all sets \citep{Mothukuri2021Survey, Brecko2022Federated, Abad2022Security}. It can facilitate training a common distributed model, such as a risk management model \citep{Brecko2022Federated}. Implementing vertical FL is more challenging than horizontal FL, as it typically involves three types of participants: active, passive, and coordinator. The active participant is the institution that wants to build the model and provide data with both the sample set and their labels. The passive participant only has the sample set, and its feature set differs from the active participant's. The coordinator is a trusted third party responsible for coordinating training and communication among the participants. In some cases, the active participant is also the coordinator \citep{Li2023VFLTaxonomies}. Furthermore, vertical FL requires encryption and a specific learning protocol, as the training process differs from horizontal FL \citep{Mothukuri2021Survey, Brecko2022Federated}. Firstly, it is essential to align the training data in order to identify common sample identifiers. After alignment, the participants begin training the model. The most common training protocol is gradient descent, which involves transmitting intermediate values (e.g., softmax after the forward computation). Passive participants send the encoded intermediate results to the active participant. The active participant then calculates the losses and combines them with its loss. The aggregated loss is sent back to the passive participants. Active and passive participants use this aggregated loss to perform backpropagation and update the gradient for their local models. The encrypted gradients are sent to the coordinator, which decrypts them using a private key and returns them to the individual participants. The participants then retrain the model, and this entire process is iteratively repeated until the convergence of the model or a predefined condition is satisfied \citep{Li2023VFLTaxonomies, Liu2024Vertical}

Finally, federated transfer learning has been inspired by the transfer learning technique. It allows the application of knowledge from a pretrained model to a different but related problem using a similar dataset. This approach yields better results than training a new model from scratch \citep{Mothukuri2021Survey, Abad2022Security}, and is especially valuable when the data from local clients has different identifiers and feature space \citep{Brecko2022Federated}.

%% ----------------------------------------------
\subsection{Data privacy}
\label{sec:data_privacy}

The basic assumption behind the FL is to keep local data private. However, sharing local model parameters with other clients or a central server can still expose data. Suppose malicious attackers gain unauthorized access to the federated learning architecture and obtain local model parameters. In that case, they can use reverse engineering techniques to recover some information about the source data on which the model was trained. Thus, additional techniques are required to increase privacy protection, such as the \textbf{Differential Privacy (DP)}, \textbf{Secure Multi-Party Computation (MPC)}, and \textbf{Homomorphic Encryption (HE)} \citep{Li2020Preserving, Mothukuri2021Survey, Truong2021Privacy, Fereidooni2021SAFELearn}.

DP adds random noise to sensitive personal characteristics, introducing some level of uncertainty in revealing individual data \citep{Dwork2006Our}. Depending on whether the clients or the central server is responsible for adding noise, a distinction can be made between local and global DP \citep{Garcia2022FLUTE}. Noise is typically introduced to the input dataset during local training or to the local model parameters before sharing them with other clients or a central server \citep{Mothukuri2021Survey, Abad2022Security}. Hence, DP protects the model against inference attacks like model inversion. Even if the data are reverse-engineered, it will not accurately represent the data of any participant. However, the DP procedure can affect the accuracy of a global model \citep{Byrd2020Differentially, Truong2021Privacy, Abad2022Security}. Therefore, it is fundamental to introduce noise at a reasonable level and use a proper noise generation mechanism, such as Gaussian \citep{Li2020fMRI, Lu2022Giga, Peng2023FedNI} or Laplacian \citep{Li2019TumSeg, Li2020fMRI}, which is typically used in MI.

\begin{figure*}[!t]
\centering
\includegraphics[width=0.95\textwidth]{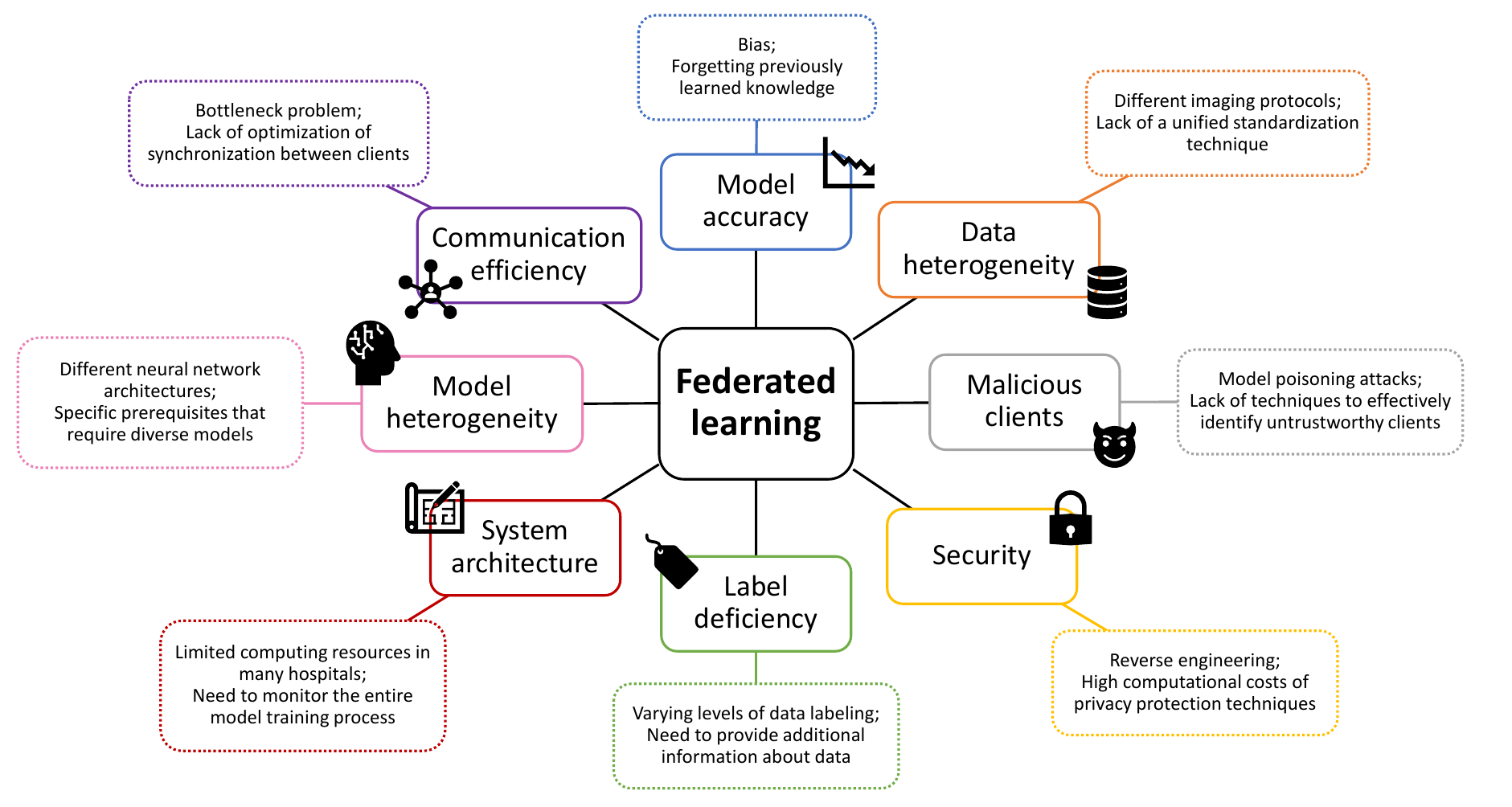}
\caption{Summary graph presenting a broad spectrum of topics related to federated learning in the context of medical imaging.}
\label{fig:FL_issues}
\end{figure*}

The MPC is an alternative to the DP method that ensures privacy while also maintaining global model accuracy \citep{Yao1986Generate}. The MPC protocol is a collaborative approach that allows multiple sites to compute a joint function of interest without revealing their private inputs. The secure MPC protocol is considered safe if the parties involved only have access to the final result and no additional information is shared \citep{Byrd2020Differentially, Mothukuri2021Survey, Truong2021Privacy}. Most MPC protocols are based on classic secret sharing techniques, such as Shamir secret sharing and verifiable secret sharing schemes \citep{Truong2021Privacy}. In FL, this methodology requires the involved sites to use a secure weighted average protocol. In other words, although clients encrypt their model weights, the server can still compute a weighted average on the encrypted data \citep{Byrd2020Differentially}. Despite its many benefits, the MPC reduces FL efficiency. This is because carrying out sequential operations on the model parameters leads to tremendous overhead. Moreover, the communication involved in secret sharing requires a significantly larger number of steps than the basic FL technique (see section~\ref{Theory}) \citep{Kanagavelu2020Two, Mothukuri2021Survey, Truong2021Privacy}.

Data privacy can also be maintained through the HE. With non-homomorphic encryption, it is nearly impossible for the server to aggregate encrypted parameters of local models without the secret decryption key \citep{Gentry2010Computing}. In contrast, HE enables computations on encrypted data without the need for a secret key. The results of these computations are also encrypted and can only be decrypted upon request. Moreover, HE guarantees that the decrypted result is the same as if the computation was accomplished using the original, unencrypted dataset. There are three variants: partial HE, somewhat HE, and fully HE, all differ from each other in the encryption schemes used and the ability to perform computational procedures on encrypted data \citep{Li2020Preserving, Fereidooni2021SAFELearn, Truong2021Privacy}. The HE can encode sum and product operations in a neural network but not the activation function. Activation functions are approximated using higher degree polynomials, Taylor series, or Chebyshev polynomials, implemented as part of HE schemes \citep{Li2020Preserving}. Although HE guarantees strong privacy, it is limited in practical scenarios due to the enormous computational overhead \citep{Truong2021Privacy}.

%% --------------------------------------------------------------------------------------------------
%% --------------------------------------------------------------------------------------------------
%% --------------------------------------------------------------------------------------------------
\section{Broader overview of federated learning}
\label{sec:broader_overview}

Despite the rapid development of FL and a growing number of diverse MI applications, one has to face the challenges that must be clearly illustrated before introducing FL into practice (see Fig.~\ref{fig:FL_issues}). Handling these issues is critical, as they cover typical characteristics of MI problems, such as data heterogeneity, data privacy and security, model accuracy, system architecture, communication efficiency, and data scarcity \citep{Patel2022Adoption, Aouedi2022Handling, Dasaradharami2023Comprehensive}. This section briefly outlines all these fundamental issues and prospects that should be considered before deciding to translate the MI problem into a federated scenario. More details on particular solutions to these problems are included in the following sections, namely, \textbf{Aggregation methods in medical imaging} (section~\ref{sec:fl_aggregation}) and \textbf{Learning methods in medical imaging} (section~\ref{sec:fl_learning}).

\subsection{Model accuracy}
\label{sec:issues:accuracy}

The bias that favors a specific client is a factor that could affect the accuracy of the global model~\citep{Darzidehkalani2022Federated}. If such a situation occurs, it can lead to an increase in the accuracy of the DL task for one client and a decrease in the accuracy for the others. The reason why the bias appears is the variability in the size and distribution of the datasets across the involved sites. The models generated by participants with larger datasets might be assigned with higher weighting factors, thus resulting in an over-representation or under-representation of specific patterns \citep{Abay2020MitigatingBias}. Besides, depending on the aggregation technique, only some clients can be involved, e.g., the institutions that conduct the entire local training within the specific timeframe. This can lead the FL algorithm to favor clients with a better computing infrastructure or faster and more reliable network connections. In clinical contexts, this usually corresponds to larger and richly endowed imaging centres, further reinforcing site-specific biases.

To conclude, the FL requires significant computational and communication resources \citep{Shahid2021Communication}. Thus, it is essential to establish an appropriate level of accuracy and versatility that satisfies the sites involved, considering each participant's hardware capabilities.

%% ----------------------------------------------
\subsection{Data heterogeneity}
\label{sec:issues:heterogeneity}

Data heterogeneity is a factor that significantly affects the accuracy of the global model in MI. Imaging data across institutions are typically collected with the equipment provided by various vendors under different acquisition protocols or even various data reconstruction procedures \citep{Gibson2018Inter, Jiang2022Harmofl, Darzidehkalani2022Federated}. For example, histological images may look different due to various staining conditions. The MRI data may exhibit variations in distributional properties due to multifarious imaging protocols or contrasts applied, such as $T_1$-/$T_2$-weighting or diffusion-weighting \citep{Jiang2022Harmofl}, or even the procedure used to reconstruct magnitude data resulting in various statistical properties of the data \citep{den2014data,pieciak2016variance,pieciak2017non,vegas2017statistical}. Consequently, the data across the institutions may be nonindependent and nonidentically distributed (non-IID).

Non-IID data can arise not only from the uneven distribution of data features caused by the use of different types of devices or acquisition techniques but also from variations in the distribution of labels \citep{Zhu2021Federated}. In FL, it is common for label distribution to differ among clients. Two main scenarios can be observed in this context. In the former, all institutions keep data representing all groups (e.g., diseased patients and healthy individuals), but the number of samples from each group varies significantly between clients. In the latter case, some centers exclusively cover data from healthy subjects, while others secure data from diseased individuals. These situations can lead to the suboptimal accuracy of standard aggregation methods presented in section~\ref{sec:optimization_and_aggregation}.

\begin{figure}[t!]
\centering
\includegraphics[width=0.5\textwidth]{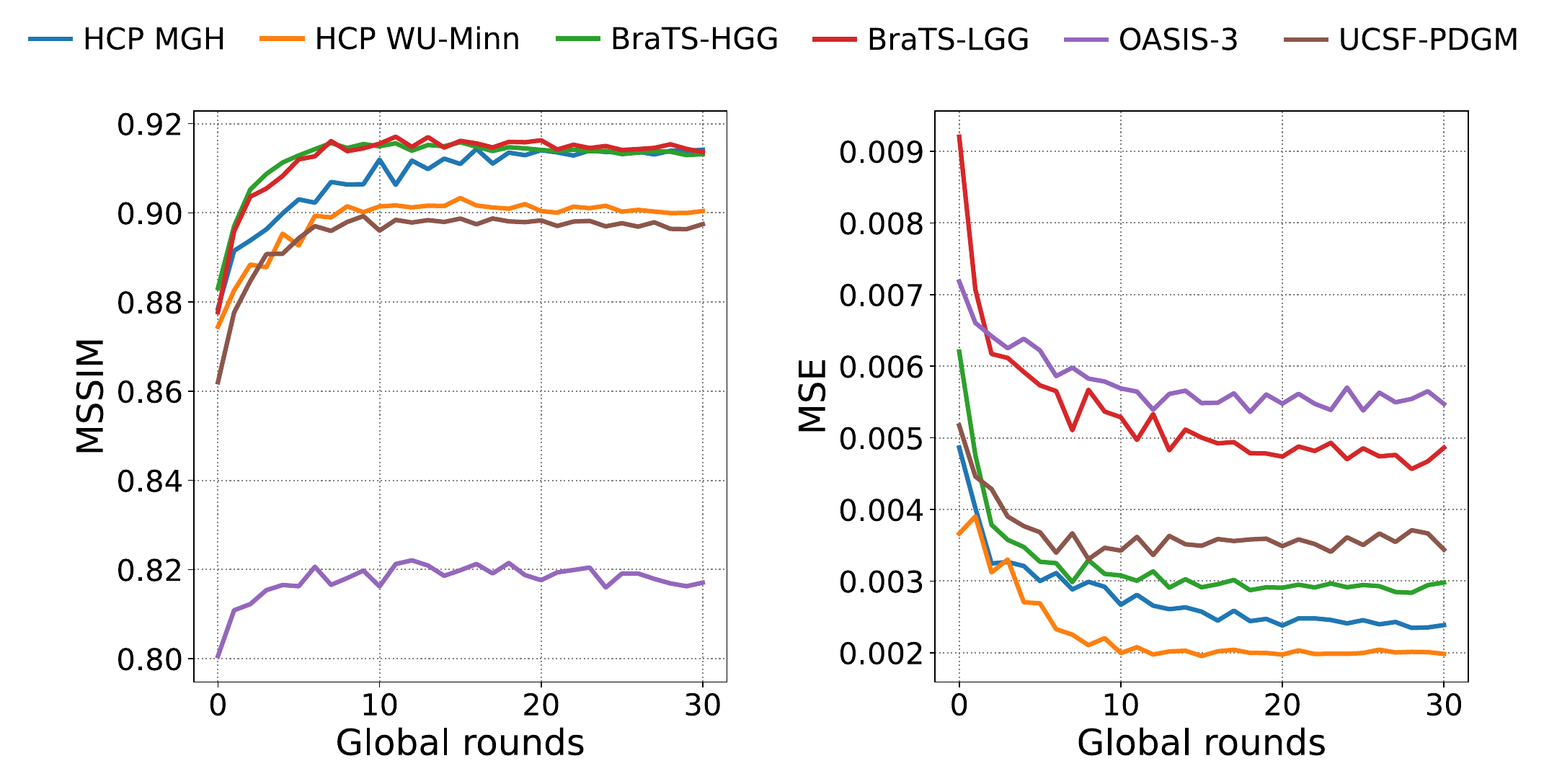}
\caption{Changes in mean structural similarity index measure (MSSIM) and the mean squared error (MSE) for individual clients as a function of the global round in $T_1$- to $T_2$-weighted MRI data synthesis task. The graph presents the results for the FedCostWAvg algorithm (see section~\ref{sec:fl_learning:accuracy} for more details on FedCostWAvg aggregation scheme). }
\label{fig:Clients_MSSE_SSIM}
\end{figure}

\begin{figure}[t!]
\centering
\includegraphics[width=0.5\textwidth]{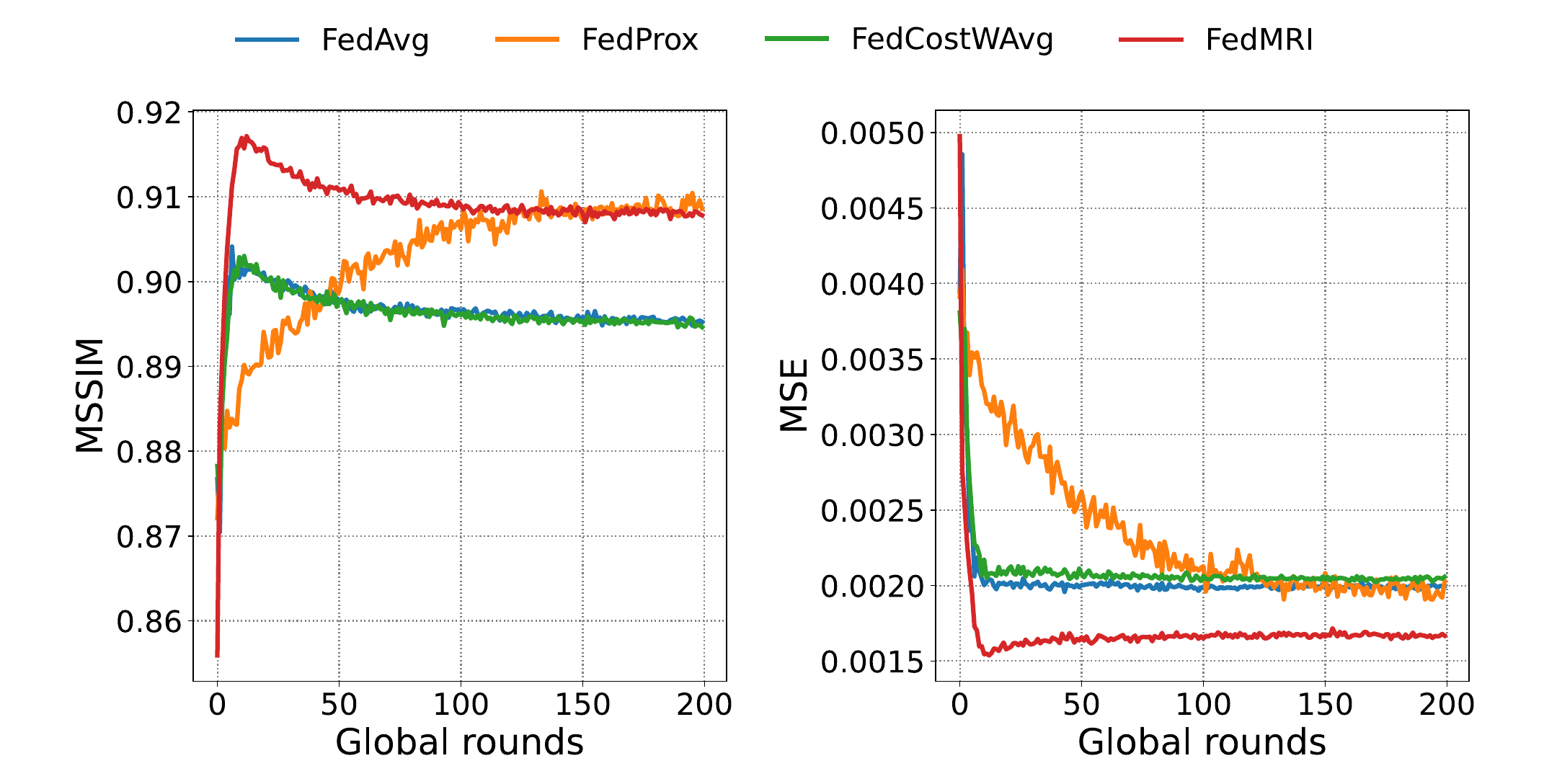}
\caption{Changes in the mean structural similarity index measure (MSSIM) and mean squared error (MSE) for the HCP WU-Minn client as a function of the global round in $T_1$- to $T_2$-weighted MRI data synthesis task. Four different aggregation methods have been used to generate global models.}
\label{fig:Methods_MSSE_SSIM}
\end{figure}

For the sake of illustration, we present two experimental results for the image synthesis task using $T_1$- and $T_2$-weighted MR data from heterogeneous sources. We follow the scheme presented recently in \cite{fiszer2025validation}. Fig.~\ref{fig:Clients_MSSE_SSIM} demonstrates differences in the accuracy of the global model obtained with the FedCostWAvg algorithm for different institutions. Here, two involved sites include healthy brain data, i.e., Human Connectome Project (HCP) MGH and HCP WU-Minn, while others cover brain tumors (BraTS, UCSF-PDGM) and cognitive decline data (OASIS-3). We observe significant deviations in terms of the mean structural similarity index measure (MSSIM) of the OASIS-3, which is supposed to be a malicious client. Fig.~\ref{fig:Methods_MSSE_SSIM} exemplifies another issue that emerges from heterogeneous data -- the rapid overfitting of the global model. The impact of non-IID data on the accuracy of the FL model has been highlighted recently in a real-world examination by \cite{fiszer2025validation}. 

As mentioned above, data heterogeneity originates from the absence of unified imaging standardization procedures. Although the involved sites use the same acquisition protocol to collect data, they may employ different software to preprocess data, leading, for example, to reduced reproducibility of experiments \citep{veraartassessing2023}. In addition to these, the number of samples and their distribution can vary over time \citep{Darzidehkalani2022Federated, Aouedi2022Handling}. For example, the number of patients affected by a particular disease in one hospital can fluctuate, including the patient's recovery, death, or transfer of care. Finally, the institutional data deviates according to the demography of the sample. As an illustration, the MRI data change to such an extent throughout life that it allows reconstruction of the patient's age \citep{cox2016ageing,peng2021accurate,pieciak2023spherical}. Finally, the diverse nature of the data can result in reduced global model accuracy due to its inability to converge. The global model may perform exceptionally well for some clients but inaccurately for others \citep{Darzidehkalani2022Federated}. According to \cite{Zhao2018Non-IID}, highly heterogeneous data can cause up to 50\% degradation in the accuracy of the FL global model, making heterogeneous data one of the most critical issues to address. In the context of MI, such a severe decrease in performance quality has not been observed. Nevertheless, the authors reported a decline of approximately five percentage points in classification  accuracy \citep{Zhao2023ADI}, a reduction in the Sørensen–Dice index ranging from about 0.05 to 0.15 for segmentation tasks \citep{Jiang2023IOP-FL, Yang2023DynAggr}, and a decrease in the MSSIM index by approximately 0.17 for data reconstruction task  \citep{Feng2023FedMRI}.

In addition to this, institutions still lack proper infrastructure for image data processing and data management systems that are compliant with current standards \citep{Darzidehkalani2022Federated}.

%% ----------------------------------------------
\subsection{Malicious clients}
\label{sec:issues:malicious}

The quality deterioration of the results generated from the global model may come from malicious clients in all institutions. Due to its distributed nature, the FL is susceptible to model poisoning attacks. These attacks occur if a malicious client sends manipulated model updates to the server \citep{Alkhunaizi2022DOS, Zhang2022FLDetector, Onsu2023Cope}. Model poisoning attacks can be classified into two categories: untargeted and targeted. In the former case, a corrupted global model generates incorrect predictions for multiple test inputs. In the latter scenario, a corrupted global model causes incorrect predictions for specific test inputs chosen by the attacker, while the accuracy of the global model for other inputs generally remains unaffected \citep{Abad2022Security, Zhang2022FLDetector}. In clinical settings, the behavior of malicious clients may be unintentional rather than adversarial. It can occur due to issues such as corrupted imaging data, misconfigured preprocessing pipelines, or systematic annotation errors at individual sites. Despite being unintentional, these issues can have a disproportionately negative impact on the aggregated model. For MI applications, the risks posed by malicious clients and their effects on the accuracy of the global model have been discussed concerning untargeted poisoning attacks \citep{Alkhunaizi2022DOS}.

%% ----------------------------------------------
\subsection{Security}
\label{sec:issues:security}

MI data are personal information that requires particular protection under specific confidentiality procedures. Any violation of the security procedure is a significant issue that might lead to the disclosure of personality \citep{Truong2021Privacy}. The FL naturally provides a privacy protection mechanism by storing institutional datasets locally within each site without sharing them with third parties. However, the FL-based system can involve participants with unknown motives, and therefore, unduly entrusted clients can compromise security by reverse engineering \citep{Li2020Preserving, Darzidehkalani2022Federated, Sohan2023Systematic}. Specifically, since a DL algorithm stores information in its weights, the untrusted client may reconstruct some portion of the data in a decentralized network \citep{Darzidehkalani2022Federated}. For example, recent studies have successfully rendered a face from MRI brain data \citep{Schwarz2019Identification, Abramian2019Refacing}. 

%% ----------------------------------------------
\subsection{Label deficiency}
\label{sec:issues:label_deficiency}

Training DL models requires reliable and large labeled datasets to achieve satisfactory results. Frequently, the institutions involved cover only partially annotated or even entirely unannotated datasets. This is not surprising, as the labeling of medical data is very time-consuming and requires sufficient knowledge and experience, making it challenging to construct necessarily labeled datasets \citep{Dercksen2019Dealing, Ji2023Emerging}. In an FL scenario, different clients may have data labeled to varying degrees. In this context, the standard FL scheme (described in section \ref{Theory}) cannot be used as it requires all clients to hold fully labeled data. To address this problem, semi-supervised \citep{Mushtaq2023FAT, Liu2023Class, Wang2024FedDUS} and self-supervised learning \citep{Park2024MS-DINO} have been introduced.

%% ----------------------------------------------
\subsection{System architecture}
\label{sec:issues:architecture}

Another critical factor in FL-based systems is the appropriate computing architecture chosen to ensure the feasibility, effectiveness, and efficiency \citep{Darzidehkalani2022Federated}. The architecture for a specific MI problem should address the issues raised in previous paragraphs while also controlling client synchronization problems \citep{Aouedi2022Handling}. More information on optimizing client synchronization will be discussed in section \ref{sec:issues:communication}, as the communication efficiency problem relates to the system architecture.
Generally, system architecture covers computing power that can be either private or cloud-based train models locally, storage systems (i.e., local storage or in the cloud), and software employed to archive and maintain the data such as the Picture Archiving and Communication System (PACS) and Radiological Information System (RIS) \citep{Darzidehkalani2022Federated,jacobs2023challenges}. A complete system architecture for the FL partnership should include additional elements that allow tracking of the model training process, including recording data access history, training structure, selection and modification of configuration, and hyperparameter tuning \citep{Dasaradharami2023Comprehensive}. Such a tracking system is necessary to meet traceability and accountability requirements, as it assists researchers in interpreting the global model by examining the data source. Finally, the tracking system enables one to assess the contribution level of individual participants \citep{Naz2022Comprehensive}.

%% ----------------------------------------------
\subsection{Model heterogeneity}
\label{sec:issues:models}

One of the key assumptions of the standard FL is that all clients participating in the training process use the same neural network architecture \citep{alam2022fedrolex}. However, this assumption often does not hold due to variations in system architecture, particularly in computational resources. Additionally, different institutions may have specific prerequisites that require diverse model architectures \citep{nezhad2025generative, weng2025fedskd}. For instance, a client may prefer traditional convolutional networks for their spatial precision and computational efficiency. In contrast, others may choose transformers to effectively capture long-range dependencies and enable context-aware processing \citep{nezhad2025generative}. Some institutions may also use other models, such as random forests \citep{afonin2022modelagnostic}. In these scenarios, employing standard FL becomes impractical, and requiring all institutions to use the same model architecture could be inefficient and hinder the performance \citep{nezhad2025generative, weng2025fedskd}.

%% ----------------------------------------------
\subsection{Communication efficiency}
\label{sec:issues:communication}

To allow institutions to train the model smoothly, a reliable way of communicating with each other must be secured. Communication involves constantly sharing the parameters of global and local models directly between clients or between clients and the global server. This leads to a bottleneck problem if the ability to disseminate information due to network bandwidth is limited \citep{Patel2022Adoption, Aouedi2022Handling}. This issue is especially significant for medical imaging models, which often require high-capacity architectures to process high-resolution volumetric data. Consequently, the communication overhead becomes a practical constraint in hospital networks that have limited or regulated bandwidth.

Communication efficiency is closely related to client synchronization. Institutions may differ in the characteristics of the computing system, including computing power and data storage systems. For instance, advanced DL models may require heavy computations and specialized hardware, such as GPU, making the slowest clients unable to finish local training before aggregation occurs. Sharing local model parameters at varying frequencies or with further delays negatively impacts the entire system's speed. In addition, the institution can stop local training before the global FL process is complete or withdraw from further collaboration. Ultimately, any transfer interruption due to poor network infrastructure or temporary maintenance may occur, leading in some cases to stop the training procedure~\citep{Aouedi2022Handling}.

%% --------------------------------------------------------------------------------------------------
%% --------------------------------------------------------------------------------------------------
%% --------------------------------------------------------------------------------------------------
\section{Aggregation methods in medical imaging}
\label{sec:fl_aggregation}

In MI the most commonly used aggregation method is the \textit{standard} FedAvg \citep{Roy2019BrainTorrent, Li2019TumSeg, Roth2020BrDen, Cetinkaya2021Commu, Parekh2021Cross-domain, Ziller2021Differentially, Liu2021FedDG, Dou2021FederatedDL, Feki2021X-ray, Stripelis2021Scali, Roth2021Prost, Stripelis2021SecureNeuro, Linardos2022FL-EV, Adnan2022Federated, Li2022Breast, He2022CosSGD, Yang2022RosFL, Subramanian2022Effect, Luo2022FedSLD, Slazyk2022CXR-FL, Zhou2022Comm-effi, Elshabrawy2022Ensemble, Pati2022BigData, Misonne2022HeartSeg, Stripelis2022Secure, Lu2022Giga, Kumar2022DataVal, Tan2023Transf, Liu2023Class, Sanchez2023Memory, Mushtaq2023FAT, Elmas2023FedGIMP, Levac2023EndToEnd, Denissen2023Transf, Kanhere2023SegViz, Kaushal2023Eye, Makkar2023SecureFedFL, Wu2023FedIIC, Wang2023FedDP, Kim2024Federated, Qi2024Simulating, Alsalman2024Federated, Mitrovska2024Secure, Yamada2024Investigation, Yan2024Cross, Xiang2024Federated, Zheng2024Federated, Zhou2024DistributedFL, Hossain2024Collaborative, Sun2024FKD, Khan2024Bilevel, Deng2024Feddbl, Babar2024Investigating, Myrzashova2024Safeguarding, Gupta2024Enhancing, Albalawi2024Integrated, Gupta2024Blockchain, Kumar2024Privacy, Deng2024Federated} or its modification with equal weights (FL-EV) \citep{Li2020fMRI, Guo2021FL-MR, Lo2021Federated, Florescu2022Pre-train, Linardos2022FL-EV, Peng2023FedNI, Naumova2024MyThisYourThat, Abbas2024Multidisciplinary, Liu2024Multi, Vo2024Comparison}. Some of the algorithms also use FedProx \citep{Elshabrawy2022Ensemble, Subramanian2022Effect, Qi2024Simulating}, FedBN \citep{Elshabrawy2022Ensemble, Kanhere2023SegViz, Kulkarni2023FedFBN} or adaptive algorithms \citep{Stripelis2022Secure, Levac2023EndToEnd, Qi2024Simulating}. Other techniques dedicated to MI are modifications or extensions of FedAvg, involving adjusting the weighting factor \citep{Shen2021DWA, Machler2022FedCostWAvg, Khan2022SimAgg, Alkhunaizi2022DOS,  Machler2023FedPIDAvg, Khan2023RegSimAgg, Wicaksana2023FedMix, Wu2023ModFed, Liu2023MultSkle, Yang2023DynAggr, Noman2023Block, Lai2024Bilateral, Zhang2024Fedsoda, Wang2024FedDUS} or performing the aggregation only on specific parameters of local models, such as those from particular layers, instead of on all parameters \citep{Gunesli2021FedDropoutAvg, Bernecker2022FedNorm, Liu2023MultSkle}.

New aggregation techniques aim to overcome the fundamental limitations of FL. The following subsections present the most important features of different aggregation methods categorized into algorithms that enhance the global model accuracy, handle malicious clients, and improve data security. These strategies operate at the level of model parameters or updates, indicating they are largely independent of specific downstream tasks, such as classification or segmentation, or network architectures. Although individual studies validate these strategies for specific MI tasks, they are generally applicable to a wide range of applications.

%% ----------------------------------------------
\subsection{Improving model's accuracy}
\label{sec:model_accuracy}

Local models trained on small or low-quality datasets can negatively impact the accuracy of the global model, leading to decreased accuracy across all clients participating in federated learning. By introducing new aggregation algorithms, we expect to increase the accuracy of the global model while preventing the accuracy reduction locally. The widely used approach is to \emph{dynamically assign weighting factor} to each involved client based on their local loss functions \citep{Shen2021DWA, Machler2022FedCostWAvg, Machler2023FedPIDAvg, Wicaksana2023FedMix, Wu2023ModFed, Liu2023MultSkle}. The methods presented in section~\ref{sec:fl} use constant weights across global rounds. This section reviews up-to-date methods that employ round-dependent weights.

We start our review with the \textbf{Dynamic Weight Averaging (DWA)} \citep{Shen2021DWA}, which aggregation weights are based on the variation of the loss values from the previous round:

\begin{equation}
w_c^r = \frac{\xi \exp (\rho_c^{r-1}/T)}{\sum_{c'=1}^C \exp (\rho_{c'}^{r-1}/T)}, \hspace{0.4em} \rho_c^{r-1} = \frac{\mathcal{L}_c(\mathbf{\Theta}_c^{r-1};\mathcal{D}_c)}{\mathcal{L}_c(\mathbf{\Theta}_c^{r-2};\mathcal{D}_c)}
\label{eq:DWA_wei}
\end{equation}

\noindent
with $w_c^r$ being the weight associated with client $c$--th at round $r$--th. The symbol $T$ in Eq.~\eqref{eq:DWA_wei} represents a temperature, $\xi$ is used to control the impact of the weights, $\rho_c^{r-1}$ is the dynamic proportion of the loss function $\mathcal{L}$ changes from two consecutive rounds $r-1$ and $r-2$.

Another strategy, known as \textbf{Federated Cost Weighted Averaging (FedCostWAvg)} \citep{Machler2022FedCostWAvg}, incorporates  the reciprocal of the factor $\rho_c^{r}$, as follows
\begin{equation}
w_c^r = \alpha \frac{n_c}{N} + (1-\alpha) \frac{1}{\rho_c^{r}\sum_{c'=1}^C (\rho_{c'}^{r})^{-1}},
\label{eq:FedCostWAvg_wei}
\end{equation}

\noindent
where $\alpha \in [0,1]$ is a hyper-parameter that needs to be selected appropriately.

An extension of this method called \textbf{Federated PID Weighted Averaging (FedPIDAvg)} \citep{Machler2023FedPIDAvg} has been inspired by the well-known proportional-integral-derivative controller mechanism in control theory. Here, the quotient in the factor $(\rho_c^{r})^{-1}$ has been changed to the difference between the loss functions in the consecutive rounds, $\mathcal{L}_c(\mathbf{\Theta}_c^{r-1};\mathcal{D}_c) - \mathcal{L}_c(\mathbf{\Theta}_c^r;\mathcal{D}_c)$, and the additional summation over the local loss function from the previous five global rounds included.

Compared to DWA and FedPIDAvg procedures that incorporate the factor built upon the local loss function defined in previous rounds, the following two methods utilize non-linearly transformed local loss functions. First, the
\textbf{Mixed Supervised Federated Learning (FedMix)} \citep{Wicaksana2023FedMix} incorporates the normalized local loss function raised to the power of $\beta$

\begin{equation}
u_c^r = \frac{ \mathcal{L}_c(\mathbf{\Theta}^r_c; \mathcal{D}_c)^{\beta}}{\sum_{c'=1}^C \mathcal{L}_{c'}(\mathbf{\Theta}^r_{c'};\mathcal{D}_{c'})^{\beta}},
\label{eq:FedMix_u}
\end{equation}

\noindent
under the weight $w_c^r$ calculated as follows:
\begin{equation}
w_c^r = \frac{n_c/N + \lambda u_c^r}{\sum_{c'=1}^C \left(n_{c'}/N + \lambda u_{c'}^r \right)}
\label{eq:FedMix_wei}
\end{equation}

\noindent
with $\lambda$ being the hyper-parameter.

\noindent
The second method that nonlinearly transforms the loss function is the \textbf{Model-based Federated learning (ModFed)} \citep{Wu2023ModFed}, which uses the exponential function, as follows:
\begin{equation}
w_c^r = \frac{\exp\left( \mathcal{L}_c(\mathbf{\Theta}_c^r; \mathcal{D}_c) \right)}{\sum_{c'=1}^C \exp\left( \mathcal{L}_{c'}(\mathbf{\Theta}_{c'}^r; \mathcal{D}_{c'}) \right)}.
\label{eq:ModFed_wei}
\end{equation}

Weighting factors can also be estimated dynamically based on local and global model parameters \citep{Khan2022SimAgg, Khan2023RegSimAgg, Yang2023DynAggr, Lai2024Bilateral} or using various metrics, such as local prediction accuracy given by \citep{Yang2023DynAggr}
\begin{equation}
\mathrm{acc}_c = \frac{\mathrm{ratio\,\,of\,correct\,\,predictions}}{\mathrm{total\,\,number\,\,of\,\,predictions}}.
\label{eq:acc_wei}
\end{equation}

Another group of methods is similarity-based techniques, which calculate the similarity between the parameters of the local and global models:
\begin{equation}
y_c^r = \frac{\text{sim}_{c,1}^r(\mathbf{\Theta}_{c}^r, \overline{\mathbf{\Theta}}^r )}{\sum_{c'=1}^{C}\text{sim}_{c',1}^r(\mathbf{\Theta}_{c'}^r, \overline{\mathbf{\Theta}}^r )},
\label{eq:SimAgg_u}
\end{equation}
where $\overline{\mathbf{\Theta}}^r$ is the non-weighted average of the parameters of all local models, calculated in round $r$.

The similarity between two models represented by $\mathbf{\Theta}_{1}^r$ and $\mathbf{\Theta}_{2}^r$ at round $r$ is given by
\begin{equation}
\text{sim}_{c,p}^r ( \mathbf{\Theta}_{1}^r, \mathbf{\Theta}_{2}^r ) = \frac{\sum_{c'=1}^{C}\|\mathbf{\Theta}_{c'}^r-\mathbf{\Theta}_2^r\|_p^p}{\|\mathbf{\Theta}_1^r-\mathbf{\Theta}_2^r\|_p^p},
\label{eq:SimAgg_general}
\end{equation}

\noindent
It is worth to remark two methods here, namely \textbf{Similarity Weighted Aggregation (SimAgg)} \citep{Khan2022SimAgg} and \textbf{Regularized Aggregation (RegAgg)} \citep{Khan2022SimAgg}. The former defines the weight as
\begin{equation}
w_c^r = \frac{y_c^r + n_c/N}{\sum_{{c'}=1}^{C}(y_{c'}^r + n_{c'}/N)}
\label{eq:SimAgg_wei}
\end{equation}
while the latter uses the multiplication between the similarity factor \eqref{eq:SimAgg_u} and a normalized number of samples at each client:
\begin{equation}
w_c^r = \frac{y_c^r n_c}{\sum_{{c'}=1}^{C} y_{c'}^r n_{c'}}.
\label{eq:RegAgg_wei}
\end{equation}

Yet another method, the \textbf{Privacy-preserving Generative Adversarial Network (MixFedGAN)} framework \citep{Yang2023DynAggr} uses the similarity \eqref{eq:SimAgg_general} in its aggregation weights combined with the local model test accuracy:
\begin{equation}
w_c^r = \alpha \frac{\text{acc}_c}{\sum_{c'=1}^C \text{acc}_{c'}} + \beta\, \left( \text{sim}_{c,2}^r ( \mathbf{\Theta}_{c}^r, \mathbf{\Theta}^r ) \right)^{-1},
\label{eq:MixFedGAN_wei}
\end{equation}

\noindent
where $\text{acc}$ is the accuracy of the predictions, $\alpha$ and $\beta$ are positive constants used to balance both terms in the weight.

Another class of methods used to enhance the efficacy of the global model compared to the standard approaches is based on aggregating only selected parameters from each vector $\mathbf{\Theta}_c^r$ \citep{Gunesli2021FedDropoutAvg, Bernecker2022FedNorm}.
The \textbf{Federated Dropout Averaging (FedDropoutAvg)} \citep{Gunesli2021FedDropoutAvg} is one of the representatives of this class that assumes that some randomly selected parameters from $\mathbf{\Theta}_c^r$ are dropped before global aggregation. This approach introduces a new parameter called Federated Dropout Rate (FDR), which determines the degree of dropout (i.e., the number of clients to be removed from the aggregation process). If the $\text{FDR} = 0$, the FedDropoutAvg algorithm reduces to the FedAvg.

The \textbf{Modality-Based Normalization in Federated Learning (FedNorm)} \citep{Bernecker2022FedNorm} based on a modification of FedBN is yet another technique that falls into the same category. This method uses the mode normalization (MN) technique. FedNorm differentiates between normalization parameters (all parameters and statistics from all MN layers) and non-normalization parameters (all other parameters). Once the training is complete locally, the server aggregates only the non-normalization parameters.

Finally, to reduce accuracy discrepancies among the involved clients, a novel objective function has been introduced \citep{Hosseini2023PropFFL}

\begin{equation}
\mathcal{L}(\mathbf{\Theta};\mathcal{D}) = (1-\lambda) \mathcal{L}_\text{red}(\mathbf{\Theta}; \mathcal{D}) + \lambda \mathcal{L}_\text{fair}(\mathbf{\Theta}; \mathcal{D}),
\label{eq:Prop-FFL_objective}
\end{equation}
\noindent
where $\lambda$ is a hyper-parameter that balances both term with $\mathcal{L}_\text{red}$ and $\mathcal{L}_\text{fair}$ being the objective functions defined as follows:
\begin{equation}
\mathcal{L}_\text{fair}(\mathbf{\Theta}; \mathcal{D})= \sum_{c=1}^C \log \left\{ \frac{\sum_{{c'}=1}^C \mathcal{L}_{c'}(\mathbf{\Theta};\mathcal{D}_{c'})}{\mathcal{L}_c(\mathbf{\Theta};\mathcal{D}_c)} \right\},
\label{eq:Prop-FFL_fair}
\end{equation}
and
\begin{equation}
\mathcal{L}_\text{red}(\mathbf{\Theta}; \mathcal{D}) =  \frac{1}{\beta+1} \sum_{c=1}^C \mathcal{L}_c(\mathbf{\Theta};\mathcal{D}_c)^{\beta+1}.
\label{eq:Prop-FFL_red}
\end{equation}
\noindent

\noindent
The loss function $\mathcal{L}_\text{fair}$ stabilizes the variations in favoring particular client(s). Thus, the same relative loss is observed through all involved sites. This method has been called the \textbf{Proportionally Fair Federated Learning (Prop-FFL)}. 

%% ----------------------------------------------
\subsection{Malicious clients}
\label{Malicious clients_Agg}

Many current FL frameworks are built with the assumption that clients are trustworthy. However, they frequently prove to be vulnerable to poisoning attacks by malicious collaborators who seek to intentionally degrade the overall accuracy of the global model.

The \textbf{Distance-based Outlier Suppression (DOS)} technique \citep{Alkhunaizi2022DOS} provides a mechanism to filter out local parameter updates from malicious clients during aggregation. DOS estimates the distance between the updates of parameters from different collaborators and utilizes the Copula-based Outlier Detection (COPOD) method to determine the outlier score. The server calculates both the Euclidean and cosine distances by comparing the local parameters submitted by different clients $c$ and $i$, $c,i \in \{1,2,...,C \}$. The normalized aggregation weights are defined as:

\begin{equation}
w_c^r = \frac{\exp(-o_c^r)}{\sum_{i=1}^C \exp(-o_i^r)},
\label{eq:DOS_wei}
\end{equation}

\noindent
where  $o^r$ is the element of outlier scores vector $\mathbf{o}^r$, $\mathbf{o}^r = (\mathbf{o}_E^r + \mathbf{o}_K^r)/2$, with $\mathbf{o}_E^r$ and $\mathbf{o}_K^r$ being the vectors of outlier scores of Euclidean and cosine distance matrices.

%% ----------------------------------------------
\subsection{Security}
\label{sec:security_agg}

The central premise of FL is to maintain privacy across the clients while generating a global model. Yet, especially in the case of heterogeneous data across the involved sites, the default FL mechanism (described in detail in section~\ref{Theory}) can use an insecure method of updating the global model.

To handle this problem, a \emph{secure aggregation} can be used \citep{Malekzadeh2021Dopamine, Makkar2023SecureFedFL, Mitrovska2024Secure, Myrzashova2024Safeguarding}. For example, \cite{Malekzadeh2021Dopamine} proposed the \textbf{Differentially Private Federated Learning on Medical Data} approach, which uses HE during global model aggregation. Each client generates a pair of keys, respectively public and private. The public key is distributed to all involved sites, while the private key is shared only with other collaborators except with the server. The server receives and aggregates encrypted model updates while every client decrypts the aggregated parameters before performing local training.

Another approach involves directly applying DP during the aggregation process \citep{Shiri2024Differential, Ahmed2024Efficient, Kong2024Federated}. An example of this approach is the \textbf{Gaussian Differentially Private Federated Averaging with Adaptive Quantile Clipping (GDP-AQuCl)}, proposed by \citep{Shiri2024Differential}. This technique combines two key strategies: adds Gaussian noise to the local model updates before sending them to the central server, and uses an adaptive quantile clipping method for aggregation:

\begin{equation}
\mathbf{\Theta}^{r+1} = \mathbf{\Theta}^r + \frac{1}{C} \sum_{c=1}^{C} \text{Clip} \left( \text{DP} \left( \mathbf{\Theta}^r - \mathbf{\Theta}_c^r \right), D^r \right),
\label{eq:GDP-AQuCl_agg}
\end{equation}

\noindent
where $\text{DP}(.)$ is the function that adds Gaussian noise to the model update, and $\text{Clip}(.)$ is the adaptive quantile clipping function, defined as follows:

\begin{equation}
\text{Clip} \left( \mathbf{\Theta}, D^r \right) = \mathbf{\Theta}/\text{max} \left( 1, \frac{||\mathbf{\Theta}||_2}{D^r} \right).
\label{eq:GDP-AQuCl_clip}
\end{equation}

An alternative solution for secure aggregation is to replace a centralized approach with \emph{peer-to-peer} \citep{Giuseppi2022FedLCon} or \emph{blockchain} schemes \citep{Kalapaaking2023Block, Noman2023Block}. These avenues often require developing dedicated aggregation functions.

In a decentralized (peer-to-peer) FL, the \textbf{Consensus-based Distributed FL (FedLCon)} \citep{Giuseppi2022FedLCon} procedure employs a consensus round to update the global model parameters $\mathbf{\Theta}^{r+1}$ without involving the central server. To reach a consensus after global round $r$, the collaborators exchange information $n_{\epsilon}$ times, starting from the initial values $\mathbf{z}_c^0 = \mathbf{\Theta}_c^{r}$:

\begin{equation}
\mathbf{z}_c^{g+1} = \mathbf{z}_c^g + \frac{\epsilon}{n_c} \sum_{c' \in \eta_c} (\mathbf{z}_{c'}^g - \mathbf{z}_c^g)
\label{eq:FedLCon_z}
\end{equation}

\noindent
with $g = 0,...,n_{\epsilon}-1$ being a consensus round, $\eta_c$ a set of neighbours connected to client $c$--th that can exchange local model parameters, and $n_c$ the number of samples from collaborator $c$, and $\epsilon$ denotes the step size. At the end of the communication, the proxy variables $\mathbf{z}_c^{n_{\epsilon}-1}$ approximate the parameters of the global averaged model $\mathbf{\Theta}^{r}$ (cf.~Eq.~\eqref{eq:FedAvg_agg}):

\begin{equation}
\mathbf{z}_c^{n_{\epsilon}-1} \approx \frac{1}{N}\sum_{c'=1}^C n_{c'} \mathbf{\Theta}_{c'}^{r}.
\label{eq:FedLCon_agg}
\end{equation}

The \textbf{Blockchain-based Secure Aggregation} \citep{Kalapaaking2023Block} uses the additive secret-sharing scheme for the secure aggregation of the model. It allows  trusted third party $\mathcal{T}$ to receive secrets $s_{c'}$ (i.e., the selected parameters of the models $\mathbf{\Theta}_{c'}$) from parties $P_1,P_2,...,P_{C'}$. 

Each party computes the sum of the shares in $\mathbf{\Theta}_{c,i}$ to obtain the average of the model parameters:

\begin{equation}
\mathcal{P}_{i} = \left( \sum_{c'=1}^{C'} \mathbf{\Theta}_{c',i} \right) \mathrm{mod} ~ Q,
\label{eq:Blockchain_P}
\end{equation}

\noindent
where $i$ is the parameter index and $Q$ denotes high prime number generated by $\mathcal{T}$. The sum of shares from all sites $P_1,P_2,...,P_{C'}$ is then added together and averaged as follows:

\begin{equation}
\mathbf{\Theta}_{i} = \frac{1}{C'}\left( \sum_{c'=1}^{C'} \mathcal{P}_{i} \right) \mathrm{mod} ~ Q,
\label{eq:Blockchain_Theta}
\end{equation}
where $\mathbf{\Theta}_{i}$ refers to $i-$th parameter index ($i=0, \ldots, p-1$) of the global model.

%% --------------------------------------------------------------------------------------------------
%% --------------------------------------------------------------------------------------------------
%% --------------------------------------------------------------------------------------------------
\section{Learning methods in medical imaging}
\label{sec:fl_learning}

\textbf{Centralized approach} is the most common learning method used in context of MI \citep{Li2019TumSeg, Roth2020BrDen, Li2020fMRI, Sheller2020Federated, Wang2020Autom, Liu2021FedDG, Lo2021Federated, Zhang2021Dynamic, Ziller2021Differentially, Huang2021CML, Lee2021Federated, Sarma2021Federated, Gunesli2021FedDropoutAvg, Ke2021Style, Cetinkaya2021Commu, Shen2021DWA, Tian2021Privacy, Stripelis2021Scali, Stripelis2021SecureNeuro, Pati2021Federated, Parekh2021Cross-domain, Feki2021X-ray, Roth2021Prost, Malekzadeh2021Dopamine, Guo2021FL-MR, Dou2021FederatedDL, Linardos2022FL-EV, Bernecker2022FedNorm, Slazyk2022CXR-FL, Subramanian2022Effect, Luo2022FedSLD, Zhang2022SplitAVG, Li2022Breast, He2022CosSGD, Yang2022FedZaCt, Alkhunaizi2022DOS, Zhou2022Comm-effi, Islam2022Effectiveness, Lu2022DistributionFree, Adnan2022Federated, Florescu2022Pre-train, Elshabrawy2022Ensemble, Yang2022RosFL, Bercea2022FedDis, Guo2022Auto-FedRL, Khan2022SimAgg, Machler2022FedCostWAvg, Dalmaz2022MRITrans, Misonne2022HeartSeg, Muthukrishnan2022MammoDL, Kumar2022DataVal, Stripelis2022Secure, Lu2022Giga, Baid2022Federated, Dalmaz2022OneModel, Kandati2023Federated, Lin2023Hyper, Machler2023FedPIDAvg, Khan2023RegSimAgg, Wicaksana2023FedMix, Tan2023Transf, Wu2023ModFed, Kulkarni2023FedFBN, Liu2023MultSkle, Liu2023Class, Sanchez2023Memory, Jiang2023IOP-FL, Hosseini2023PropFFL, Yang2023DynAggr, Zhao2023ADI, Mushtaq2023FAT, Feng2023FedMRI, Zhou2023FedFTN, Elmas2023FedGIMP, Levac2023EndToEnd, Rajagopal2023Federated, Denissen2023Transf, Makkar2023SecureFedFL, Peng2023FedNI, Kaushal2023Eye, Wu2023FedIIC, Wang2023FedDP, Kim2024Federated, Naumova2024MyThisYourThat, Qi2024Simulating, Alsalman2024Federated, Mitrovska2024Secure, Yamada2024Investigation, Yan2024Cross, Lai2024Bilateral, Xiang2024Federated, Abbas2024Multidisciplinary, Zheng2024Federated, Liu2024Multi, Zhou2024DistributedFL, Hossain2024Collaborative, Sun2024FKD, Deng2024Feddbl, Babar2024Investigating, Ahmed2024Efficient, Kong2024Federated, Shiri2024Differential, Gupta2024Enhancing, Zhang2024Fedsoda, Albalawi2024Integrated, Vo2024Comparison, Liang2024ACFL, Wang2024FedDUS, Deng2024Federated}, although a \textbf{decentralized federated learning}, \textbf{peer-to-peer} \citep{Roy2019BrainTorrent, Wu2021FCL, Giuseppi2022FedLCon, Huang2022Continual, Kalra2023Decen, Khan2024Bilevel} or \textbf{blockchain-based approach} \citep{Nguyen2022GenAdv, Kalapaaking2023Block, Noman2023Block, Myrzashova2024Safeguarding, Gupta2024Blockchain, Kumar2024Privacy}, can also be found in some articles.

In recent years, several modern solutions have emerged to overcome the significant challenges of federated learning, such as data heterogeneity \citep{Li2020fMRI, Guo2021FL-MR, Liu2021FedDG, Guo2021FL-MR, Huang2021CML, Yang2022FedZaCt, Bercea2022FedDis, Zhu2022Federated, Zhang2022SplitAVG, Dalmaz2022MRITrans, Dalmaz2022OneModel, Luo2022FedSLD, Zhao2023ADI, Elmas2023FedGIMP, Feng2023FedMRI, Wu2023ModFed, Jiang2023IOP-FL, Sanchez2023Memory, Yang2023DynAggr, Lin2023Hyper}, or to improve the data security \citep{Li2019TumSeg, Ziller2021Differentially, Malekzadeh2021Dopamine, Yang2022RosFL, Nguyen2022GenAdv, Kalra2023Decen}. This section presents the learning algorithms addressing limitations specific to MI, including model accuracy, data heterogeneity, label deficiency, and security.

\begin{figure*}[!t]
\centering
\includegraphics[scale=0.111]{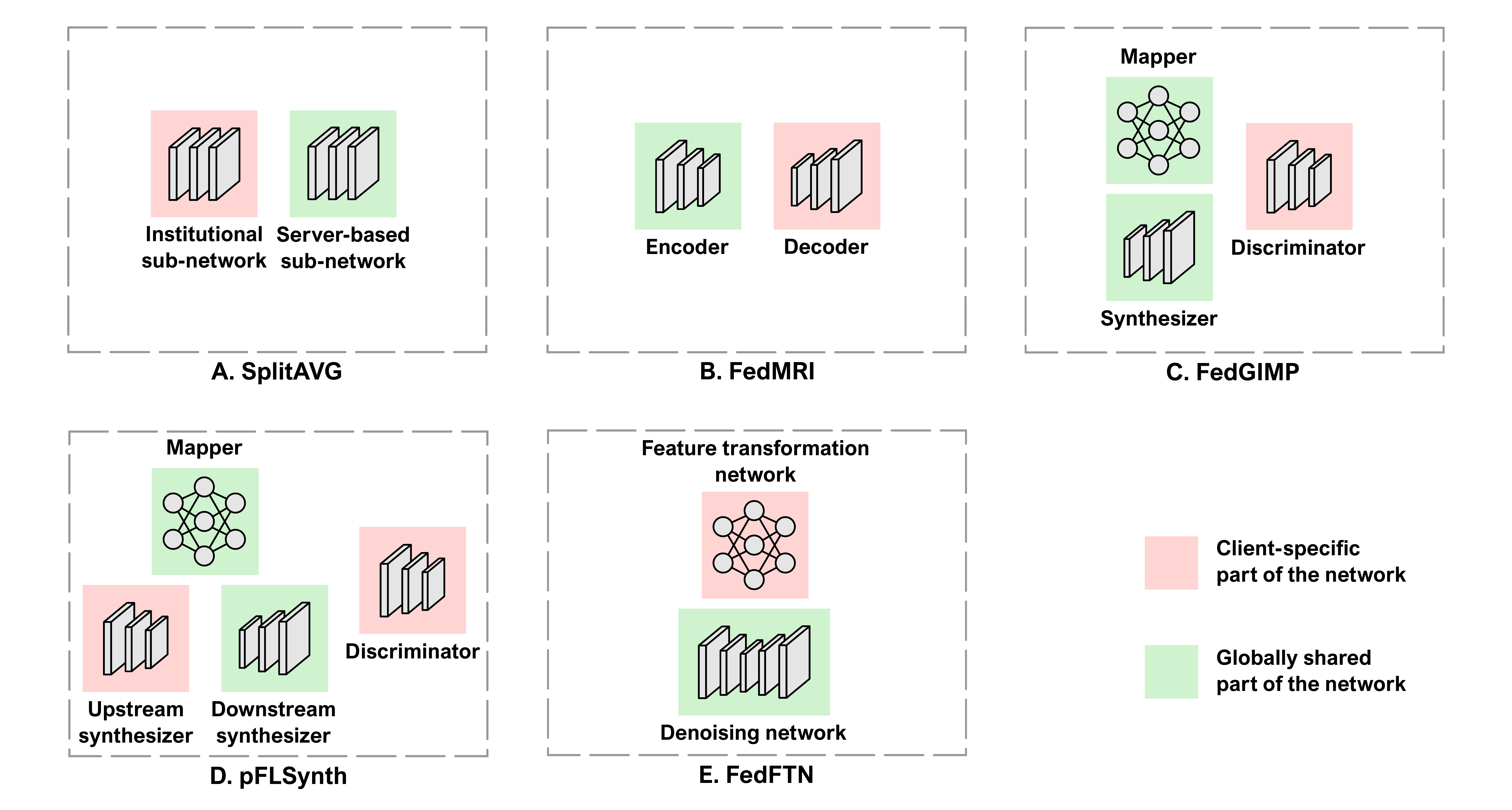}
\caption{Schematic representation of split learning techniques: \textbf{A.} SplitAVG \citep{Zhang2022SplitAVG}, \textbf{B.}~FedMRI \citep{Feng2023FedMRI}, \textbf{C.}~FedGIMP \citep{Elmas2023FedGIMP}, \textbf{D.}~pFLSynth \citep{Dalmaz2022OneModel}, \textbf{E.}~FedFTN \citep{Zhou2023FedFTN}. Each scheme represents a local model of a single client participating in the FL-based procedure. The model is divided into two components: the client-specific part and the globally shared part. The client-specific part (highlighted in red) is available exclusively to the client and remains private. In contrast, the globally shared part (highlighted in green) is sent to the central server for the aggregation procedure.}
\label{fig:Split_learning}
\end{figure*}

%% ----------------------------------------------
\subsection{Improving model accuracy}
\label{sec:fl_learning:accuracy}
The objective of novel learning schemes is, in general, to enhance the global model accuracy after the aggregation phase. The global model accuracy may decline after aggregating individual model parameters once evaluated by some or all clients involved in the FL-based procedure. An effective approach to address this issue is to employ \emph{curriculum learning}, which involves presenting training data to a neural network in a specific sequence. The method has been effective in classifying breast cancer using mammography data.

In the following, we briefly review current approaches used to improve the accuracy of the global model.

The \textbf{Memory-aware Curriculum Federated Learning} \citep{Sanchez2023Memory} penalizes inconsistent predictions caused by the \textit{forgotten samples}, i.e., the inputs for which the predictions become incorrect after the global model aggregation phase. The method uses a data scheduler with a scoring function, curriculum probabilities, and a permutation function. The scoring function assigns each sample a score based on local and global predictions. The normalized score values comprise the curriculum probability used to sample the training set. Higher probabilities indicate samples that are presented earlier to the optimizer. The permutation function can be used to modify the order of the data fed to the optimizer.

An alternative approach used to improve the accuracy of a global model is to follow \emph{continual learning} \citep{Sheller2020Federated} with a particular focus on the use of \emph{synaptic intelligence (SI)} \citep{Huang2022Continual}. Synapses accumulate task-relevant information over time, enabling them to efficiently create and retain new memories without forgetting old ones.
An example of SI is the \textbf{Continual Learning for Peer-to-Peer Federated Learning} approach \citep{Huang2022Continual}. This method introduces a slightly different peer-to-peer scheme than the one discussed in section \ref{sec:FL_intro_learning}. The clients do not share their trained local models for the aggregation procedure. Instead, a single model is trained and transferred successively from one institution to another, repeating this process multiple times. This approach penalizes alternations in model parameters that are important for gaining knowledge from the previous client's data.

%% ----------------------------------------------
Yet another approach to handle the problem of global model deterioration is to use \emph{dynamic weight correction strategy}. 
An example method of this kind is the \textbf{Robust Split Federated Learning (RoS-FL)} \citep{Yang2022RosFL}. Here, the global model in round $r$ is used as the anchor model and the correction model parameters are formulated accordingly:
\begin{equation}
\mathbf{\Theta}_\mathrm{cor}^r = \mathbf{\Theta}^r + \eta \nabla \mathcal{L}_\mathrm{cor}(\mathbf{\Theta}^r,\mathbf{\Theta}^{r-1}),
\label{eq:RoS-FL_cormodel}
\end{equation}
where $\mathcal{L}_\mathrm{cor}$ is the correction loss function that stabilizes the training process and diminishes the model drift. The robust model is then obtained as a linear combination of the anchor and the corrected model $\mathbf{\Theta}_\mathrm{cor}^r$

\noindent
\begin{equation}
\mathbf{\Theta}_\mathrm{robust}^r = (1-\alpha) \mathbf{\Theta}^r + \alpha \mathbf{\Theta}_\mathrm{cor}^r.
\label{eq:RoS-FL_robmodel}
\end{equation}

%% ----------------------------------------------
\subsection{Data heterogeneity}
\label{sec:fl_learning:data_heterogeneity}
In real-world scenarios, medical data typically comes from multiple sites and may have significantly different distributions, resulting in data heterogeneity. Consequently, it can lead to substantial accuracy degradation of the global model, as reported in the previous section.

\underline{Personalized federated learning:\\}
The simplest method to personalize the models is to use a domain-specific such as \emph{personalized FL} \citep{Jiang2023IOP-FL, Wang2023FedDP}. This procedure involves fine-tuning the global model during global rounds or after the local training, using local data to make it more suitable for each client.

Specifically, the \textbf{Inside and Outside model Personalization in FL (IOP-FL)} framework \citep{Jiang2023IOP-FL} employs a locally adapted model for inside personalization (i.e., clients involved in the joint training of the global model) and applies test-time routing for outside personalization (i.e., for noninvolved clients), enabling institutions that are not part of the FL process to use the trained model effectively. By combining local gradients (client-specific optimization) and global gradients (common knowledge) during local model training, the framework generates a model that can learn the common pattern and fit the local data distribution. 
The IOP-FL approach allows for a flexible combination of training knowledge and information from test data.

The \textbf{Federated Learning Scheme with Dual Personalization (FedDP)} \citep{Wang2023FedDP} improves model personalization from both feature and prediction perspectives. FedDP covers two stages: dependency personalization and inter-site prediction inconsistencies. In the former, the clients send their globally shared parameters to the central server. The parameters are aggregated and then sent back to the institutions, which combine them with their locally personalized parameters to generate a set of preliminary local models. In the latter stage, the clients collect all their local models and compute their inconsistencies with local datasets. The phrase ``inconsistencies" refers here to the differences between the output of a model personalized for a specific client and the outputs of models for other clients. The parameters of each model are eventually adjusted according to the identified inter-site inconsistencies using a specialized loss function. This process leads to more accurately calibrated models.

%% ----------------------------------------------
\underline{Split learning:}
An important group of personalized methods constitutes the \emph{split learning} procedure, which divides the parameters into local and global parts \citep{Zhang2022SplitAVG, Dalmaz2022MRITrans, Dalmaz2022OneModel, Elmas2023FedGIMP, Feng2023FedMRI, Zhou2023FedFTN} (see Fig.~\ref{fig:Split_learning} for comparison). In the following, we briefly characterize the methods following this paradigm.

The \textbf{Split Averaging (SplitAVG)} by \cite{Zhang2022SplitAVG} (see Fig.~\ref{fig:Split_learning}A) divides a DL network into two sub-networks, respectively, local and global at a specific cut layer. Model training involves forward and backpropagation. First, intermediate feature maps are obtained from the local sub-network and sent with corresponding labels for concatenation to the central server. These concatenated feature maps are then propagated into the global sub-network. Second, gradients are backpropagated from the last layer to the first layer of the global model, with the first global layer gradients transferred back to all clients for further backpropagation. Finally, each client and the server update their sub-networks. An improved version of the SplitAVG technique is \textbf{SplitAVG-v2} \citep{Zhang2022SplitAVG}, which allows the preservation of locally confidential labels without sharing them with the server.
%% ----------------------------------------------
\textbf{Specificity-preserving FL algorithm for MR Image Reconstruction (FedMRI)} \citep{Feng2023FedMRI} is another method based on dividing the model into two parts: a globally shared encoder and a client-specific decoder (see Fig.~\ref{fig:Split_learning}B). In addition to the standard loss function, the FedMRI employs an extra specialized weighted contrastive regularization loss that enables further correction for local updates and provides the model with global recognition ability.
\begin{equation}
\mathcal{L}_{\mathrm{con}_c}(\mathbf{\Theta}_{E}^r,\mathbf{\Theta}_{E_c}^r,\{\mathbf{\Theta}_{E_{c'}}^{r-1}| c' \in C \}) = \frac{||\mathbf{\Theta}_{E_c}^r-\mathbf{\Theta}_E^r||_1}{\sum_{c'=1}^C ||\mathbf{\Theta}_{E_{c'}}^{r-1}-\mathbf{\Theta}_{E_c}^r||_1}
\label{eq:FedMRI_loss}
\end{equation}

\noindent
with $\mathbf{\Theta}_{E}^r$ and $\mathbf{\Theta}_{E_c}^r$ being the parameters of global and local encoders at round $r$.

The \textbf{Federated Learning of Generative Image Priors (FedGIMP)} \citep{Elmas2023FedGIMP} employs an unconditional adversarial model divided into a shared generator and a local discriminator (see Fig.~\ref{fig:Split_learning}C). After every epoch, the local generators are sent to the server and aggregated similarly to the FedAvg method. A model trained using this technique can be employed for inference, allowing it to adapt to a new dataset not included during training. During the inference, the trained MRI prior, which is represented as an adversarial model capable of synthesizing MRI images using site-specific latent variables, is combined with a subject-specific imaging operator and adjusted to minimize data consistency loss.
Another method based on split learning is the \textbf{Personalized FL Method for MRI Syntheses (pFLSynth)} by \cite{Dalmaz2022OneModel}. Here, the adversarial model is divided into a globally shared downstream synthesizer and mapper and a client-specific upstream synthesizer and discriminator (see Fig.~\ref{fig:Split_learning}D). Finally, only the downstream synthesizer and mapper parameters are aggregated on the central server
\begin{equation}
\mathbf{\Theta}^{r+1}_\mathrm{syn} = \sum_{c=1}^{C} w_c \mathbf{\Theta}^r_{\mathrm{syn}_{c}}, \ \ \ \mathbf{\Theta}^{r+1}_\mathrm{map} = \sum_{c=1}^{C} w_c \mathbf{\Theta}^r_{\mathrm{map}_{c}},
\label{eq:pFLSynth_agg}
\end{equation}

\noindent
where $\mathbf{\Theta}_\mathrm{syn}^{r+1}$/$\mathbf{\Theta}_\mathrm{map}^{r+1}$ are the aggregated parameters of the downstream synthesizer/mapper in round $(r+1)$ and $\mathbf{\Theta}^r_{\mathrm{syn}_{c}}$/$\mathbf{\Theta}^r_{\mathrm{map}_{c}}$ are client-specific upstream synthesizer/mapper parameters.

Another method under the umbrella of split learning discussed here is the \textbf{Personalized Federated Learning Method Based on Deep Feature Transformation Networks (FedFTN)} \citep{Zhou2023FedFTN}. This model consists of two distinct networks: the globally shared denoising network and a local deep feature transformation network that modulates the shared network's feature outputs (see Fig.~\ref{fig:Split_learning}E). After two initial global epochs, the global weight constraint loss is incorporated into the standard loss function to stabilize the parameter update during local training:
\begin{equation}
\mathcal{L}_{\mathrm{gwc}_c}(\mathbf{\Theta}_\mathrm{den}^r,\mathbf{\Theta}_{\mathrm{den}_c}^r) = ||\mathbf{\Theta}_\mathrm{den}^r - \mathbf{\Theta}_{\mathrm{den}_c}^r||_2^2
\label{eq:FedFTN_loss}
\end{equation}

\noindent
with $\mathbf{\Theta}_\mathrm{den}^r$ and $\mathbf{\Theta}_{\mathrm{den}_c}^r$ being the parameters of the global, averaged by the central server, and local denoising network, respectively.

In addition to dividing the model parameters into local and global parts, it is also possible to \emph{disentangle the parameter space into shape and appearance} and share only the shape parameter between the clients. This separation uses shared structural information from various clients and reduces domain shifts by enabling personalized appearance models. 
This kind of approach is used by \textbf{Federated Disentanglement (FedDis)} algorithm \citep{Bercea2022FedDis}. The FedDis assumes that the anatomical structure of the data representing the brain is similar across all collaborators and that the heterogeneity is mainly due to differences in data collection devices and their parameters.

\begin{figure}[t!]
\centering
\includegraphics[scale=0.12]{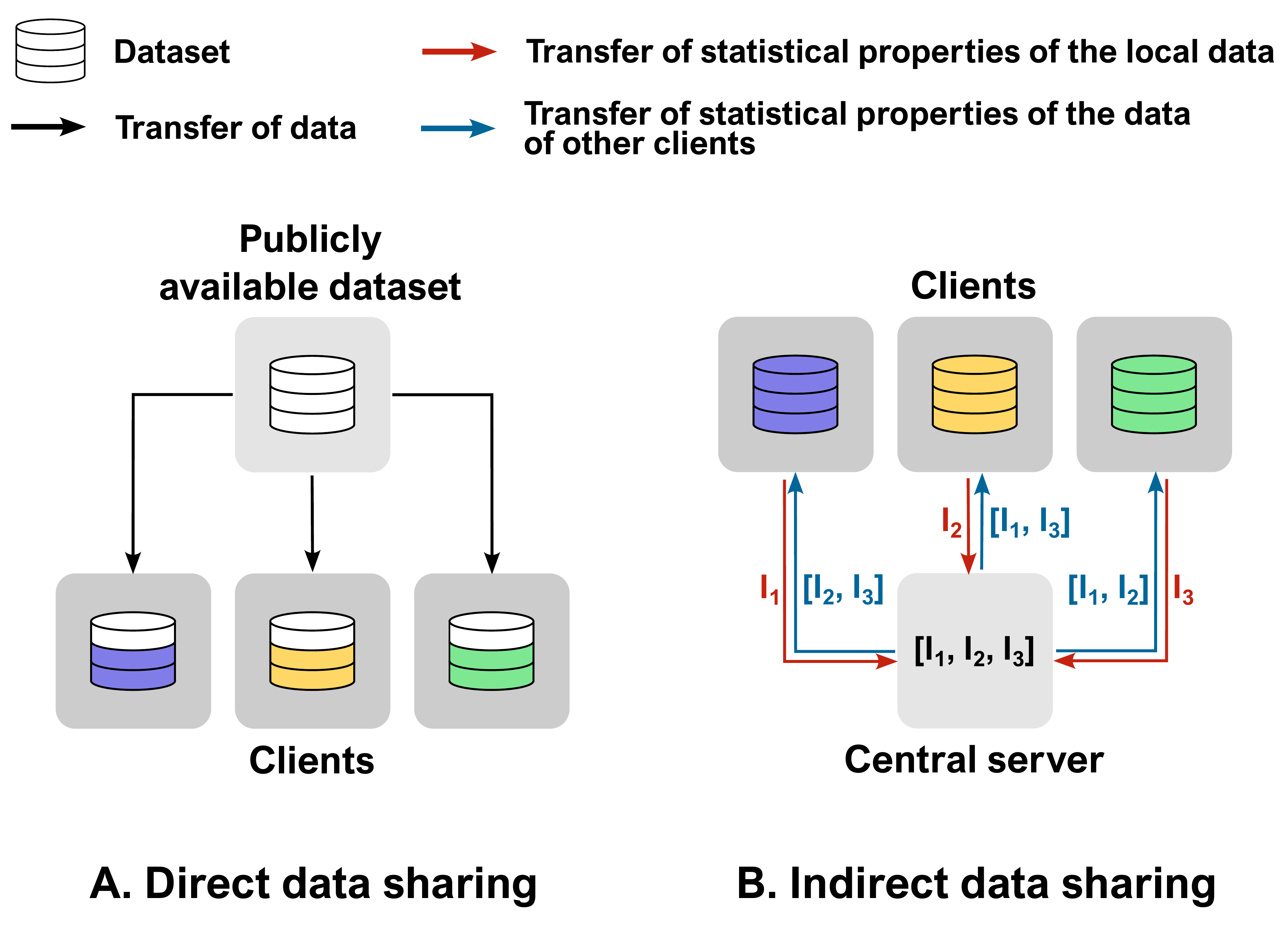}
\caption{Comparison between two types of data sharing-based techniques. \textbf{A.} Direct data sharing assumes the existence of a publicly available dataset that all customers can access. \textbf{B.} In indirect data sharing, each client sends statistical properties of their data to a central server, which are then shared with other clients.}
\label{fig:Data_sharing}
\end{figure}

%% ----------------------------------------------
\underline{Data sharing:} 
The next group of methods handling data heterogeneity follows \emph{data sharing} scheme, which can be direct \citep{Guo2021FL-MR} or indirect \citep{Liu2021FedDG, Luo2022FedSLD, Zhao2023ADI, li2025retrospective}. Direct data sharing involves distributing the publicly available dataset among all clients, while indirect data sharing indicates that clients exclusively exchange details about the data, such as its distribution, instead of the actual data (see Fig.~\ref{fig:Data_sharing} for a graphical explanation).

The first indirect data sharing method mentioned here is the \textbf{Episodic Learning in Continuous Frequency Space (ELCFS)} by \cite{Liu2021FedDG}, which enables multiple clients to benefit from data distributions from different sources. The ELCFS employs a continuous frequency space interpolation mechanism to share distribution information between clients. Initially, low-level distributions, such as style, are extracted from the frequency domain of local participant data as an amplitude spectrum. These distributions are then compiled into a centralized distribution bank that is accessible to all clients. Consequently, this approach enables the generation of images with modified appearances that reflect the distribution characteristics of other clients. Additionally, ELCFS employs an episodic meta-learning approach that learns generalizable model parameters by simulating train and test domain shifts. Combined with boundary-oriented meta-optimization, which focuses on distinguishing boundary-related features from background-related ones, it enhances domain generalization.

Another example of an indirect data sharing method is \textbf{Federated Learning with Shared Label Distribution (FedSLD)} \citep{Luo2022FedSLD}. This method assumes that the number of samples from every class for all clients is known. This information is then used in the local objective function to ensure that each class contributes proportionally, relative to the total number of samples from that class across all clients within the FL architecture.

Yet another indirect method called \textbf{Distribution Information Sharing Federated Learning approach (FedDIS)} by \cite{Zhao2023ADI} enables the involved clients to share their unique local data distributions with each other to create an independent and identically distributed (IID) dataset. The method trains a variational autoencoder (VAE) and subsequently maps local medical data to a hidden space via the encoder. The distributional information about the mapped data in the hidden space is then estimated and shared among the clients. Clients can generate a new dataset based on the received distribution information using the VAE decoder.

Another method of indirect data sharing is \textbf{Hepa-FedBoost} \citep{li2025retrospective}. Hepa-FedBoost sends anonymized data summaries, known as prototypes, along with model parameters to the server. Each client generates two types of prototypes during local training: 1) class-specific prototypes, computed as the mean feature vector across samples of a specific class, and 2) domain-agnostic prototypes, which aggregate global appearance cues across all data patches regardless of the class. These prototypes enable the central server to group clients with similar data characteristics. Client clusters are updated regularly, and model parameter aggregation occurs separately for each cluster. Hepa-FedBoost utilizes a technique called global-local contrastive boosting to distinguish between positive and negative samples. This approach helps refine decision boundaries, which are the lines, curves, or hyperplanes in a feature space that separate different classes.

%% ----------------------------------------------
\underline{Domain adaptation:} \emph{Domain adaptation} is the next presented approach to reduce domain shift between medical image datasets and increase model generalizability \citep{Li2020fMRI, Guo2021FL-MR, Sanchez2023Memory}.

The first of two domain adaptation approaches introduced here is the \textbf{Multi-site fMRI Analysis Framework} \citep{Li2020fMRI}. The technique uses two domain adaptation methods: a mixture of experts (MoE) and adversarial domain alignment. The MoE is a gating network that dynamically assigns weight values during the global aggregation of local models. This approach is used for adaptation near the output layer. 
On the other hand, adversarial domain alignment is a technique employed for adaptation on the data knowledge representation level.

The second method, previously mentioned under the umbrella of data sharing techniques, is the \textbf{FL-MR with Cross-site Modeling (FL-MRCM)} \citep{Guo2021FL-MR}. The FL-MRCM aims to solve the domain shift problem by using the data of the target site, along with the adversarial domain identifier, and the reconstruction network consisting of an encoder and a decoder. The encoder projects the input from the source and target sites into the latent space, and the identifier aligns the distribution of the latent space between each pair of source-target domains.

%% ----------------------------------------------
\underline{Knowledge distillation:} Distillation-based approaches often involve clients' models to teach the global model their data knowledge \citep{Huang2021CML, Yang2022FedZaCt, Yang2023DynAggr}.

\begin{figure}[t!]
\centering
\includegraphics[width=0.5\textwidth]{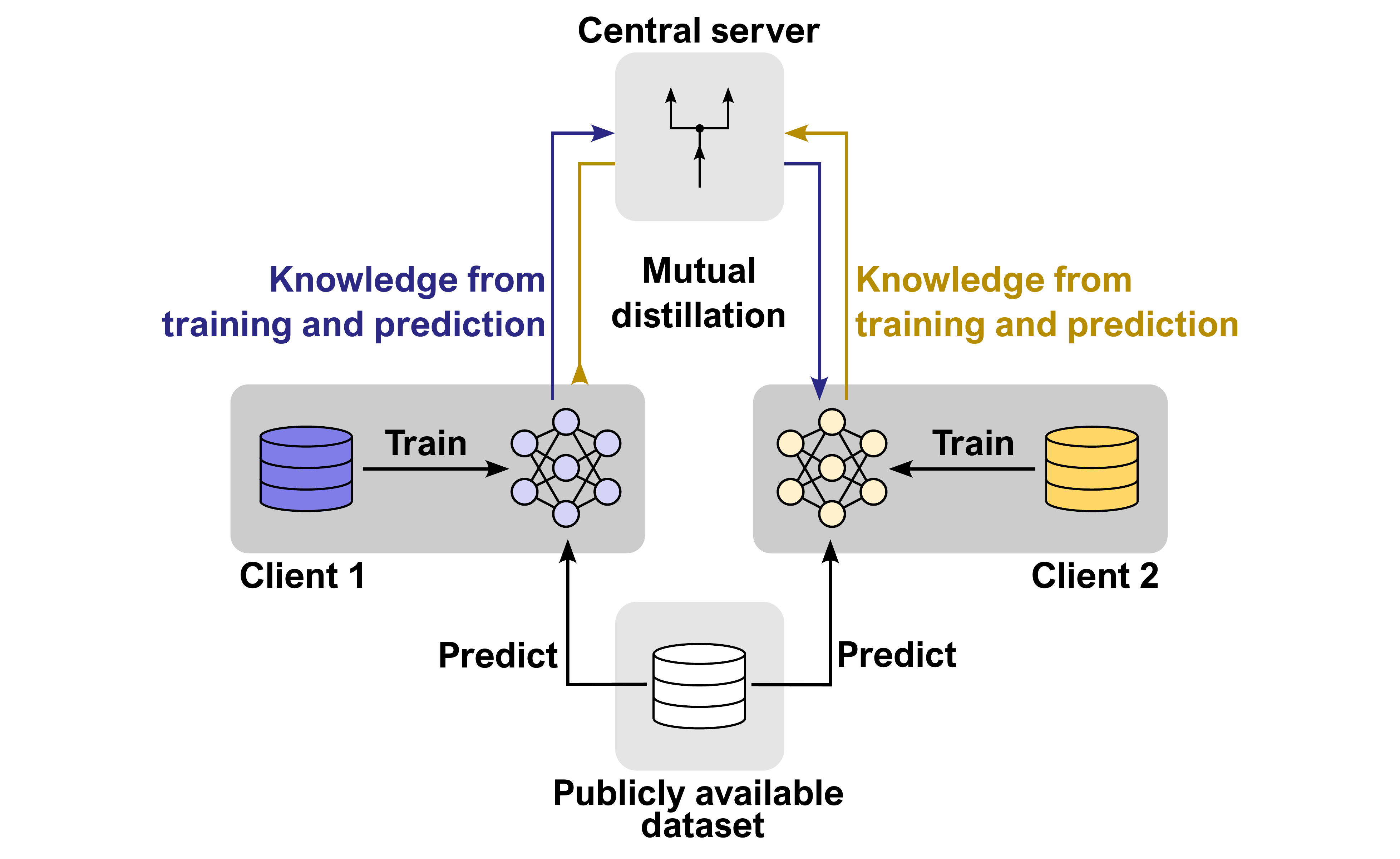}
\caption{A schematic representation of knowledge distillation-based FedCM technique \citep{Huang2021CML}. Clients upload predicted logits from public data and cross-entropy loss from private test sets to a central server. The server sends back the logits and loss to the remaining clients. Each client updates its knowledge through mutual distillation.}
\label{fig:Knowledge_distillation}
\end{figure}

The \textbf{Federated Conditional Mutual Learning (FedCM)} by \cite{Huang2021CML} uses the local performance of the clients and the similarity to improve the overall accuracy of the global model. The client performance is evaluated based on the cross-entropy loss calculated on their private test set, while similarity is assessed using predicted logits (class scores) from public data. Each client periodically uploads the loss value and the predicted logits to the server. The server then sends back the set of losses and logits of the remaining clients (see Fig.~\ref{fig:Knowledge_distillation}). The client updates its knowledge by mutual distillation and fine-tuning its local dataset for personalization.

\textbf{Federated Learning with Z-average and Cross-teaching (FedZaCt)} \citep{Yang2022FedZaCt} is another method based on knowledge distillation that uses the Z-average method to aggregate models from multiple collaborators. This approach first calculates normalized cross-evaluation metrics of Z score $Z_{c,c'}$, which are obtained by evaluating the local model of the client on the local data with all other clients. These metrics are then used to aggregate Z-average models that maintain diverse knowledge:
\begin{equation}
\mathbf{\Theta}^{r+1}_{Z_c} = \sum_{c'=1}^{C} w_{c'} \overline{Z}_{c,c'} \mathbf{\Theta}^r_{c'},
\label{eq:FedZaCt_agg}
\end{equation}

\noindent
where $\overline{Z}_{c,c'}$ is the average value of the normalized cross-evaluation metrics of Z score for clients $c$ and $c'$
\begin{equation}
\arraycolsep=5pt
\overline{Z}_{c,c'} = \left\{\begin{matrix}
0.5\left( Z_{c,c'} + Z_{c',c} \right) & \text{if} \ \ c \neq c', \\ 
0.5 & \text{otherwise.}
\end{matrix}\right.
\end{equation}

\noindent
Next, FedZaCt employs a cross-teaching method, which enables local clients to use the Z-average models to train their models. To achieve this, an additional component is included in the overall loss function. This component measures the difference between the predictions given by the local model and those of the Z-average models.

The \textbf{Privacy-preserving Generative Adversarial Network (MixFedGAN)} by \cite{Yang2023DynAggr} incorporates knowledge distillation into the local model. The method introduces a distillation regularization loss, which employs a Kullback-Leibler divergence to minimize the difference between the local and global models.

\underline{Contrastive learning:} The goal of contrastive learning is to learn low-dimensional representations of data by distinguishing between similar (positive) and dissimilar (negative) samples. This approach aims to bring the positive samples closer together in the representation space while moving the negative samples away from each other. To achieve this, standard distance metrics such as Euclidean distance or cosine similarity are often employed.

The \textbf{FedIIC} \citep{Wu2023FedIIC} is an example of a contrastive learning-based approach to address the challenges of class imbalance. FedIIC combines feature and classifier learning. Firstly, feature learning covers two levels of contrastive learning: intra- and inter-client. 
On the one hand, intra-client contrastive learning extracts class-specific features by contrasting representations of images within the same class against those from different classes. On the other hand, inter-client contrastive learning involves contrasting features across different clients. This procedure employs class-wise prototypes, which are generated from the global model weights and shared among all clients. By aligning features from various clients, the model achieves better generalization. Secondly, classifier learning enables the dynamic adjustment of the classifier in response to real-time difficulty. These difficulties are assessed by the average cross-entropy loss of all samples within a specific class using a global model. Such an adjustment helps to prevent the model from becoming biased towards majority classes, ensuring fair performance across all classes.

\underline{Synthesizing additional data:} This approach has been designed to deal with non-IID data and involves synthesizing additional virtual data. The \textbf{Federated Medical Image Analysis with Virtual Sample Synthesis} \citep{Zhu2022Federated} synthesizes training samples using local and global models. To improve the generalization of the local model, virtual adversarial training is applied. Unlike adversarial training, this method generates adversarial neighbors that closely resemble the selected input sample in terms of raw pixels but produce different predictions. In addition, the global model is used to generate high-confident samples, which enables the alignment of the local distribution to the global one.

\underline{Replacing the architecture:} The \textbf{Federated Learning with Hyper‑network} \citep{Lin2023Hyper} enables hyper-network training instead of global model training. The hyper-network is a neural network that produces parameters for a larger target model. The hyper-network uses a set of inputs that contain information about the structure of local model parameters and generates parameters for each layer of the target network.

%% ----------------------------------------------
\subsection{Label deficiency}
\label{sec:fl_learning:label_deficiency}
Supervised DL requires a large amount of labeled data to achieve satisfactory accuracy. In a real-world scenario, each client may cover insufficient labeled data due to the tedious and high cost of the data labeling procedure. As a result, the local model may be poorly trained and, therefore, negatively impact other clients involved in global training \citep{Liu2023Class, Kim2024Federated}. We can categorize the methods used to handle label deficiency into four groups, namely \emph{contrastive learning}, \emph{semi-supervised learning}, \emph{synthetic data generation} and \emph{knowledge distillation}.

\underline{Contrastive learning:} Contrastive learning, mentioned in subsection \ref{sec:fl_learning:data_heterogeneity}, has also been applied to address label deficiency problem \citep{Wu2021FCL, Dong2021Federated}.

In \textbf{Federated Contrastive Learning (FCL)} framework \citep{Wu2021FCL}, the clients exchange features with each other during the pre-training procedure to provide more diverse data. Local contrastive learning involves using two encoders: the main encoder, which is learned and used for initialization during fine-tuning, and the momentum encoder, a slowly evolving version of the main encoder, generating features to discriminate and intended for sharing. Then, the FCL aligns similar features among collaborators using the structural similarity of the data.

The \textbf{Robust Federated Contrastive Learning (FedMoCo)} framework \citep{Dong2021Federated} combines contrastive learning and metadata transfer, facilitating the augmentation of inter-node statistical data. During contrastive learning, the data augmentation procedure creates positive pairs by generating two random versions of the same image through various transformations, such as flipping and rotating. In contrast, negative pairs are formed using different versions of two distinct images. Eventually, the metadata transfer involves transferring the statistical information of the transformed features between clients.

\begin{table*}[!t]
\centering
\footnotesize
\begin{tabular}{p{1.4cm}| p{3.6cm}| p{1.5cm}|  p{4cm}|  p{5cm}}
\hline\hline
\multicolumn{1}{c|}{\multirow{2}{*}{\bf{Name}}} & \multicolumn{1}{c|}{\multirow{2}{*}{\bf{Application}}} & \multicolumn{1}{c|}{\bf{Programming }} & \multicolumn{1}{c|}{\bf{Information}} & \multicolumn{1}{c}{\multirow{2}{*}{\bf{Website}}}  \\
 & & \multicolumn{1}{c|}{\bf{language}}  & \multicolumn{1}{c|}{\bf{privacy}} & \\ \hline
APPFL & Comprehensive framework and benchmarking tool & Python & global and local differential privacy & \url{https://appfl.ai/}\\\hline
FATE & Comprehensive framework & Python & Secure computation protocols, secure multi-party computation, task security scheduling, homomorphic encryption & \url{https://github.com/FederatedAI/FATE}\\\hline
FEDML & Research and production integrated edge-cloud platform & Python & Secure aggregation, homomorphic encryption, differential privacy & \url{https://www.fedml.ai/}\\\hline
Fed-BioMed & Comprehensive framework & Python & Secure aggregation, differential privacy & \url{https://fedbiomed.org/}\\\hline
FedTree & Tree-based models & C++ & Secure aggregation, homomorphic encryption, differential privacy & \url{https://fedtree.readthedocs.io/}\\\hline
Flower & Comprehensive framework & Python & Secure aggregation, symmetric encryption, differential privacy & \url{https://flower.dev/}\\\hline
FLUTE & Comprehensive framework & Python &  Privacy accounting, differential privacy & \url{https://github.com/microsoft/msrflute}\\\hline
LEAF & Comprehensive framework & Python &  N/A & \url{https://leaf.cmu.edu/}\\\hline
NVIDIA FLARE & Comprehensive framework & Python & Federated authentication and authorization, privacy policy, homomorphic encryption, differential privacy & \url{https://developer.nvidia.com/flare}\\\hline
OpenFed & Comprehensive framework & Python &  N/A & \url{https://github.com/FederalLab/OpenFed}\\\hline
OpenFL & Comprehensive framework & Python &  Privacy loss reports, differential privacy & \url{https://github.com/securefederatedai/openfl}\\\hline
PaddleFL & Comprehensive framework & Python, C++  &Secure aggregation, secure multi-party computation, differential privacy &  \url{https://github.com/PaddlePaddle/PaddleFL}\\\hline
PySyft & Comprehensive framework & Python & Secure multi-party computation, homomorphic encryption, differential privacy & \url{https://openmined.github.io/PySyft/}\\\hline
Substra & Healthcare research & Python & Secure enclaves, secure multi-party computation, differential privacy & \url{https://www.owkin.com/substra}\\\hline
TFF & Comprehensive framework & Python &  Differential privacy & \url{https://www.tensorflow.org/federated}\\
\hline\hline
\end{tabular}
\caption{Summary of actively developed open-source federated learning frameworks.}
\label{table:Frameworks}
\end{table*}

%% ----------------------------------------------
\underline{Semi-supervised learning:} This approach can be used in two scenarios: 1) only a few clients have completely unlabeled data \citep{Mushtaq2023FAT}, 2) all clients have unlabeled data, but the server can provide them with labels \citep{Liu2023Class}.

\textbf{Federated Alternate Training (FAT)} \citep{Mushtaq2023FAT} alters the training procedure between annotated and unannotated data. On the one hand, the method uses labeled datasets to fine-tune the global model parameters. On the other hand, clients with unlabeled datasets use the global model as a target model to generate pseudo-labels for self-supervised learning. Model aggregation is performed alternately. First, the global model parameters are aggregated for several rounds for clients with annotated data. Then, for the following few rounds, the aggregation is carried out for clients with unannotated data.

In \textbf{Class Imbalanced Medical Image Classification Based on Semi-Supervised Federated Learning} \citep{Liu2023Class}, local clients do not have labels at all, while the labeled data is sent from the central server to each collaborator, allowing them to participate in semi-supervised training. Each client combines the labeled and unlabeled data to construct an extended dataset used to generate pseudo-labels. This method considers the class imbalance factor, which is determined based on annotated data.

\underline{Generative methods:} \textbf{Federated Graph Learning with Network Inpainting for Population-Based Disease Prediction (FedNI)} \citep{Peng2023FedNI} synthesizes data for graph convolutional neural networks (GCN). This method employs a network inpainting module that includes a missing node generator. It aims to predict missing nodes and augment edge connections. As a result, a larger fused local graph is generated, which serves as input to the GCN node classifier.

\underline{Knowledge distillation:} \textbf{Federated Learning with Knowledge Distillation for Multi-organ Segmentation with Partially Labeled Datasets} \citep{Kim2024Federated} involves global and local versions of knowledge distillation. The global knowledge distillation is designed to prevent catastrophic forgetting by preserving predictions for organs with missing labels. In this context, catastrophic forgetting refers to the loss of knowledge about the segmentation of specific organs during local training, which occurs once clients own only partially annotated datasets. In local knowledge distillation, each client trains a model for a specific organ before starting the FL procedure and shares its parameters with collaborators. The client uses predictions from another participant's randomly selected organ-specific model during the local training.

%% ----------------------------------------------
\subsection{Security}
\label{sec:fl_learning:security}
Attacks on privacy can lead to extracting sensitive information from confidential patient data. To enhance data security during the learning process, one can use privacy-enhancing techniques such as \emph{differential privacy}, \emph{split learning} or \emph{selective parameter sharing}.

\underline{Differential privacy:} An example of implementing differential privacy is the \emph{Differentially-private Stochastic Gradient Descent (DP-SGD)} algorithm used by clients during local learning \citep{Ziller2021Differentially, Malekzadeh2021Dopamine, Nguyen2022GenAdv, Kalra2023Decen}. The DP-SDG algorithm injects a Gaussian distributed noise to the averaged minibatch gradients before performing an optimization step. The privacy loss is ultimately calculated based on the information maintained by the accountant. The accountant is a mechanism for tracking and managing the privacy budget across multiple training rounds. It ensures that the total privacy budget remains within acceptable limits while training decentralized clients. In another differential privacy algorithm, namely the \textbf{Differentially Private Federated Deep Learning for Multi-site Medical Image Segmentation} by \cite{Ziller2021Differentially}, the R\'enyi Differential Privacy Accountant technique is used instead of the standard moments accountant.

%% ----------------------------------------------
\underline{Split learning:} This approach involves dividing the model into parts stored separately by different participants. \textbf{Robust Split Federated Learning (RoS-FL)} \citep{Yang2022RosFL} is an example technique that splits the model into three parts: head, tail, and body network. The head and tail models are stored with clients' devices, while the computation server hosts the computationally intensive body networks. This server stores $C$ body models, each belonging to a different client. On top of these, the aggregation server covers only the parameters of head and tail networks, while the computation server aggregates the body model parameters.

%% ----------------------------------------------
\underline{Selective parameter sharing:} Privacy can also be violated by overfitting a locally trained model. The selective sharing of parameters is a solution to handle this problem. As an example, the \textbf{Privacy-preserving Federated Brain Tumor Segmentation} \citep{Li2019TumSeg} selectively shares local information, that is, model components $\Delta{\boldsymbol\Theta}_c$ whose absolute values are greater than the chosen threshold are transferred to the server. In addition, elements of $\Delta {\boldsymbol\Theta}_c$ are clipped to a specific range of $[-\gamma,\gamma]$.

%% ----------------------------------------------
\subsection{Model heterogeneity}
\label{sec:fl_learning:model}

Model heterogeneity can significantly restrict the application of classic FL in medical centers. This limitation arises because it requires all clients to use the same neural network architecture, which is often not feasible due to varying computational resources or differences in the data they handle. One potential solution to this challenge is \emph{model-agnostic FL}, which imposes no specific requirements on client models. This approach typically relies on \emph{knowledge distillation} or \emph{generative methods}.

\underline{Knowledge distillation:} \textbf{Federated Similarity Knowledge Distillation (FedSKD)} \citep{weng2025fedskd} is an aggregation-free method and works in a peer-to-peer manner, enabling bidirectional knowledge exchange among clients through multi-dimensional similarity knowledge distillation (SKD). FedSKD categorizes models into two types: 1) local models, $\mathbf{\Theta}_{c}$, which adapt to local data distribution, and 2) models received from other participants, $\tilde{\mathbf{\Theta}}_{c'}$, that carry cross-client knowledge. In each global round, a client either trains its model locally or receives a model from another institution, selected in a predetermined order, and then performs mutual distillation. The mutual distillation process has two stages. First, domain-specific knowledge injection, in which the local model distils its expertise into the received model using SKD. Second, cross-client knowledge absorption, in which the received model enhances the local model's generalisation capabilities. These steps are repeated until the parameters of all clients' models converge.
The total loss of SKD training for the $r$-th round on a client can be defined as follows:
\begin{multline}
    \mathcal{L}(\mathbf{\Theta}_c^r, \tilde{\mathbf{\Theta}}_{c'}^r; \mathcal{D}_c) ={}
    \mathcal{L}_{\mathrm{CE}}(\mathbf{\Theta}_c^r; \mathcal{D}_c) + \mathcal{L}_{\mathrm{CE}}(\tilde{\mathbf{\Theta}}_{c'}^r; \mathcal{D}_c) \\
    + \lambda
    \mathcal{L}_{\mathrm{SKD}}(\mathbf{\Theta}_c^r, \tilde{\mathbf{\Theta}}_{c'}^r; \mathcal{D}_c),
    \label{eq:FedSKD_loss}
\end{multline}
\noindent
where $\mathcal{L}_{\mathrm{CE}}$ and $\mathcal{L}_{\mathrm{SKD}}$ are the cross-entropy loss and multi-dimensional SKD loss, respectively, and $\lambda$ is a balancing hyperparameter.

\underline{Generative methods:} An example of a generative method is the \textbf{Federated Generative Autoregressive Transformer (FedGAT)} \citep{nezhad2025generative}. This approach facilitates collaboration among clients using different reconstruction model architectures through a two-tier strategy. In the first tier, a GAT prior captures the distribution of MRI data from multiple sites. The GAT prior includes a variational autoencoder that encodes MRI images into discrete token maps at various spatial scales, which are then modeled by an autoregressive transformer to predict feature maps at higher scales.
In the second tier, each institution trains its own preferred reconstruction model on local data and fine-tunes on a hybrid dataset generated by augmenting local MRI data with synthetic images from the global GAT prior.

%% --------------------------------------------------------------------------------------------------
%% --------------------------------------------------------------------------------------------------
%% --------------------------------------------------------------------------------------------------
\section{Federated learning frameworks}
\label{sec:frameworks}
This section reviews publicly available actively developed FL computing frameworks. These software tools provide application interfaces for developing FL algorithms in C++/Python programming languages employing various resource types such as multi-CPU, GPU or cloud-based environments. We briefly characterize these frameworks below and summarize them in Table \ref{table:Frameworks}.

\textbf{APPFL} (Advanced Privacy-Preserving Federated Learning) \citep{li2024advances} is a high-performance framework and benchmarking tool that handles various communication protocols, data transfer and compression methods, and privacy preservation strategies. APPFL supports the Message Passing Interface standard for single-machine and cluster servers and gRPC/Globus Compute for distributed training.

\textbf{FATE} (Federated AI Technology Enabler) \citep{Liu2021FATE} is an open-source tool that provides a secure computing framework to support a federated AI ecosystem. It delivers multiple secure computation protocols for collaboration between the sites involved over large datasets while complying with data protection regulations. FATE supports numerous ML algorithms, including tree-based algorithms, deep learning, and transfer learning. The framework comprises several components, such as an end-to-end platform, a visual tool for exploring FL models, and support for high-performance and cloud computing.

\textbf{FEDML} (Foundational Ecosystem Design for Machine Learning) \citep{He2020FedML} is a cloud service dedicated to developers working with large language models and generative AI. It allows for launching complex model training, deployment, and FL algorithms on decentralized GPUs, multi-clouds, edge servers, and smartphones in a secure and cost-effective way.

\textbf{Fed-BioMed} \citep{cremonesi2025fed} is an open-source research and development initiative aimed at translating collaborative learning, including FL, into practical medical applications. It offers a variety of features, including a demonstrated framework for deploying FL within hospital networks and user-friendly tools for data management and client participation. Additionally, Fed-BioMed facilitates the easy implementation of state-of-the-art methods while ensuring compliance with data providers' privacy and governance requirements.

\textbf{FedTree} \citep{Li2023FedTree} is a specialized tool for FL systems using tree-based models. It is designed to be highly efficient, effective, and secure. It delivers a range of features, including parallel computing on multi-core CPUs and GPUs, support for HE, secure aggregation, and DP.

\textbf{Flower} \citep{Beutel2022Flower} is a highly customizable framework with many components that can be used to develop new or state-of-the-art systems. Flower supports multiple ML libraries, including PyTorch, Keras, TensorFlow, and Scikit-learn. It is a platform-agnostic framework that supports Android, iOS, mobile, desktop, and cloud computing.

\textbf{FLUTE} (Federated Learning Utilities for Testing and Experimentation) \citep{Garcia2022FLUTE} is a high-performance PyTorch-based tool for rapid prototyping and validation. FLUTE features standard optimizers and aggregation methods and handles local/global differential privacy. The framework supports single-/multi GPU and cloud computing.

\textbf{LEAF} \citep{Caldas2019LEAF} is a comprehensive benchmarking tool designed for learning in federated settings. The LEAF can be used in various domains, such as multi-task learning, meta-learning, and on-device learning. The LEAF has several open-source datasets, and it also provides a comprehensive evaluation framework and a set of reference implementations.

\textbf{NVIDIA FLARE} (NVIDIA Federated Learning Application Runtime Environment) \citep{Roth2022NVIDIA_FLARE} is an open-source and versatile software development kit for FL. NVIDIA FLARE supports ML libraries such as PyTorch, TensorFlow, Scikit-learn, and XGBoost. It also supports horizontal and vertical FL and has built-in algorithms like FedAvg, FedProx, FedOpt, and Scaffold. NVIDIA FLARE offers several server and client-controlled training workflows, including scatter and gather, cyclic, and validation workflows such as global model evaluation and cross-site validation. It also incorporates privacy preservation techniques like DP, HE, and private set intersection to ensure data privacy and security.

\textbf{OpenFed} \citep{Chen2023OpenFed} is a library that makes FL easier for researchers and downstream users by addressing existing challenges. OpenFed provides a framework for researchers to quickly implement new methods and evaluate them against various benchmarks. For downstream users, OpenFed enables FL to be seamlessly integrated into different subject contexts without requiring in-depth expertise in the field.

\textbf{OpenFL} (Open Federated Learning) \citep{Foley2022OpenFL} is a library designed to be a flexible, extensible, and easy-to-learn tool for data scientists. OpenFL allows the setup of an FL experiment using one of three workflows: director-based (recommended for FL research), aggregator-based (helpful in defining and distributing experiments manually), or workflow interface (for complex experiments). The library also provides tutorials covering various frameworks, models, and datasets.

\textbf{PaddleFL} \citep{PaddlePaddle2023PFL} is an open-source framework that allows researchers to replicate and compare various FL algorithms easily. PaddleFL provides several FL strategies for computer vision, natural language processing, and recommendation systems.

\textbf{PySyft} \citep{Ziller2021PySyft} is an open-source stack built on top of PyTorch. PySyft provides a numpy-like interface and integrates with DL frameworks, allowing researchers to continue working with their current workflows while benefiting from enhanced privacy features.

\textbf{Substra} \citep{Owkin2024Substra} is open-source and ready-to-use FL software to train and validate ML models on distributed datasets. Substra offers a flexible Python interface and a web application for FL training. This framework is mainly used in production environments and has already been successfully deployed and used by biotech companies and hospitals for clinical research, drug discovery, and development tasks.

\textbf{TFF} (TensorFlow Federated) \citep{Google2024TFF} is an open-source framework for ML and other computations on decentralized data. The framework offers starting points and complete examples for various types of research. The building blocks provided by TFF can also be exploited for non-learning computations, such as federated analytics.

%% --------------------------------------------------------------------------------------------------
%% --------------------------------------------------------------------------------------------------
%% --------------------------------------------------------------------------------------------------
%% --------------------------------------------------------------------------------------------------
%% --------------------------------------------------------------------------------------------------
%% --------------------------------------------------------------------------------------------------
\section{Real-world implementations of federated learning in medical imaging}
\label{sec:real-world}

The application of FL in MI is predominantly limited to simulations rather than real-world deployments. In these scenarios, data from a single source is typically experimentally divided to mimic the local datasets of individual clients \citep{Malekzadeh2021Dopamine, Stripelis2021SecureNeuro, Islam2022Effectiveness, gupta2025applying}. Alternatively, data from various institutions or publicly available datasets is handled to simulate multi-sites \citep{Jiang2023IOP-FL, Dalmaz2022OneModel, Tan2024Joint, Wang2024FedDUS}. However, the clients in these simulations are not authentic medical infrastructures – they are virtual nodes or independent computing processes operating on a centralised research cluster. This means that the federation is simulated in a controlled manner, bypassing the complexities of real clinical systems.

Recently, several real-world implementations of FL in clinical settings have appeared. These developments provide an opportunity for a more thorough analysis and understanding of the practical challenges associated with implementing FL in clinical environments, particularly those related to the General Data Protection Regulation (GDPR) and the technical aspects of hospital infrastructure.

%% ----------------------------------------------
\subsection{Clinical cases of federated learning}
\label{sec:real-world:applications}

\uline{Prediction of COVID-19 clinical outcomes:} \cite{dayan2021federated} \cite{dayan2021federated} developed EXAM, a global FL model to predict the future oxygen requirements of symptomatic COVID-19 patients. The model combines two types of data: 1) chest X-ray images and 2) Electronic Medical Record (EMR) data. The FL architecture employs a client-server architecture, with the model implemented in TensorFlow (\url{https://www.tensorflow.org/}) using the NVIDIA Clara Train SDK (\url{https://docs.nvidia.com/clara/index.html}). This project involved 20 institutions across four continents. Due to variations in device manufacturers, imaging protocols and demographics, the datasets were heterogeneous, concerning both X-ray images and EMR feature distributions. The final EXAM model achieved an average Area Under the Curve (AUC) greater than 0.92, indicating a 16\% improvement in AUC and a 38\% increase in generalizability compared to models trained locally at individual sites, with significant benefits for institutions with small or unbalanced datasets.

\underline{Rare cancer boundary detection:} \cite{Pati2022BigData} implemented an FL framework to detect the sub-compartment boundaries of glioblastoma using multi-parametric MRI scans.
The FL architecture features a central aggregation server at the University of Pennsylvania, securely connected to all collaborating sites. Communication during the training process is protected using the Transport Layer Security cryptographic protocol, while datasets are aggregated within Trusted Execution Environments using the Intel Software Guard Extensions (SGX; \url{https://www.intel.com/content/www/us/en/developer/tools/software-guard-extensions/overview.html}) for confidentiality and integrity. This project includes 71 sites across six continents, with a standardised data preprocessing pipeline that accounts for variations in scanner hardware and acquisition protocols. The final consensus model demonstrated significant improvements over the initially publicly trained model, with 33\% improvement in delineating the surgically targetable tumour and 23\% improvement in defining the complete tumour extent.

\underline{Lung pathology segmentation:} \cite{bujotzek2025real} demonstrated a proof-of-concept training of segmentation models for lung pathology detection using CT data in an FL environment.
The datasets contained a variety of lung pathologies and exhibited significant differences in annotation procedures, scanner manufacturers, and acquisition protocols. The results consistently demonstrated that FL approaches outperformed local models in all evaluation scenarios.
The experiment was conducted across six university hospitals in Germany: Charité Berlin, Technical University of Munich, University Medicine Essen, and the hospitals in Frankfurt am Main, Cologne, and Kiel. 

\underline{Breast cancer classification:} \cite{tzortzis2025towards} implemented an end-to-end FL framework for breast cancer classification, specifically targeting BIRADS classification (Breast Imaging Reporting and Data System) with mammography data. This project infrastructure, developed under the INCISIVE European Project (\url{https://incisive-project.eu/}), consists of a central cluster and local clusters at hospitals. Clients operate on-site using the infrastructure that meets specified minimum requirements. Built on Kubernetes, the framework uses a standard API for communication, enabling centralized control and simplifying maintenance for hospital staff. This study involved three hospital institutions: the Hellenic Cancer Society in Greece, the University of Novi Sad in Serbia, and Aristotle University of Thessaloniki in Greece. The datasets were heterogeneous and contained artefacts, including breast implants, surgical clips, and annotation labels.
As a result, the model performance improved by approximately 35\% at the first participating hospital, aligning the FL-based results with those of a centralized model.

\underline{Cardiac CT imaging:}  \cite{tolle2025real} implemented an FL system to enhance the analysis of cardiac CT imaging for patients undergoing transcatheter aortic valve implantation. The project uses a centralized FL approach with a central server located at Heidelberg University. Different institutions used the same physical machine for training, although their network configurations varies.
This study involved eight cardiology and radiology departments, all affiliated with the German Center for Cardiovascular Diseases (DZHK). Datasets varied in sample count and acquisition protocols, with only a small fraction labelled, necessitating a semi-supervised federated knowledge distillation strategy. The resulting model showed improved generalizability in downstream tasks, such as coronary artery segmentation, indicating a better understanding of the global context.

%% ----------------------------------------------
\subsection{Challenges in deploying federated learning in a clinical scenario}
\label{sec:real-world:issues}

Implementing FL-based infrastructure in a real clinical practice demonstrated numerous organisational, technical, and legal challenges, requiring significant effort and specific mitigation strategies.

Data heterogeneity has emerged as the major challenge for all studies attempting to deploy FL in clinic settings. This issue arises from differences in patient demographics, variations in imaging protocols and acquisition equipment (such as different scanner manufacturers) \citep{tzortzis2025towards, tolle2025real}. Additionally, data quality inconsistencies and a lack of data acquisition standardization across institutions contribute to this problem \citep{dayan2021federated, Pati2022BigData, bujotzek2025real}. As a result, these factors lead to decreased accuracy and limit the network's generalizability. \cite{tzortzis2025towards} observed that data heterogeneity significantly reduced the performance of the global model at one hospital, noting a nearly 40\% decrease in predictive performance, as explained by the F1 score, compared to the standard deep learning.
In most cases, the authors sought to minimise this problem by focusing on data preparation, such as detailed data specifications \citep{bujotzek2025real}, targeted pre-processing procedures \citep{tolle2025real} or data harmonization \citep{tzortzis2025towards}. The inability to access the raw training data directly complicated the debugging procedure for unsatisfactory outcomes \citep{dayan2021federated, Pati2022BigData}. As a result, the authors recommended appointing a Data Manager or a Curator at each institution to handle local dataset curation and ensure compliance with data-related requirements \citep{tzortzis2025towards}.

Another significant challenge involves the infrastructure and logistics. Implementing an FL-based system often requires installing a specialised platform within highly restricted clinical information technology environments, enabling local model training while maintaining secure, compliant communication with a central server \cite{bujotzek2025real}.
The authors therefore emphasise the importance of designing frameworks that are easy to deploy, potentially leveraging tools such as Kubernetes, and that are sufficiently intuitive for routine use by hospital staff \citep{tzortzis2025towards}. Other identified challenges include disparities in institutional resources, including hardware capabilities and network bandwidth, on-site computational power and storage management, and the integration with third-party systems, such as PACS and annotation tools \citep{bujotzek2025real, tolle2025real}. To address some of these issues, \cite{tolle2025real} standardized the infrastructure by providing identical hardware to all participating institutions. However, differences in network configurations and software installations across sites continued to create difficulties. Infrastructure instability is also a challenge, with sites dropping out due to issues such as straggling, crashes, and disconnected nodes. These issues require experiment restarts, leading to idle machines at other locations \citep{bujotzek2025real}. Asynchronous FL is therefore recommended to effectively manage nodes with heterogeneous resources \citep{tzortzis2025towards}.

Security, confidentiality, and legal compliance have posed significant barriers to implementing FL in clinical settings. Healthcare applications, especially in the European Union and the United States, must adhere to strict regulations such as the GDPR and the Health Insurance Portability and Accountability Act (HIPAA) \citep{Pati2022BigData, tzortzis2025towards}.
While the FL principle generally aligns well with data protection laws, FL-based systems are not automatically compliant with the GDPR or similar regulations. Simply keeping data locally does not provide sufficient privacy warranties \citep{Truong2021Privacy}. Besides, the model parameters shared among participants can retain sensitive information that may be exploited in advanced attacks, such as model inversion \citep{Truong2021Privacy, tolle2025real, tzortzis2025towards}. To address these issues in practice, collaborative networks use signed data-sharing agreements among trusted partners \citep{tzortzis2025towards}. Technical mitigations include employing Transport Layer Security, username/password authentication, and server-side IP address whitelisting to ensure secure communication \citep{tolle2025real}. 
Security-enhancing features, such as partial weight sharing, to minimise the risk of information leakage from model gradients, have also been examined \citep{dayan2021federated}. Likewise, clients are required to perform privacy impact assessments before joining FL networks \citep{tzortzis2025towards}.

%% --------------------------------------------------------------------------------------------------
%% --------------------------------------------------------------------------------------------------
%% --------------------------------------------------------------------------------------------------

%% --------------------------------------------------------------------------------------------------
%% --------------------------------------------------------------------------------------------------
%% --------------------------------------------------------------------------------------------------

\begin{figure*}[!t]
\centering
\includegraphics[scale=0.111]{./fig/Issue-method-effect_all.pdf}
\caption{Issue–method–effect diagram illustrating how different aggregation strategies (top graph) and learning methods (bottom graph) address key challenges in federated learning. The color-encoded lines indicate the relationships between specific issues, the methods used to mitigate them, and the resulting benefits. Methods that address multiple challenges are highlighted with multiple colors and connections.}
\label{fig:Issue_method_effect_aggregation}
\end{figure*}

\section{Discussion}
\label{sec:discussion}
We have been observing an enormous interest in FL over the past decade, resulting in a full gamut of new aggregation and learning techniques, as previously mentioned in sections~\ref{sec:fl_aggregation} and \ref{sec:fl_learning}. We analyzed distributions of these methods, categorized them by imaging modalities and deep learning tasks (see Fig. \ref{fig:Data_task_aggregation}). These graphs lead to some interesting conclusions. 1)~The standard FedAvg and FL-EV are leading the way to aggregate the models, accounting for more than half of medical applications across modalities and tasks. 2)~The same aggregation schemes have been used in many contexts, which highlights their universal applicability, although it should be noted that not every algorithm might be suitable for every task and/or modality. 3)~Concerning the learning procedures employed in the federated learning system, aside from fully supervised training, the most frequently used strategies across modalities and tasks are split learning and knowledge distillation. 4)~Interestingly, the usability of various techniques may be (potentially) limited in some particular applications. For instance, in a translation/synthesis task, fully supervised training and split learning have been used so far.

Figure \ref{fig:Issue_method_effect_aggregation} provides a concise overview of aggregation strategies and learning methods, highlighting the problems they aim to solve and summarizing their main effects. However, these approaches often only partially address the problems or, not surprisingly, create new ones that require an in-depth examination. In this section, we thoroughly discuss these issues before MI can fully leverage the benefits of FL.

%% ----------------------------------------------
\subsection{Model accuracy}
\label{sec:discussion:accuracy}
We start our discussion with some important comments about the accuracy of the global model. Once training the global model is initiated, one has to ensure that the involved sites do not negatively affect the aggregation function \citep{Darzidehkalani2022Federated}. Usually, this is achieved by dynamically assigning a weighting factor to each client based on its local loss function \citep{Shen2021DWA, Machler2022FedCostWAvg, Machler2023FedPIDAvg, Wicaksana2023FedMix, Wu2023ModFed, Liu2023MultSkle} or accuracy-related metrics \citep{Yang2023DynAggr}. The weighting factor may also cover the similarity between the parameters of transferred local and global models \citep{Khan2022SimAgg, Khan2023RegSimAgg, Yang2023DynAggr}. Another approach is to abandon specific parameters of the local models during the aggregation process. The parameters to be withdrawn might be randomly selected \citep{Gunesli2021FedDropoutAvg} or chosen from specific layers (e.g., normalization layer) \citep{Bernecker2022FedNorm}.

To prevent global model deterioration, it is fundamental to avoid forgetting previously learned knowledge. One can achieve this through the curriculum learning \citep{Sanchez2023Memory}, where inconsistent predictions caused by forgotten data are penalized. An alternative approach is continuous learning, which introduces and manages new memories without forgetting the previously gained knowledge \citep{Sheller2020Federated, Huang2022Continual}. The consequence of catastrophic interference is even more critical as the training might introduce a bias towards the most recent clients involved in the training procedure. To handle the bias favoring new clients, one can use the cyclic weight transfer approach, which trains the model at the client and then transfers model parameters to the next one \citep{Darzidehkalani2022Federated}. Dynamic weight correction is another technique that stabilizes the training process and avoids a drift between the global model and local models \citep{Yang2022RosFL}.

Lastly, some aggregation and learning methods require computing additional parameters during local training and transferring them to the global server. Examples include MixFedGAN \citep{Yang2023DynAggr} and Memory-aware Curriculum Federated Learning \citep{Sanchez2023Memory}. These prerequisites might be problematic for hospitals with limited computing resources and/or weaker communication infrastructure. Therefore, finding alternative methods to ensure the federated-based DL system properly works for all clients involved is essential in the near future.

%% ----------------------------------------------
\subsection{Data heterogeneity}
\label{sec:discussion:heterogeneity}
The most commonly used method to aggregate MI data is the FedAvg \citep{McMahan2017FedAvg}. However, the FedAvg-derived results strongly depend on the homogeneity of the data between the sites. If the data is heterogeneous, the accuracy of the global model decreases, and the model performs satisfactorily only for some clients \citep{Zhao2018Non-IID}. However, although the FedProx is a typical option for MI \citep{Li2020FedProx}, it does not work adequately in all scenarios and therefore does not necessarily always show significantly better results than the FedAvg \citep{Luo2022FedSLD, Zhou2023FedFTN, Yang2023DynAggr}. FedBN \citep{Li2021FedBN} is often considered a baseline strategy for handling data heterogeneity by excluding the parameters of batch normalisation layers during aggregation. This method can reduce distributional shifts across clients. However, its effectiveness may vary depending on the amount of local data available, as batch normalisation statistics are computed from local samples.
To ensure the best possible results, participants, before training starts, must find a consensus on the criteria for selecting the best possible global model, such as the metric chosen at a satisfactory level or the number of global rounds that must be passed \citep{Darzidehkalani2022Federated}.

Nowadays, a popular way to handle data heterogeneity is to slightly alter the concept of FL by training many personalized local models simultaneously instead of a single global model. This is usually achieved by splitting the local model into two parts: a local part being specific to each client and a global one, transmitted to the server and used in the aggregation procedure (see Fig.~\ref{fig:Split_learning}E;  \cite{Zhang2022SplitAVG, Dalmaz2022MRITrans, Dalmaz2022OneModel, Elmas2023FedGIMP, Feng2023FedMRI, Zhou2023FedFTN}). Alternatively, it is possible to separate the parameter space into shape and appearance features and share shape parameters with the server only \citep{Bercea2022FedDis}. These strategies effectively address the data heterogeneity problem resulting from engaging different devices and protocols for data acquisition. We note that training personalized local models can result in challenges in generalizing these models to other datasets not included in the training procedure. A more straightforward approach to handling data heterogeneity is fine-tuning the global model according to the client's needs using local data \citep{Jiang2023IOP-FL}.

Another possible solution to handle data heterogeneity is to harmonize the data or standardize the pre-processing pipeline across the collaborators \citep{Jiang2022Harmofl}. These may require sharing metadata between the institutions or even exemplary acquisitions to set up a harmonization protocol. Precisely, data harmonization protocols typically require selecting a target institution to which the data is normalized, and hence, sharing exemplary acquisitions with other clients is favorable. This might be challenging to achieve, considering the privacy constraints. Therefore, the cooperation of clinicians and technicians is essential to standardize the way data is handled in different institutions \citep{Darzidehkalani2022Federated}. It is also beneficial to create model datasets that can serve as a target site for harmonization and to establish universal metrics used to determine the quality of harmonized data \citep{pinto2020harmonization}.

Yet another practical approach to deal with data heterogeneity is direct \citep{Guo2021FL-MR} or indirect \citep{Liu2021FedDG, Luo2022FedSLD, Zhao2023ADI, li2025retrospective} data sharing (see Fig.~\ref{fig:Data_sharing}). This strategy addresses the problem from a statistical point of view by sharing the distribution of classes in local data or a small amount of data that contains examples from each class. It enables clients to augment their datasets and create an independently and identically distributed (IID) dataset.

However, these kinds of solutions are computationally expensive and result in significant communication overheads \citep{Aouedi2022Handling}. In addition, the size and distribution of classes in the data may change over time, demanding frequent modifications to local datasets to remain independently and identically distributed in relation to other institutions. Finally, more efforts could be put into decreasing the data heterogeneity by integrating demographic factors such as gender, age or intelligence \citep{bedford2020large}.

The current literature still lacks in-depth analyses of new techniques that involve heterogeneous datasets. Numerous reports rely on a single dataset acquired from one institution, which is then ``virtually" shared among clients, creating a simulated FL scenario. In the future, it will be vital to standardize validation methods for assessing new aggregation and learning algorithms, together with collecting standardized datasets for testing procedures.

%% ----------------------------------------------
\subsection{Malicious clients}
\label{sec:discussion:malicious_clients}
Most FL-based techniques assume that all clients involved in the training process are credible and do not intentionally modify the parameters of transmitted local models. If not, the global modal is susceptible to poisoning attacks. Defenses against model poisoning attacks mainly rely on either Byzantine-robust methods or provably robust methods. In general, these methods aim to train a well-functioning global model even in the presence of malicious clients \citep{Zhang2022FLDetector}.

Theoretically, Byzantine-robust FL methods can minimize the changes to the global model parameters caused by malicious clients. They employ a technique that involves correlating the local updates from each client and filtering out any statistical outliers before the aggregation. However, these methods treat every client as a potential threat since there is no reliable way to determine which local updates are suspicious \citep{Alkhunaizi2022DOS, Cao2022FLTrust, Zhang2022FLDetector}. Therefore, the methods are only effective for a limited number of malicious clients. Provably robust methods, in turn, can guarantee the minimum level of testing accuracy. However, these methods require a clean validation dataset available on a server with data distribution properties similar to the overall training data distribution. In a typical FL scenario, the server may have restricted access to such a validation dataset. Thus, the global model may still be eroded by malicious clients \citep{Zhang2022FLDetector, Onsu2023Cope}. 

In conclusion, new methods must identify and reject malicious clients to protect FL-based learning procedures against poisoning attacks, which are particularly dangerous in a clinical setting.

%% ----------------------------------------------
\subsection{Security}
\label{sec:discussion:security}
The most often exploited methods to preserve data privacy are DP, secure MPC, and HE. The first method, the DP, introduces random noise to the dataset to reduce the risk of revealing sensitive personal information. The commonly used algorithm for training local models is the DP-SGD \citep{Ziller2021Differentially, Malekzadeh2021Dopamine, Nguyen2022GenAdv, Kalra2023Decen}. However, this technique may also reduce the global model accuracy \citep{Truong2021Privacy, Abad2022Security}. The trade-off between security and neural network performance vastly depends on the model architecture and the privacy regime. According to studies by \cite{Ziller2021Differentially} and \cite{Nguyen2022GenAdv}, the reduction in accuracy typically ranges from about 5 to 15 percentage points. Such a decrease may be unacceptable for models intended for clinical practice. In contrast to DP, the MPC preserves high global model accuracy. It enables multiple sites to compute a function with their inputs without revealing them to other parties \citep{Mothukuri2021Survey, Truong2021Privacy}. HE is yet another technique that can improve data privacy and is often used for secure aggregation \citep{Malekzadeh2021Dopamine, Makkar2023SecureFedFL}. Here, the computations are performed on encrypted data, and the results can only be decrypted by the institution that requests it \citep{Fereidooni2021SAFELearn, Truong2021Privacy}. However, the MPC and HE increase the computational burden and complicate the training algorithm, consequently significantly increasing the local training duration \citep{Mothukuri2021Survey, Truong2021Privacy}. Research on the application of HE in FL for MI shown that the time required for joint model training can increase up to six times compared to FL without HE \citep{lessage2024secure}. Additionally, even small changes to the model architecture, such as adding a single layer, can increase inference time by up to eight times when HE is used \citep{dutil2021application}.

Contrary to the DP, secure MPC and HE mentioned above, there are also other solutions that can be used to preserve data security. One of these approaches is to transfer local model parameters selectively, thus preventing reverse engineering \citep{Li2019TumSeg}. Another solution is to split the local model into separate parts and store them independently. Hence, since the model is only partially stored by the global server and the involved clients, the ability to retrieve its parameters by other sites or an unwanted external party is limited \citep{Yang2022RosFL}.

However, this technique increases the learning period as more parties are involved in the training procedure. Other solutions are also no less relevant: peer-to-peer \citep{Giuseppi2022FedLCon} and blockchain \citep{Kalapaaking2023Block, Noman2023Block}.

Due to the high prerequisites of privacy protection techniques, it is essential to consider whether it is necessary or not to implement them in the FL-based system. Profitability depends largely on the level of trust between the parties involved and the scale of the project \citep{Darzidehkalani2022Federated}. If the institutions trust each other or the training involves a small group of clients who can sign appropriate data protection agreements, employing the above methods might be unfounded. Nevertheless, this does not exempt the clients from protecting their data from other parties, i.e., clarifying the required level of trust among clients is paramount before introducing a federated learning architecture \citep{Dasaradharami2023Comprehensive}.

To sum up, there is a need to test privacy protection methods on highly heterogeneous datasets, like those observed across multi-institutional medical modalities. Current requirements for security preservation are directed toward developing techniques that can provide security without significantly increasing the computational cost.

%% ----------------------------------------------
\subsection{Label deficiency}
\label{sec:discussion:label_deficiency}
Three main approaches address the problem of insufficient data labeling: semi-supervised, self-supervised, and transfer learning \citep{Jin2023Federated}.

Semi-supervised learning aims to use both annotated and unannotated data owned by clients. This approach is practical only if some of the clients involved have unlabeled data \citep{Mushtaq2023FAT} or all the clients have unannotated data, but the labels are available on the central server \citep{Liu2023Class}. In such cases, the central server shares the labeled data to fine-tune the global model by each client. Therefore, the fine-tuned model can generate pseudo-labels for unannotated data on the clients' side \citep{Jin2023Federated}.

Self-supervised learning is a dedicated strategy used when the server cannot access labeled data and the institutions own unannotated data only. There are two types of self-supervised learning: generative and contrastive learning. However, in FL, attention has been focused mainly on the latter type \citep{Wu2021FCL, Dong2021Federated}. Contrastive learning involves training a model to differentiate between similar and different pairs of data collection by maximizing their similarity within the same class and minimizing similarity between different classes. However, playing with heterogeneous data is challenging in contrastive learning, as there are no globally consistent labels, and local contrastive objectives can vary significantly \citep{Jin2023Federated}.

Lastly, the concept of federated transfer learning (Fig. ~\ref{fig:HFL,VFL,FTL}C.) can be categorized into two cases: homogeneous and heterogeneous. Homogeneous transfer learning can be further classified into single- and multi-source architectures, depending on the number of source and target domains involved \citep{Jin2023Federated}. In a single-source architecture, there is a central server that maintains labeled data and clients with unannotated data. Since each institution has different data distribution (i.e., the data is heterogeneous across the involved sites), it creates a unique target domain, requiring adaptation methods to tackle multiple targets. 
The multi-source setting is much more complicated as it involves multiple source domains with labeled data transferring knowledge to a single unlabeled target domain. In this case, selecting a source domain with valuable knowledge is required without directly observing the source data. On the other hand, to make the heterogeneous transfer learning work, there must be cross-domain links between data to bridge the gap between the different feature spaces (e.g., cross-domain links between imaging modalities used for scanning the patient).

%% ----------------------------------------------
\subsection{System architecture}
\label{sec:discussion:architectures}
We discuss here two main architectures commonly used in FL: centralized and decentralized (see Fig.~\ref{fig:CFLvsDFL}). The former is a frequent choice in MI imaging but has weaknesses, i.e., even minor failures, like network instability, can cause a bottleneck and halt the entire global training procedure \citep{Brecko2022Federated}. Besides, this architecture has a security risk, as clients must trust each other. The latter architecture might obey a peer-to-peer or blockchain approach. The peer-to-peer architecture eliminates the risk of a single-point failure, but the global model accuracy depends heavily on the network topology. The blockchain, in turn, significantly enhances data security but is computationally expensive as it additionally requires data validation and block generation \citep{Brecko2022Federated, Aouedi2022Handling, Gammar2023Securing}.

Many institutions lack solid computing resources and reliable network infrastructure \citep{Darzidehkalani2022Federated}. Hence, redundant computing facilities and data centers are critical to prevent data loss. An effort in architecture planning should be made to handle the failure of one or more centers without disrupting the overall training process. The FL system must be adaptable to operate with various centers in case the institution is no longer involved in the training and maintaining the global model to solve a particular MI task.

%% ----------------------------------------------
\subsection{Model heterogeneity}
\label{sec:discussion:model}

The infrastructures involved in a FL-based system typically provide different computational resources, leading to the choice of neural network architectures tailored to specific hardware capabilities or application needs \citep{afonin2022modelagnostic, nezhad2025generative, weng2025fedskd}. Attempting to average the parameters of models with different architectures using standard FL methods, such as FedAvg, often fails. Even when models enclose similar structures, challenges can arise due to the non-convex loss landscapes \citep{afonin2022modelagnostic}. Consequently, there is a demand to shift the paradigm from merely transmitting raw parameters to sharing derived knowledge. The two most commonly used strategies for handling this issue are knowledge distillation \citep{weng2025fedskd} and generative methods \citep{nezhad2025generative}.

Knowledge distillation protocols employ the model-agnostic nature of prediction outputs to transfer knowledge between models. This involves training a student model on the predictions made by a teacher model rather than aligning weights \citep{afonin2022modelagnostic, weng2025fedskd}. However, some methods require extra access to publicly available datasets to facilitate knowledge distillation, which may not always be accessible \citep{wu2022communication}. Further, knowledge distillation techniques suffer from sequential information loss, potentially leading to reduced model accuracy \citep{afonin2022modelagnostic}. In contrast, methods that mitigate catastrophic forgetting often come with significantly higher computational costs \citep{weng2025fedskd}.

An alternative is to employ generative priors collaboratively trained to capture the data distribution across multiple sites. Each client can use this prior locally to generate synthetic data. This approach enables clients to train their selected heterogeneous networks independently while benefiting from globally informed and augmented datasets \citep{kang2025gefl, nezhad2025generative}. However, generative methods, such as adversarial or diffusion models, often face challenges, such as unstable training, high inference or sampling costs \citep{nezhad2025generative}. Additionally, these methods may encounter scalability issues, as achieving stable convergence becomes increasingly complex with a higher number of clients \citep{kang2025gefl}.

To effectively address model heterogeneity, it is necessary to develop methods that are accessible to hospitals and medical centres, which often work with limited computational resources and may lack access to extensive external datasets. Data heterogeneity, discussed in section \ref{sec:discussion:heterogeneity}, further complicates model-agnostic FL by limiting the transferable knowledge between institutions \citep{afonin2022modelagnostic}.

%% ----------------------------------------------
\subsection{Communication efficiency}
\label{sec:discussion:communication}
Communication with clients reporting delayed results is one of the most consequential problems to consider in the FL-based system. The direct and practical strategy is to ignore such clients if they cannot provide the model's parameters in a given time regime, as done in the FedAvg algorithm \citep{McMahan2017FedAvg}. However, this can lead to significant biases in the learning process. Contrary to FedAvg, FedProx \citep{Li2020FedProx} allows clients to perform only fragmentary work within system constraints, facilitating partial update to be safely incorporated via a proximal term \citep{Aouedi2022Handling}. However, there is still a lack of comprehensive analysis in the literature on how removing specific clients involved in the aggregation procedure affects the accuracy of the global model. 
Another approach to address the issue of out-of-sync and client dropout is to randomly select clients from a large pool of participants for each training round \citep{Wang2020Autom, Shen2021DWA, Feki2021X-ray, Malekzadeh2021Dopamine, Slazyk2022CXR-FL, Zhang2022SplitAVG, Khan2022SimAgg, Khan2023RegSimAgg, Machler2023FedPIDAvg, Lin2023Hyper}, or choose the subset of clients with the best training results \citep{Zhang2021Dynamic} or those with updates that are the most informative \citep{Zhou2022Comm-effi}. The server can also accept model parameters submitted by a single client that has achieved the best value for a specific measure, such as accuracy or loss function \citep{Kandati2023Federated}. In more advanced techniques, the central server collects additional information about the client's resources, such as wireless channel states, computational capabilities, and dataset size. The server then uses this knowledge to determine which clients can participate in the training process \citep{Aouedi2022Handling}. However, there is still a risk of significant deterioration in the accuracy of the global model, especially when working with highly heterogeneous data. 
Eventually, to improve communication efficiency, a compression framework can be employed to reduce the size of the messages exchanged in each round \citep{He2022CosSGD}. Various model compression schemes, such as sparsification, quantization, or subsampling, can significantly minimize the size of model parameters while achieving high accuracy at low communication costs. Nevertheless, these methods may require more computational resources as the clients must perform the compression procedure locally in addition to model training.

%% ----------------------------------------------
\section{Conclusions}
\label{sec:conclusions}
This article thoroughly reviewed the latest developments in federated learning, expressly focusing on medical image analysis applications. The paper classified newly introduced aggregation and learning techniques specifically designed to address the challenges attributed to federated learning that differentiate it from standard deep learning pipelines. These challenges are incredibly significant in the case of multi-institutional medical imaging data delivered with different equipment and/or vendors and cover issues such as accuracy of the jointly trained model, data and model heterogeneities, label deficiency, the occurrence of malicious clients, data security, system architecture, and communication efficiency across the involved sites. The study has inspected open-source software packages that allow for rapid prototyping of a deep learning algorithm in a federated way using various computational supplies. Finally, we explore real-word implementations of FL-based systems and discuss the challenges in deploying these systems in clinics. We believe that the federated learning concept is a giant step in demarcating the future of deep learning involving multi-institutional medical imaging data.

%---------------------------------------------------------------------------------------------------------
%---------------------------------------------------------------------------------------------------------
\section*{Competing interests}
None Declared.

%---------------------------------------------------------------------------------------------------------
%---------------------------------------------------------------------------------------------------------
\section*{Declaration of generative AI in scientific writing}
The authors did not use generative AI in writing of this paper.

%---------------------------------------------------------------------------------------------------------
%---------------------------------------------------------------------------------------------------------
\section*{CRediT authorship contribution statement}
\textbf{Dominika Ciupek:} Conceptualization, Methodology, Investigation, Resources, Software, Writing - Original Draft, Writing – review \& editing, Visualization; \textbf{Maciej Malawski:} Conceptualization, Writing - Original Draft, Writing – review \& editing, Supervision, Funding acquisition; \textbf{Tomasz Pieciak:} Conceptualization, Methodology, Investigation, Resources, Writing - Original Draft, Writing – review \& editing, Supervision, Project administration.

%---------------------------------------------------------------------------------------------------------
%---------------------------------------------------------------------------------------------------------
\section*{Acknowledgements}
This work is supported by the European Union’s Horizon 2020 research and innovation programme under grant agreement No 857533 and the International Research Agendas Programme of the Foundation for Polish Science No MAB PLUS/2019/13. This publication was created within the project of the Minister of Science and Higher Education "Support for the activity of Centers of Excellence established in Poland under Horizon 2020" on the basis of the contract number MEiN/2023/DIR/3796. Tomasz Pieciak acknowledges the Polish National Agency for Academic Exchange for grant PPN/BEK/2019/1/00421 under the Bekker programme and the Ministry of Science and Higher Education (Poland) under the scholarship for outstanding young scientists (692/STY/13/2018).

\bibliographystyle{elsarticle-harv}
\bibliography{bibliography}

\end{document}